\documentclass[preprint, 5p]{elsarticle}
\title{Machine-learning potentials enable predictive \emph{and} tractable\\high-throughput screening of random alloys}
\author[]{M.~Hodapp\fnref{fn1}\corref{cor1}}
\author[]{A.~Shapeev\fnref{fn2}}
\fntext[fn1]{\textit{E-mail address:} m.hodapp@skoltech.ru}
\fntext[fn2]{\textit{E-mail address:} a.shapeev@skoltech.ru}
\cortext[cor1]{Corresponding author}
\address{\vspace{0.5em}Skolkovo Institute of Science and Technology (Skoltech), Center for Energy Science and Technology, Moscow (RU)}
\journal{Physical Review Materials}
\usepackage{amssymb}
\usepackage{mathrsfs}
\usepackage{amsthm}
\usepackage[nonumberlist,nopostdot,sort=def,nowarn,acronym]{glossaries} 
\usepackage{glossary-mcols} 
\usepackage{color,upgreek}
\usepackage[ruled,lined]{algorithm2e}
\usepackage{setspace}
\usepackage{pifont}
\usepackage{array}
\usepackage{multirow}
\usepackage{multicol}
\usepackage{arydshln}
\usepackage{titlesec}
\usepackage{appendix}
\usepackage{fancybox,framed} 
\usepackage{manfnt} 
\usepackage{scrextend}
\usepackage{float}
\usepackage{enumitem}
\usepackage{mathtools} 
\usepackage{textcomp} 
\usepackage[bottom]{footmisc}
\usepackage{xcolor}
\usepackage{hyperref}
\newcommand{\sty}[1]{\boldsymbol{#1}}

\newcommand{\ubar}[1]{\mkern 0.5mu\underline{\mkern-0.5mu#1\mkern-0.5mu}\mkern 0.5mu}
\newcommand{\uubar}[1]{\ubar{\ubar{#1}}}
\let\mepsilon\epsilon
\let\epsilon\varepsilon
\let\mtheta\theta
\let\theta\vartheta

\let\rho\varrho

\let\phi\varphi
\let\Gamma\varGamma
\let\Delta\varDelta
\let\Theta\varTheta
\let\Lambda\varLambda
\let\Xi\varXi
\let\Pi\varPi
\let\Sigma\varSigma
\let\Upsilon\varUpsilon
\let\Phi\varPhi
\let\Psi\varPsi

\let\Omega\varOmega

\newcommand{\abs}[1]{\vert #1 \vert}

\newcommand{\bmk}{\sty{k}}

\newcommand{\bmr}{\sty{r}}

\newcommand{\clE}{\mathcal{E}}

\newcommand{\scT}{\mathscr{T}}

\newcommand{\rma}{\mathrm{a}}

\newcommand{\rme}{\mathrm{e}}
\newcommand{\rmf}{\mathrm{f}}

\newcommand{\rmx}{\mathrm{x}}

\newcommand{\uc}{\ubar{c}}

\newcommand{\umtheta}{\ubar{\mtheta}}

\newcommand{\uuA}{\uubar{A}}

\makeglossaries

\glsdisablehyper 
\renewcommand{\glossarysection}[2][]{} 
\glssetwidest[0]{\hspace{3.5em}}

\newproof{prf}{Proof}

\newtheorem*{rem2*}{Important remark}

\titleformat*{\section}{\bfseries}

\makeatletter
\def\blfootnote{\gdef\@thefnmark{}\@footnotetext}
\makeatother

\newcommand{\mrm}{\mathrm}
\newcommand{\Etot}{\ensuremath{\Pi}}
\newcommand{\Esite}{\ensuremath{\clE}}
\newcommand{\force}{\ensuremath{f}}
\newcommand{\bforce}{\ensuremath{\sty{\force}}}

\newcommand{\atom}{\ensuremath{r}}
\newcommand{\Atom}{\ensuremath{\sty{\atom}}}
\newcommand{\Atoms}{\ensuremath{\{\Atom_i\}}}

\newcommand{\Neigh}{\ensuremath{\{\Atom_{ij}\}}}

\newcommand{\bulk}{\ensuremath{{\rm bulk}}}
\newcommand{\usf}{\ensuremath{{\rm sf}}}

\newcommand{\curveA}{{Mo\,$\rightarrow$\,Nb$_{0.5}$Ta$_{0.5}$}\xspace}
\newcommand{\curveB}{{Nb\,$\rightarrow$\,Mo$_{0.5}$Ta$_{0.5}$}\xspace}
\newcommand{\curveC}{{Ta\,$\rightarrow$\,Mo$_{0.5}$Nb$_{0.5}$}\xspace}

\begin{document}

\begin{frontmatter}
 \begin{abstract}
  We present an automated procedure for computing stacking fault energies in random alloys from large-scale simulations using moment tensor potentials (MTPs) with the accuracy of density functional theory (DFT).
  To that end, we develop an algorithm for training MTPs on random alloys.
  In the first step, our algorithm constructs a set of $\sim$\,10\,000 or more training candidate configurations with 50--100 atoms that are representative for the atomic neighborhoods occurring in the large-scale simulation.
  In the second step, we use active learning to reduce this set to $\sim$\,100 most distinct configurations---for which DFT energies and forces are computed and on which the potential is ultimately trained.
  We validate our algorithm for the MoNbTa medium-entropy alloy by showing that the MTP reproduces the DFT $\frac{1}{4}[111]$ unstable stacking fault energy over the entire compositional space up to a few percent.
  
  Contrary to state-of-the-art methods, e.g., the coherent potential approximation (CPA) or special quasi-random structures (SQSs), our algorithm naturally accounts for relaxation, is not limited by DFT cell sizes, and opens opportunities to efficiently investigate follow-up problems, such as chemical ordering.
  In a broader sense, our algorithm can be easily modified to compute related properties of random alloys, for instance, misfit volumes, or grain boundary energies. Moreover, it forms the basis for an efficient construction of MTPs to be used in large-scale simulations of multicomponent systems.
 \end{abstract}
\end{frontmatter}

\section{Introduction}

Over the past 15--20 years, the class of random alloys experienced an exponentially increasing interest in the materials science community due the promising perspective in pushing forward the bounds for mechanical properties, e.g., strength, ductility, fracture toughness, etc., set by conventional alloys \citep{george_high_2020}.
Random alloys break with the established convention of just having one principal element, instead allowing for several components at close-to equi-atomic composition with, in principle, arbitrary ordering.
For example, the arguably most popular subclass of random alloys, the \emph{high-entropy alloys}, typically contains five or more principal elements.
Hence, random alloys offer a vast ``compositional playground'' that holds promise for the discovery of many new, even better and tailor-made alloys.

Overall, the mechanics community seems to have settled down to the general consensus that
the physical mechanisms are largely similar to the ones observed in conventional alloys, but it is the local disorder that can heavily modify the activation barriers for the mechanisms \citep{ma_unusual_2020,george_high_2020}.
This local disorder may strongly deviate from the average properties of the random alloy, possibly invoking a new combined occurrence of mechanisms.
Therefore, to assess the potential capability of a random alloy, those mechanisms must be studied using atomic-scale models---and eventually tuned, e.g., by varying the alloy composition, in favor of the desired mechanical properties for a specific application.

An efficient way to investigate the behavior of random alloys on the atomic level would thus be atomistic simulations using empirical interatomic potentials (EIPs).
However, with few exceptions, EIPs have been found to be inadequate for modeling random alloys \citep{rao_atomistic_2017,ghafarollahi_solutescrew_2019,maresca_mechanistic_2020,george_high_2020}.
More precisely, predicting the mechanical properties of metals requires modeling crystalline defects, such as dislocations, cracks, or grain boundaries, and for precisely this purpose reliable EIPs are lacking.
Hence, the realm for making quantitative predictions with EIPs remains very limited.

In order to make quantitative predictions thus requires using ab initio methods, such as density functional theory (DFT).
For problems without pronounced long-range elastic fields which do, in principle, not require more than a few tens of atoms, the state-of-the-art DFT-based methods are the coherent potential approximation \citep[CPA][]{yonezawa_coherent_1973,vitos_anisotropic_2001} and the special quasi-random structure (SQS) method \citep{zunger_special_1990}.
The CPA is a computationally very efficient mean-field method which has been employed to predict average properties of random alloys, for instance, thermodynamic properties (equation of state etc., e.g., \citep{ma_ab_2015}) or material properties, such as elastic constants and stacking fault energies (see, e.g., \citep{zaddach_mechanical_2013}).
However, the CPA cannot take into account relaxation which can have a decisive impact on the properties of random alloys (cf., \citep{ikeda_ab_2019}). Moreover, it assumes a perfectly random solid solution and can, therefore, not be used to investigate effects due to chemical ordering.
The SQS method, on the other hand, explicitly models the random atomistic configuration, allowing for relaxation and arbitrary chemical disorder. Additionally, the SQS concept optimizes the lattice site occupations with respect to the averaged multisite correlation functions of an infinite random alloy. This reduces the number of supercells, typically required to compute accurate average properties, to $\sim$\,30.
Nevertheless, even if we optimistically assume 30--40 energy/force computations per relaxation, we still easily require a total number of $\approx$\,1\,000 DFT calculations---for only one single composition.
Clearly, the SQS method is not efficient for screening over a large compositional space and, on top, performing Monte Carlo simulations to study chemical short-range order.
Moreover, as DFT scales cubically with the number of atoms, performing those studies for extended defects (e.g., dislocations), which require supercells with 100--500 atoms or more, appears out-of-scope with the SQS method given the vast compositional space to be discovered.

To overcome these limitations, we present a new method based on machine-learning interatomic potentials (MLIPs) since they, contrary to EIPs, provably approximate local DFT energies and forces with arbitrary accuracy.
The main idea behind using MLIPs is their ability to avoid repetitive/redundant DFT calculations by interpolating between a relatively small set of training configurations \cite{behler_generalized_2007,bartok_gaussian_2010,thompson_spectral_2015,shapeev_moment_2016,smith_ani-1:_2017,schutt_schnet_2017,pun2019-pinn,jinnouchi2019-kresse-on-the-fly,park2020-gnn,lysogorskiy2021-PACE} (several orders of magnitude smaller than a pure DFT-based study using SQSs would require).
The computational cost for computing material properties is thus feasible---even when large supercells with more than 10\,000 atoms are required---since computing energies and forces with MLIPs is negligibly small compared to DFT.

Yet, the main challenge in the construction of MLIPs is the selection of the \emph{right} training configurations.
This is in particular crucial for random alloys due to the vast amount of atomic neighborhoods.
For example, given a configuration of $N^\rma$ atoms, the amount of possibilities for distributing the elements to the atoms at a specific composition grows \emph{superexponentially} with the number of elements $N^\rme$, more precisely,
$\prod_{i=1}^{N^\rme-1} \, \frac{\left(N^\rma - \sum_{j=1}^{i-1} N^\rme_j\right)!}{N^\rma!\,\left(N^\rma - \sum_{j=1}^{i} N^\rme_j\right)!}$,
where $N^\rme_j$ is the number of ``element $j$'' atoms.
That is, for a configuration of 16 atoms and two elements there are 12\,870 possibilities at equi-atomic composition; for four elements already 63 million.
Of course, this number includes equivalent/symmetric configurations---but we need an algorithm to detect those.
Therefore, our method makes use of a state-of-the-art active learning algorithm \citep{podryabinkin_active_2017,gubaev_accelerating_2019} which selects the most distinct configurations from a large pool of candidate configurations (typically of the order of 10\,000 or higher).
We will show that the number of configurations on which we inevitably need to perform DFT calculations is typically just $\sim$\,50--100.

As a prototypical test problem, we compute the $\frac{1}{4}[111]$ unstable stacking fault energy of the refractory MoNbTa medium-entropy alloy using moment tensor potentials \citep[MTP,][]{shapeev_moment_2016}, a class of MLIPs, in order to validate our method.
We show in particular that, using our training algorithm, we are able to construct an MTP which predicts the DFT stacking fault energy over the entire compositional space with an error of not more than a few percent.
Further, we illustrate how the method can be implemented in practice to compute other material properties of random alloys, e.g., elastic constants, misfit volumes or surface/grain boundary energies.


\section{Machine-learning based approach for predicting material properties of random alloys}
\label{sec:method}

\subsection{\texorpdfstring{$\frac{1}{4}[111]$}{} unstable stacking fault energy}
\label{sec:method.usfe}

In the following we denote the position of some atom $i$ by $\Atom_i$ and a configuration of several atoms by $\Atoms$.
Further, we assume that the potential energy of a configuration $\Atoms$ is given by $\Etot = \Etot(\Atoms)$.

In what follows we assume a body-centered-cubic (bcc) lattice with lattice constant $a_0$.
To compute the unstable stacking fault energy (SFE) in bcc crystals, we first create a rectangular prismatic configuration with an orientation where the $\rmx_1$-, $\rmx_2$-, and $\rmx_3$-axes correspond to the [11$\bar{2}$], [$\bar{1}$10], and [111] directions, respectively.
We denote this configuration by $\Atoms^\bulk$.
We then translate half of the crystal by $\frac{a_0}{2}$[111] in the [111] direction, which corresponds to half a Burgers vector, to create the configuration with the stacking fault.
This configuration is denoted by $\Atoms^\usf$.

\begin{figure}[t]
 \centering
 \includegraphics[width=0.7\textwidth]{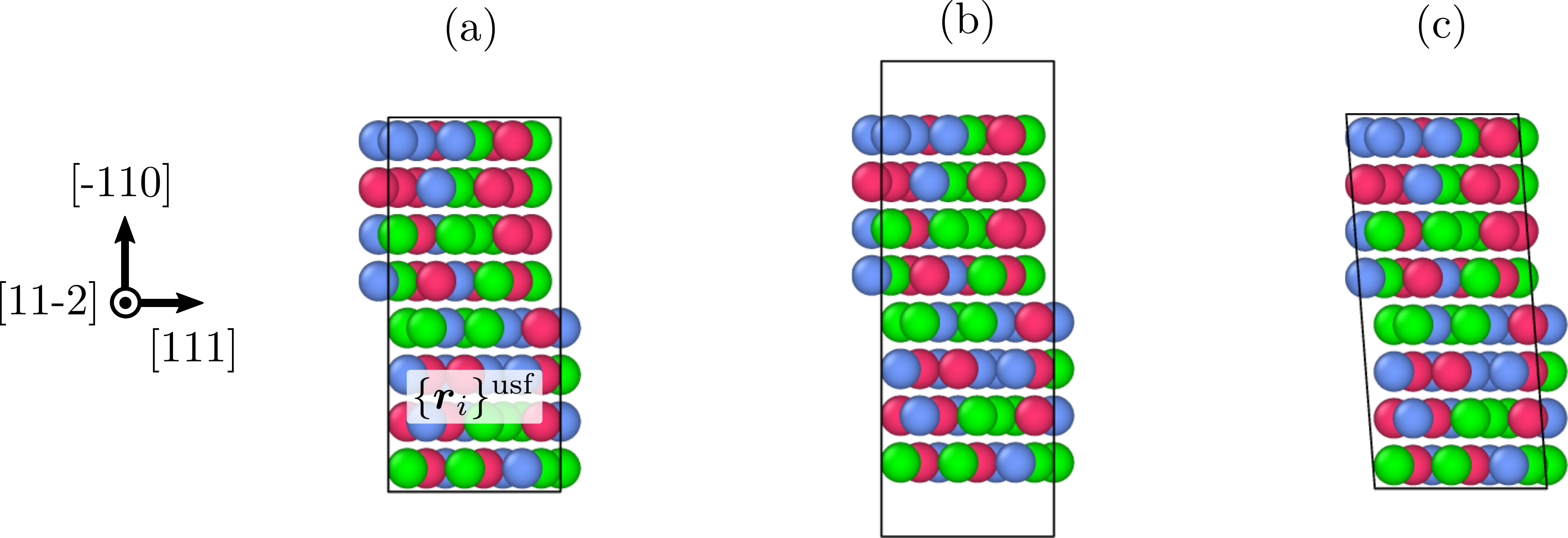}
 \caption{Typical supercell types for computing the $\frac{1}{4}$[111] unstable stacking fault energy}
 \label{fig:supercell_types}
\end{figure}

Several types of common supercell boundary conditions are discussed in the following (see, e.g., \citep{ghafarollahi_solutescrew_2019,xu_atomistic_2020,hu_screening_2021}.
One possibility is to apply periodic boundary conditions to the supercells containing $\Atoms^\bulk$ and $\Atoms^\usf$.
This procedure creates a second stacking fault as shown in Figure \ref{fig:supercell_types} (a).
If strong electronic interactions between the stacking faults are expected, another popular choice is to add a vacuum buffer to the supercell between the periodic configurations in the $\rmx_3$-direction (Figure \ref{fig:supercell_types} (b)).
This, in turn, creates additional computational cost when using DFT, so, yet another option is to apply a shear displacement of half a Burgers vector to the supercell without buffer, as shown in Figure \ref{fig:supercell_types} (c). In this case every atom at the periodic boundaries sees a perfect crystalline environment.
The latter type of boundary conditions therefore appears the preferable choice to us.

Having the configurations well-defined, we let the atoms in $\Atoms^\bulk$ and $\Atoms^\usf$ relax in the $\rmx_2$-direction (a relaxation in the lateral directions might, in general, also be required, but we found the effects negligibly small for our test problem in Section \ref{sec:results} and do therefore not consider it here).
The energy difference between both relaxed configurations divided by the area of the slip plane $A$ then gives the SFE
\begin{equation}\label{eq:usfe}
 \Etot^\usf = \frac{\Etot(\Atoms^\usf) - \Etot(\Atoms^\bulk)}{A},
\end{equation}
where, in case of type (a) boundary conditions, we additionally need to divide the energy by two since there are two stacking faults in the supercell.

For a random alloy we typically create a set of configurations with the atom types distributed randomly according to the desired alloy composition. For each configuration from this set we compute \eqref{eq:usfe}, and then the average.

\subsection{A few remarks on computing the SFE with DFT}
\label{sec:method.sqs-dft}

Typical configurations $\Atoms^\bulk$ and $\Atoms^\usf$ that can be handled with DFT in a reasonable amount of time contain 72 atoms.
One energy/force computation for a 72-atom configuration takes about 10--20 minutes using the plane-wave DFT code VASP \citep{kresse_efficient_1996} (already assuming parallelization over 72 cores).
Relaxing the two configurations $\Atoms^\bulk,\Atoms^\usf$ takes, optimistically, $\approx$\,50 energy/force computations.
We then need to repeat the calculations for different random configurations in order to obtain the average SFE of the random alloy.
The ``magic'' minimum number, i.e., the number mostly reported in other works \citep{xu_atomistic_2020,hu_screening_2021}, of required 72-atom configurations is 28 when the atom types are distributed using the SQS method (cf., e.g., \citep{xu_atomistic_2020,hu_screening_2021}).
Additionally assuming that we are interested in the average SFE at 5--10 different compositions of the random alloy, we already require more than 10\,000 DFT calculations corresponding to \textbf{two months} of absolute computing time.
Newer, possibly more advanced alternatives to the SQS method, such as the partial occupation method \citep{yang_modeling_2016}, of sampling relevant alloy structures may further reduce the number of required random samples per composition, but the computational burden of relaxing them still remains.
Clearly, a high-throughput screening of random alloys is thus not feasible within a reasonable amount of time using a pure DFT-based approach.

\subsection{Moment tensor potentials and active learning}

In order to overcome the limitations of pure DFT-based approaches, we propose an alternative method that uses machine-learning interatomic potentials (MLIPs).
The idea is to train a MLIP on all \emph{relevant} neighborhoods that are representative of all neighborhoods that \emph{appear} in any random configuration $\Atoms^\bulk,\Atoms^\usf$.
Then, our hope is that a relatively small number of single-point DFT calculations on 72-atom (or less) configurations of the order of 100 will suffice to construct a MLIP that predicts the SFE of the random alloy at any composition.
This is much more tractable than using DFT-SQS methods that require more than 10\,000 single-point DFT calculations on 72-atom configurations for solving the same problem (cf. Section \ref{sec:method.sqs-dft}).

We begin by stating our main assumption which justifies the use of interatomic potentials, namely the separation of the total energy $\Etot(\Atoms)$ of some configuration $\Atoms$ into per-atom contributions such that
\begin{equation}\label{eq:Epartition}
 \Etot(\Atoms) = \sum_i \Esite(\Neigh;\umtheta),
\end{equation}
where $\Neigh$ is the neighborhood of atom $\Atom_i$ and $\umtheta$ is the vector of fitting parameters.
To model the per-atom energies \Esite(\Neigh), we use the moment tensor potentials \citep[MTPs,][]{shapeev_moment_2016,gubaev_accelerating_2019} defined as follows
\begin{equation}\label{eq:Eatom}
 \Esite(\Neigh; \umtheta) = \sum_{\alpha=1}^m \mtheta_\alpha B_\alpha(\Neigh;\{\mtheta_\beta\}),
\end{equation}
where the $\mtheta_\alpha$'s are scalar fitting parameters, independent of $\Neigh$, and the $B_\alpha$'s are the Basis functions. The basis functions have a nonlinear dependency on an additional set of parameters $\{\mtheta_\beta\}$ which, e.g., depend on the element type. For further details the reader is referred to \ref{sec:appdx.mtp}.

The cornerstone of our method to assess whether a configuration is adequately represented by the MTP is \emph{active learning} \citep{settles_active_2010}.
That is, given a new configuration $\Atoms^\ast$ not contained in the training set, active learning checks if a new DFT calculation on this configuration is required to ensure transferability of the MTP.
Active learning thus allows us to conveniently bypass the major bottleneck of the DFT-SQS methods since we perform structural relaxation using MTPs---and trigger new single-point DFT calculations only in rare cases when the MTP becomes inaccurate.

Recently, a number of active learning algorithms have been developed for MLIPs based on, e.g., query-by-committee \citep{zhang_active_2019}, Bayesian inference \citep{vandermause_--fly_2020}, and D-optimality \citep{podryabinkin_active_2017}.
Here, we use D-optimal active learning which has already successfully been applied to predict crystal structures of ternary alloys in tandem with MTPs \citep{gubaev_accelerating_2019}.

To briefly review the basic functioning of D-optimal active learning, assume, for simplicity, an active set containing $m$ neighborhoods.
Then, D-optimal active learning assesses a new neighborhood $\Neigh^\ast$, contained in $\Atoms^\ast$, based on the maximum change in the determinant of the $m \times m$ Jacobian
\begin{equation}
 \uuA =
 \begin{pmatrix}
  \frac{\partial \Esite(\Neigh_1;\umtheta)}{\partial \mtheta_1} & \cdots & \frac{\partial \Esite(\Neigh_1;\umtheta)}{\partial \mtheta_m} \\
  \vdots & \ddots & \vdots \\
  \frac{\partial \Esite(\Neigh_m;\umtheta)}{\partial \mtheta_1} & \cdots & \frac{\partial \Esite(\Neigh_m;\umtheta)}{\partial \mtheta_m}
 \end{pmatrix}
\end{equation}
if we would replace one of the $\Neigh$'s from the training set with $\Neigh^\ast$.
We call this maximum change in the determinant the \emph{extrapolation grade} $\gamma$ and compute it as follows
\begin{equation}
 \gamma = \underset{i}{\max\,}{\abs{c_i}}, \qquad \text{with} \quad
 \uc =
 \begin{pmatrix}
   \frac{\partial \Esite(\Neigh^\ast;\umtheta)}{\partial \mtheta_1} & \cdots & \frac{\partial \Esite(\Neigh^\ast;\umtheta)}{\partial \mtheta_m}
 \end{pmatrix}
 \uuA^{-1}.
\end{equation}
Following \citep{novikov_mlip_2021},
\[
 \begin{aligned}
  &\gamma < 1 && \qquad \text{indicates no extrapolation,} \\
  1 \le &\gamma < 2 && \qquad \text{indicates accurate extrapolation,} \\
  2 \le &\gamma < 10 && \qquad \text{indicates still reliable extrapolation,} \\
  10 \le &\gamma && \qquad \text{indicates risky extrapolation.}
 \end{aligned}
\]

In practice we thus set a threshold on $\gamma$ below which we tolerate extrapolation (see next section).
If $\gamma$ exceeds this threshold for some neighborhood $\Neigh^\ast$ contained in configuration $\Atoms^\ast$, we add $\Atoms^\ast$ to the training set and refit the potential.
Since the training set now contains more than $m$ neighborhoods, we replace those $\Neigh$'s with $\Neigh^\ast$'s using the maxvol algorithm \citep{olshevsky_how_2010} so that the determinant of $\uuA$ is maximal (that is, we maximize linear independency between the row vectors of $\uuA$).

\subsection{Active learning algorithm}
\label{sec:method.algo}

\begin{figure}[H]
 \centering
 \includegraphics[width=\textwidth]{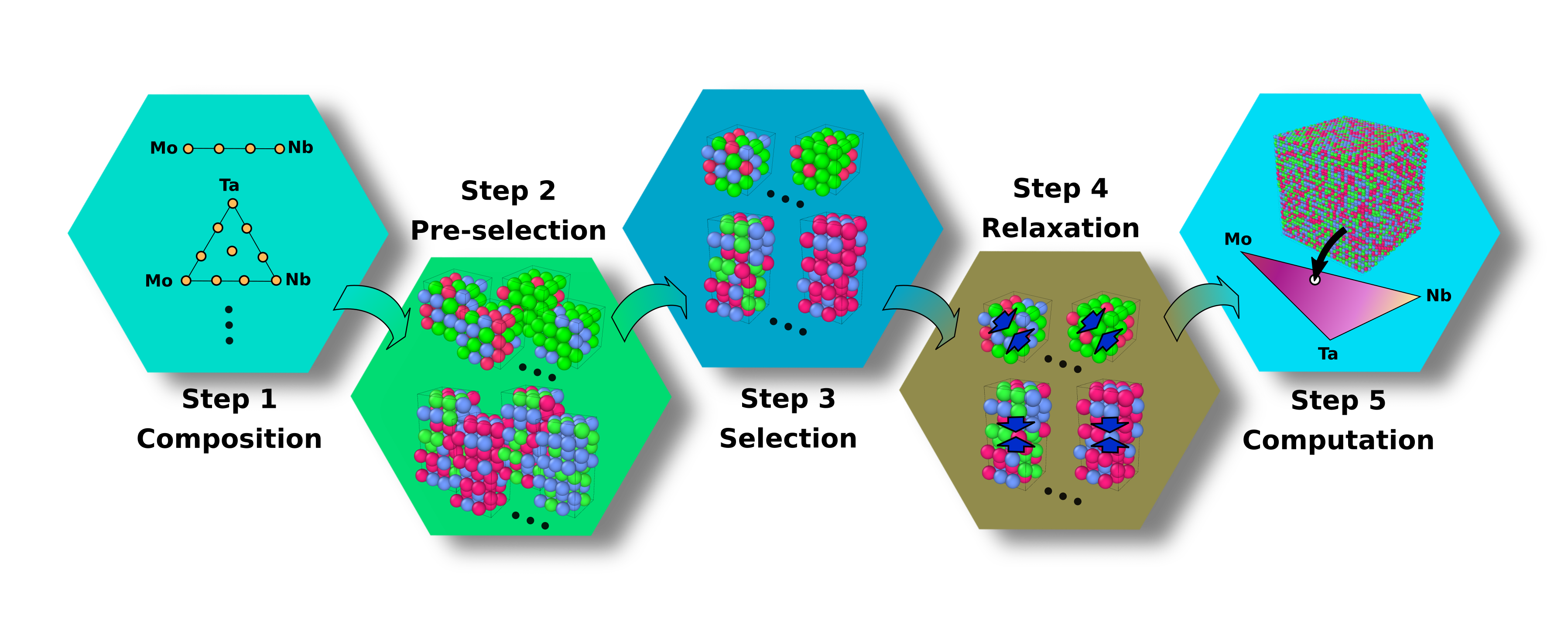}
 \caption{Schematic illustration of the active learning algorithm presented in Section \ref{sec:method.algo}}
 \label{fig:algo}
\end{figure}

Our goal is now to use MTPs and active learning to develop an algorithm that accurately predicts the SFE for an $n$-element alloy at arbitrary composition.
The individual steps of the algorithm that we propose in the following are schematically depicted in Figure \ref{fig:algo}.

\begin{itemize}[leftmargin=1.7cm]
 \item[\textbf{Step\;1}]
 First, we choose the number of elements and the compositional domain, i.e., a subset of the compositional simplex, in which we seek to compute the SFE (in the Figure a regular grid is used over the entire simplex).
 \item[\textbf{Step\;2}]
 Next we construct the set of training candidates. Therefore, we select the types of configurations which are representative for our problem (here $\Atoms^\bulk,\Atoms^\usf$). For each grid point in the compositional domain we then create a number of samples of each configuration with the atom types randomly distributed.
 For each of these samples we create additional copies with varying cell volume between the smallest and the largest lattice constant of the chosen elements.
 \item[\textbf{Step\;3}]
 In the third step, we then select the most distinct training candidates from the set of samples created in \textbf{Step\;2}, compute those selected configurations with DFT, and train the MTP on them.
 For this purpose, we may, e.g., first randomly select a small number of configurations (5--10 for each type of configuration) to be computed with DFT in order to initialize the MTP.
 We then use active learning to populate the training set with extrapolative configurations from the set of training candidates.
 Since D-optimality computes the extrapolation of the MTP around a linearized state we do this iteratively: we first add only configurations with a very large $\gamma$ (e.g., $>$\,10\,000), then the ones with $\gamma$\,$>$\,1\,000, 100, 10, and finally some value $\gamma_\mrm{min}$ between 1 and 2 (corresponding to little or no extrapolation).
 \item[\textbf{Step\;4}]
 We now consecutively relax all configurations from the training set using the MTP trained on all configurations selected in \textbf{Step\;3}.
 Thereby, we leave active learning switched on to collect any remaining training configurations ``on-the-fly''.
 Again, we emphasize that no structural relaxation with DFT needs to be performed.
 \item[\textbf{Step\;5}]
 In the final step we use the trained potential to compute the SFE.
 Since we are interested in the average SFE of the random alloy, we may, for simplicity, compute \eqref{eq:usfe} for one sufficiently large random configuration.
 This is computationally feasible because even relaxing a configuration of 10\,000 atoms or more using MTPs is orders of magnitudes more efficient than computing the training set with DFT. 
 Yet we remark that using MTPs in combination with the SQS method is also possible, but not explicitly required, thus simplifying our algorithm by circumventing the need for creating SQSs (as this procedure becomes increasingly expensive for larger supercells with $>$\,100 atoms).
\end{itemize}

At this point we remark that each step of the proposed algorithm offers room for various adjustments and improvements.
We have decided to leave the description of the algorithm as applied in the following section for binary and ternary random alloys in order to explain its basic functioning in the most understandable setting.
We will comment on further modifications and potential applications to other material properties in Section \ref{sec:discussion}.
As such, the algorithm presented in this section should therefore be considered as a proof-of-concept method.

\subsection{Comparison with on-lattice methods}

Another popular class of data-driven methods for simulating metallic alloys are on-lattice methods, e.g., the cluster expansion method \citep{sanchez_generalized_1984,natarajan_linking_2020}, or the low-rank potential method \citep{shapeev_accurate_2017}.
The main difference, compared to our method using MTPs, is that the per-atom energies in on-lattice methods disregard atomic positions and only depend on the atomic types of neighboring atoms.
Hence, the training set of on-lattice methods usually only consists of fully-relaxed configurations.
The reduced atomic interaction model is the main advantage of on-lattice methods, allowing for much more efficient energy evaluations compared to MTPs.
Atomic positions, however, are inevitable for using active learning during relaxation in order to overcome the immense computational cost of relaxing configurations solely with DFT (cf. Section \ref{sec:method.sqs-dft}).
Moreover, we point out that for our test problem of computing average SFEs most time is still spend on the DFT calculations and, so, a faster atomic interaction model than MTP is not required here.

We nevertheless remark that, if energy evaluations become the bottleneck (for example, when performing Monte Carlo simulations), our algorithm can also be used for training on-lattice models, thus bypassing the need for relaxing configurations with DFT.

\section{Computational results}
\label{sec:results}

\subsection{Setup}

Our implementation of the active learning algorithm presented in Section \ref{sec:method.algo} uses two configuration types to construct the set of training candidates in \textbf{Step\;2}: bulk configurations of 54 atoms with equal supercell dimensions, and the 72-atom stacking fault configurations shown in Figure \ref{fig:supercell_types} (c).
In addition, we have applied random displacements to each training candidate which are drawn from a normal distribution with standard deviation $0.0033\cdot a_0$;
we have found this to improve the stability of our algorithm.
The minimal extrapolation grade above which we add configurations to the training set in \textbf{Step\;3} and \textbf{Step\;4} is chosen to be $\gamma_\mrm{min}$\,=\,2, unless stated otherwise.
In order to compute the SFE in \textbf{Step\;5}, we use one very large configuration of $\approx$\,10\,000 atoms with the atom types randomly distributed according to the desired composition.
For the relaxation in \textbf{Step\;4} and \textbf{Step\;5} we have used the damped dynamics solver FIRE \citep{bitzek_structural_2006}, implemented within the ASE library \citep{hjorth_larsen_atomic_2017}.
The relaxation terminates when the maximum force on all atoms is less than 0.001\,eV/\AA.

In the following, we validate the active learning algorithm for the MoNbTa ternary random alloy.
The computation of energies and forces for all training configurations use DFT with plane-wave basis sets and the projector-augmented wave pseudopotential method \citep{blochl_projector_1994,kresse_ultrasoft_1999} as implemented in VASP \citep{kresse_efficient_1996}.
The simulation parameters that we have used in our calculations are given in \ref{sec:appdx.vasp}.
With this DFT setup, we have computed the lattice constants and the relaxed SFEs for the pure materials.
Our results are in agreement with recent results reported in \citep{xu_frank-read_2020}, obtained with a DFT setup similar to ours, as shown in Table \ref{tab:a0_&_SFE_pure}.

\begin{table}[hbt]
 \centering
 \begin{tabular}{|c||c|c|c|c|}
  \hline
  \multirow{2}{*}{Element}
  & \multicolumn{2}{c|}{Lattice constant [\AA]}
  & \multicolumn{2}{c|}{$\Etot^\mrm{sf}$ [mJ/m$^2$]} \\
  & This work & Ref. \citep{xu_frank-read_2020} & This work & Ref. \citep{xu_frank-read_2020} \\
  \hline\hline
  Mo & 3.149 & 3.16 & 1468.82 & 1443.39 \\ \hline
  Nb & 3.316 & 3.324 & 648.29 & 676.78 \\ \hline
  Ta & 3.307 & 3.32 & 747.25 & 724.46 \\ \hline
 \end{tabular}
 \caption{Lattice constants and SFEs for the pure materials considered in this work}
 \label{tab:a0_&_SFE_pure}
\end{table}

\subsection{Binary systems MoTa, MoNb, NbTa}
\label{sec:results.binary}

We first apply our active learning algorithm to all the binaries of the MoNbTa ternary.

To analyze the influence of the input parameters on the training set, generated by our active learning algorithm, we investigate the choice of the grid type (\textbf{Step\;1}) and
\begin{itemize}[leftmargin=2.7cm,labelsep=1cm]
 \item[$n_\mrm{rand}$,]
 the number of random configurations \emph{per grid point} created in \textbf{Step\;2}.
\end{itemize}
We further denote by,
\begin{itemize}[leftmargin=2.7cm,labelsep=1cm]
 \item[$n_\mrm{cand}$,]
 the \emph{total} number of candidate configurations created in \textbf{Step\;2},
 \item[$n_\mrm{ts}$,]
 the size of the training set after \textbf{Step\;3},
 \item[$n_\mrm{ts}^\mrm{relax}$,]
 the size of the training set after the relaxation (\textbf{Step\;4}).
\end{itemize}
Using the example of MoTa, we created three sets of candidate configurations, two with grid type 3 and $n_\mrm{rand}$\,=\,100,\,1000, and one with grid type 5 and $n_\mrm{rand}$\,=\,100 (see Table \ref{tab:grids}).
To generate the training sets, denoted in the following by $\scT_\mrm{MoTa}^1$--$\scT_\mrm{MoTa}^3$, we have used a level-16 MTP (see \ref{sec:appdx.mtp}).
In all three cases the training set consists of 48 configurations, as shown in Table \ref{tab:ts_binaries}.
This result is encouraging as it implies that an MTP is able to interpolate over the entire configurational space of a binary random alloy using only a rather small number training configurations.

\begin{table}[H]
 \newcolumntype{M}[1]{>{\centering\arraybackslash}m{#1}}
 \centering
 \begin{tabular}{|M{20mm}|M{50mm}|M{50mm}|}
  \hline
  Grid type & Two elements \textbf{A}, \textbf{B} & Three elements \textbf{A}, \textbf{B}, \textbf{C} \\ \hline\hline
  3 &
  \begin{minipage}{0.2\textwidth}
   \includegraphics[width=0.7\textwidth]{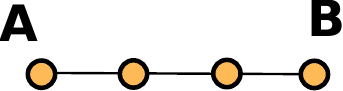}
  \end{minipage}
  &
  \begin{minipage}{0.2\textwidth}%
   \vspace{0.5em}
   \includegraphics[width=0.7\textwidth]{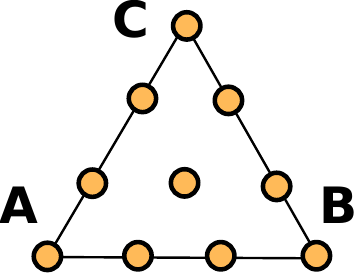}%
   \vspace{0.5em}
  \end{minipage}
  \\ \hline
  5 &
  \begin{minipage}{0.2\textwidth}
   \includegraphics[width=0.7\textwidth]{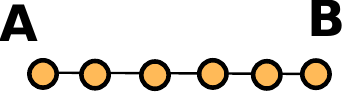}
  \end{minipage}
  &
  \begin{minipage}{0.2\textwidth}%
   \vspace{0.5em}
   \includegraphics[width=0.7\textwidth]{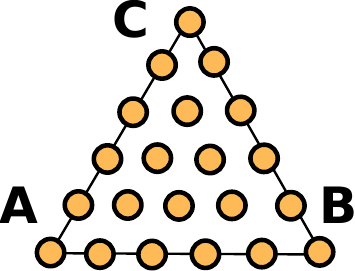}%
   \vspace{0.5em}
  \end{minipage}
  \\ \hline
 \end{tabular}
 \caption{Grid types used to create the training candidate set in \textbf{Step\;2}}
 \label{tab:grids}
\end{table}

\begin{table}[H]
 \centering
 \begin{tabular}{|c|c|c|c|c|c|c|}
  \hline
  Name & Elements & Grid type & $n_\mrm{rand}$ & $n_\mrm{cand}$ & $n_\mrm{ts}$ & $n_\mrm{ts}^\mrm{relax}$ \\ \hline\hline
  $\scT_\mrm{MoTa}^1$ & MoTa & 3 & 100 & 4\,000 & 36 & 48 \\ \hline
  $\scT_\mrm{MoTa}^2$ & MoTa & 3 & 1000 & 40\,000 & 36 & 48 \\ \hline
  $\scT_\mrm{MoTa}^3$ & MoTa & 5 & 100 & 6\,000 & 35 & 48 \\ \hline
  $\scT_\mrm{MoNb}$ & MoNb & 5 & 100 & 7\,200 & 31 & 44 \\ \hline
  $\scT_\mrm{NbTa}$ & NbTa & 5 & 100 & 2\,400 & 31 & 38 \\ \hline
  $\scT_\mrm{MoNbTa}^1$ & MoNbTa & 5 & 100 & 25\,200 & 72 & 83 \\ \hline
 \end{tabular}
 \caption{Training set sizes for the binary systems MoTa, MoNb, NbTa, and the MoNbTa ternary, generated using the algorithm from Section \ref{sec:method.algo} for different grid types and $n_\mrm{rand}$'s}
 \label{tab:ts_binaries}
\end{table}

We have then trained an ensemble of eleven MTPs (one with a uniform, and ten with a random initialization of the fitting parameters) to the training sets and selected the best five of them, i.e, those with the smallest residual, for further investigation.
The errors with respect to the training sets, reported in Table \ref{tab:errors_MoTa}, are already close to the well-known limits that can be obtained with MLIPs trained with respect to DFT data, i.e., $\approx$\,1\,meV for per-atom energies, and $\approx$\,0.1\,eV/{\AA} for per-atom forces (cf., e.g., \citep{zuo_performance_2020}).
Hence, the MTP should be able to accurately predict the SFE, provided that our active learning algorithm could select the right configurations that are representative for the large-scale problem.

\begin{table}[H]
 \centering
 \begin{tabular}{|c|c|c|c|c|}
  \hline
  Error & MTP$_{16}$($\scT_\mrm{MoTa}^1$) & MTP$_{16}$($\scT_\mrm{MoTa}^2$) & MTP$_{16}$($\scT_\mrm{MoTa}^3$) & MTP$_{20}$($\scT_\mrm{MoTa}^3$) \\ \hline\hline
  $\mepsilon_\mrm{ave}^\mrm{atom}$ [meV]     & 1.56\,$\pm$\,0.24   & 1.26\,$\pm$\,0.12   & 1.29\,$\pm$\,0.21   & 0.82\,$\pm$\,0.10 \\ \hline
  $\mepsilon_\mrm{rms}^\mrm{atom}$ [meV]     & 2.12\,$\pm$\,0.34   & 1.65\,$\pm$\,0.20   & 1.72\,$\pm$\,0.24   & 1.16\,$\pm$\,0.09 \\ \hline
  $\mepsilon_\mrm{ave}^\mrm{force}$ [eV/\AA] & 0.072\,$\pm$\,0.002 & 0.075\,$\pm$\,0.002 & 0.074\,$\pm$\,0.003 & 0.061\,$\pm$\,0.003 \\ \hline
  $\mepsilon_\mrm{rms}^\mrm{force}$ [eV/\AA] & 0.082\,$\pm$\,0.002 & 0.086\,$\pm$\,0.003 & 0.085\,$\pm$\,0.004 & 0.070\,$\pm$\,0.003 \\ \hline
 \end{tabular}
 \caption{Mean training errors and deviations between the MTPs with different parameter initializations for the MoTa binary system. The errors, including the deviations, are very small (of the order of 1\,meV for per-atom energies and of the order of 0.1\,eV/{\AA} for per-atom forces), indicating overall reliability of the training}
 \label{tab:errors_MoTa}
\end{table}

Therefore, we now compute the SFEs predicted by the MTPs trained on the previously constructed training sets.
We compare our results to our DFT values from Table \ref{tab:a0_&_SFE_pure} for the pure materials and to the DFT values for some intermediate compositions computed by \citet{hu_screening_2021}.
The DFT values in \citep{hu_screening_2021} were obtained with the SQS method and a DFT setup similar to the one we use in the present work.

\begin{figure}[t]
 \centering
 \begin{minipage}{0.333333\textwidth}
  \centering
  (a)
 \end{minipage}
 \begin{minipage}{0.333333\textwidth}
  \centering
  (b)
 \end{minipage}\\[0.1em]
 \begin{minipage}{0.333333\textwidth}
  \centering
  \includegraphics[width=0.97\textwidth]{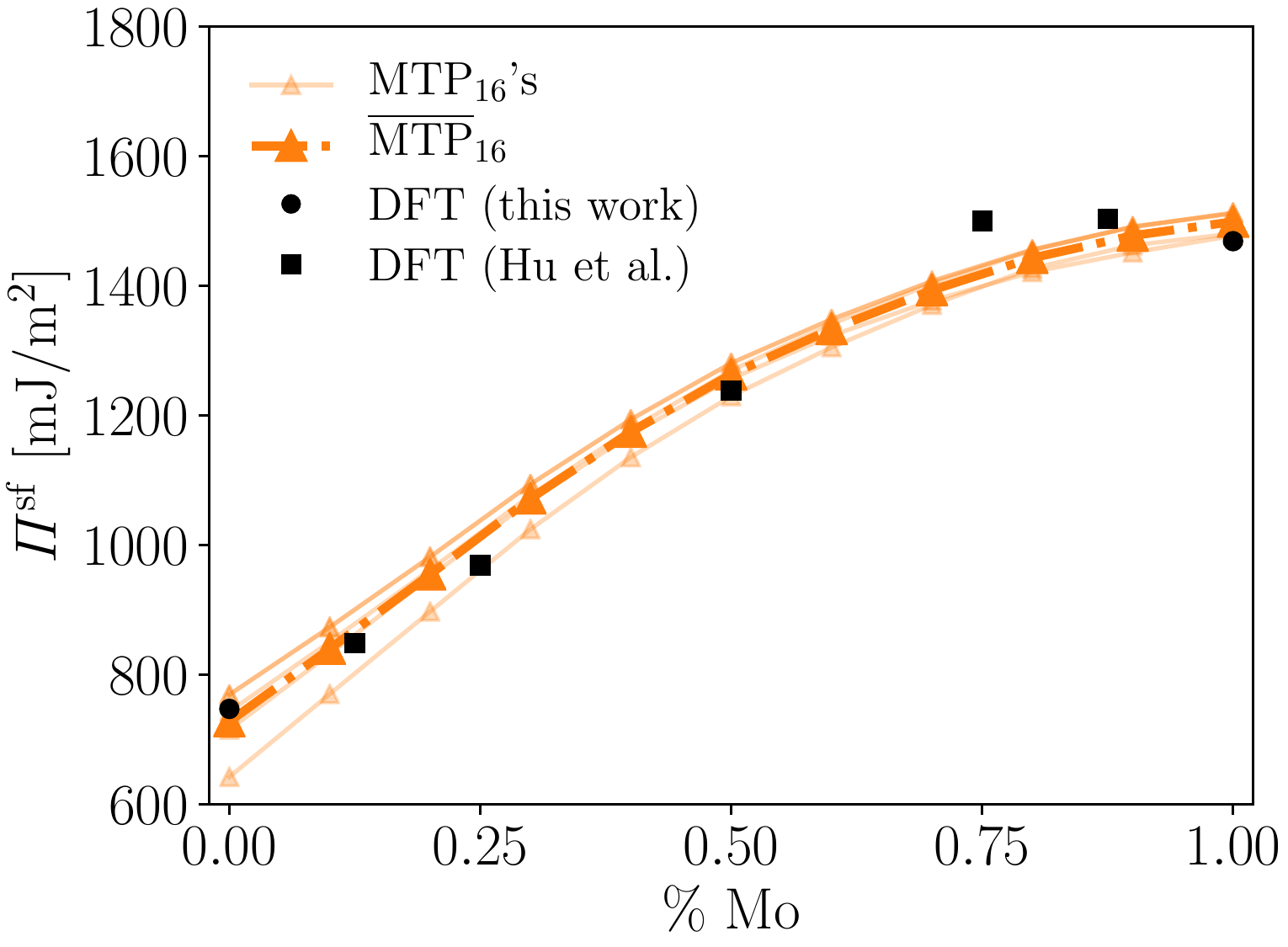}
 \end{minipage}
 \begin{minipage}{0.333333\textwidth}
  \centering
  \includegraphics[width=0.97\textwidth]{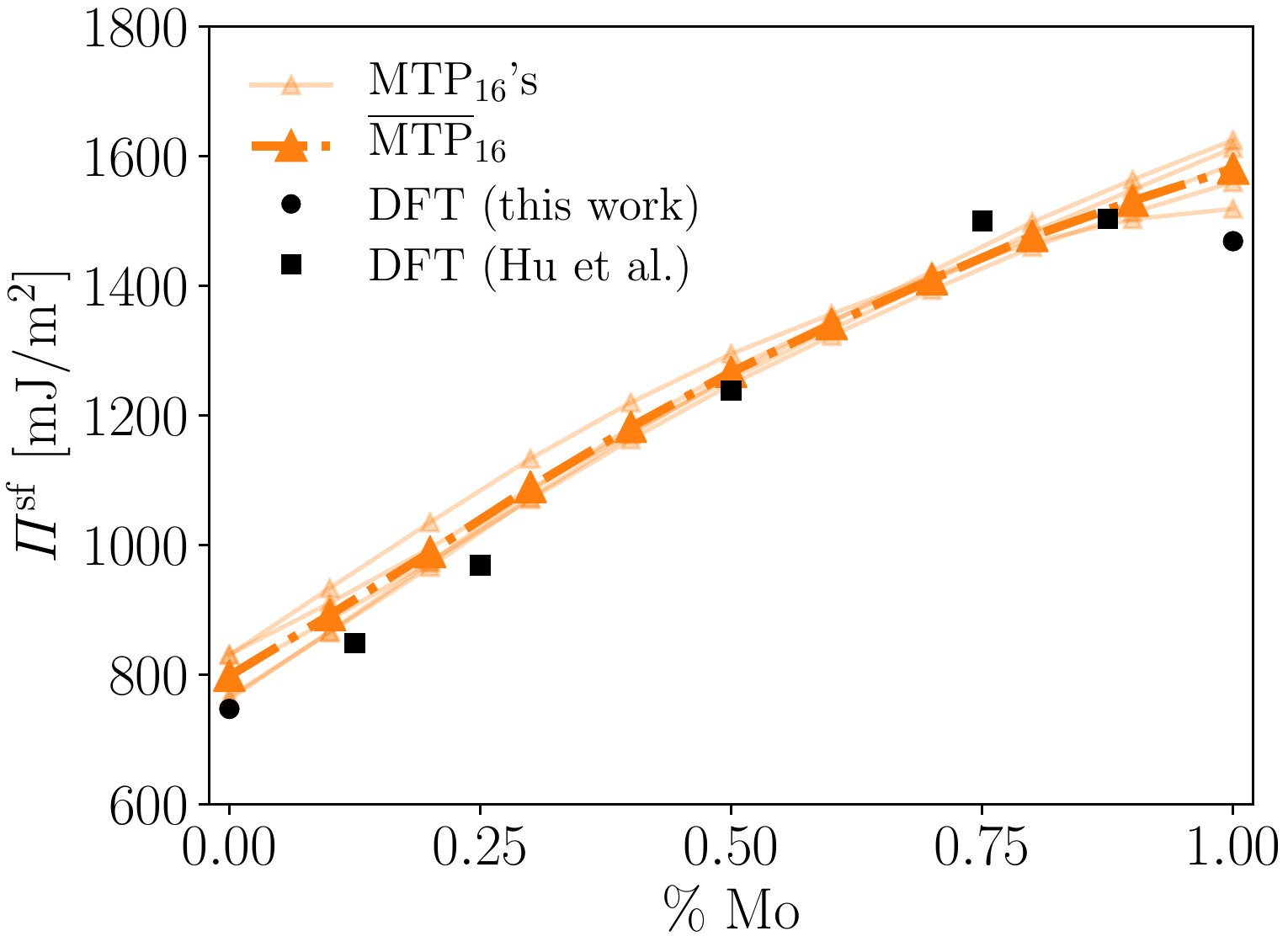}
 \end{minipage}\\[0.1em]
 \begin{minipage}{0.333333\textwidth}
  \centering
  (c)
 \end{minipage}
 \begin{minipage}{0.333333\textwidth}
  \centering
  (d)
 \end{minipage}\\[0.1em]
 \begin{minipage}{0.333333\textwidth}
  \centering
  \includegraphics[width=0.97\textwidth]{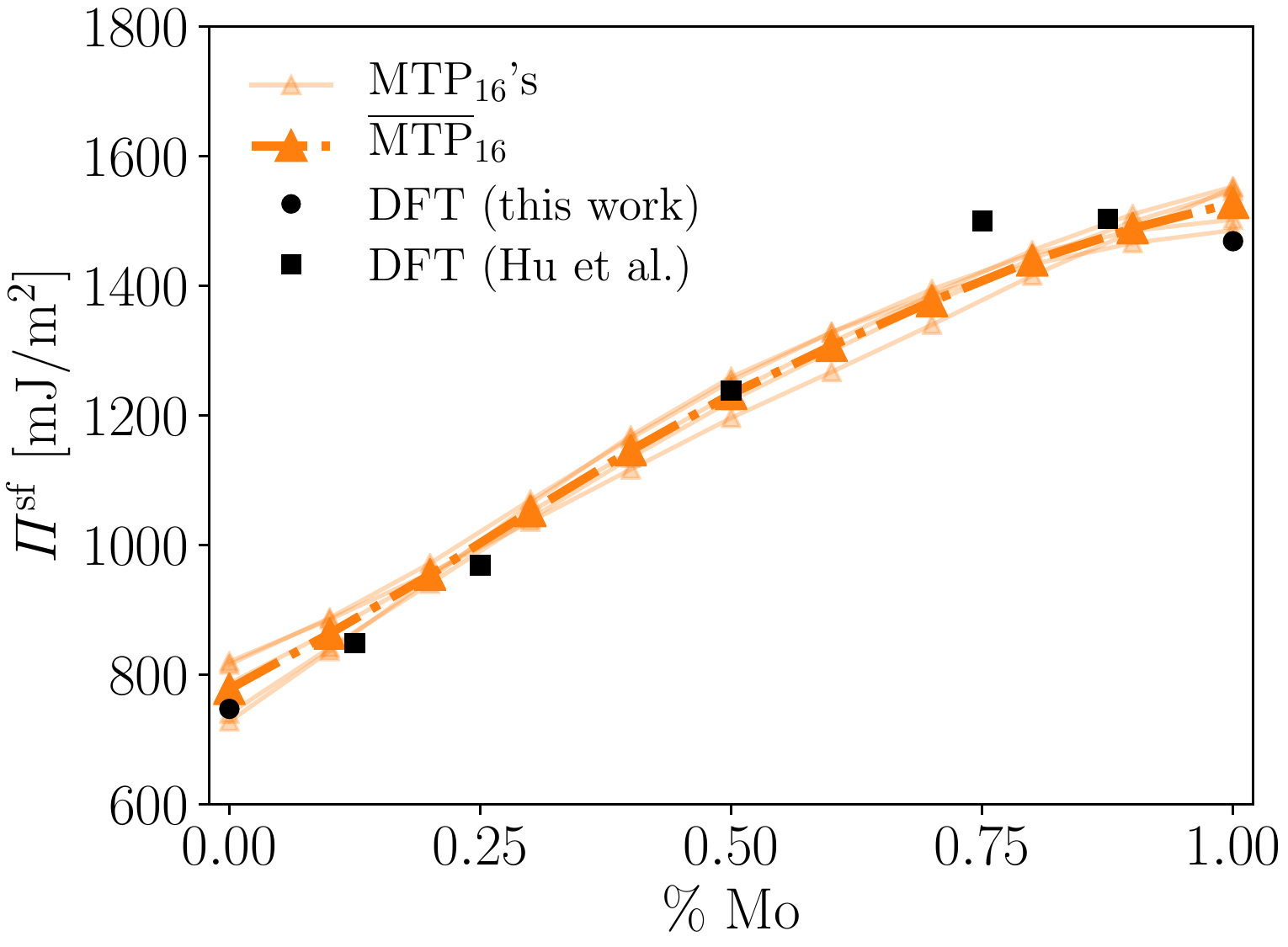}
 \end{minipage}
 \begin{minipage}{0.333333\textwidth}
  \centering
  \includegraphics[width=0.97\textwidth]{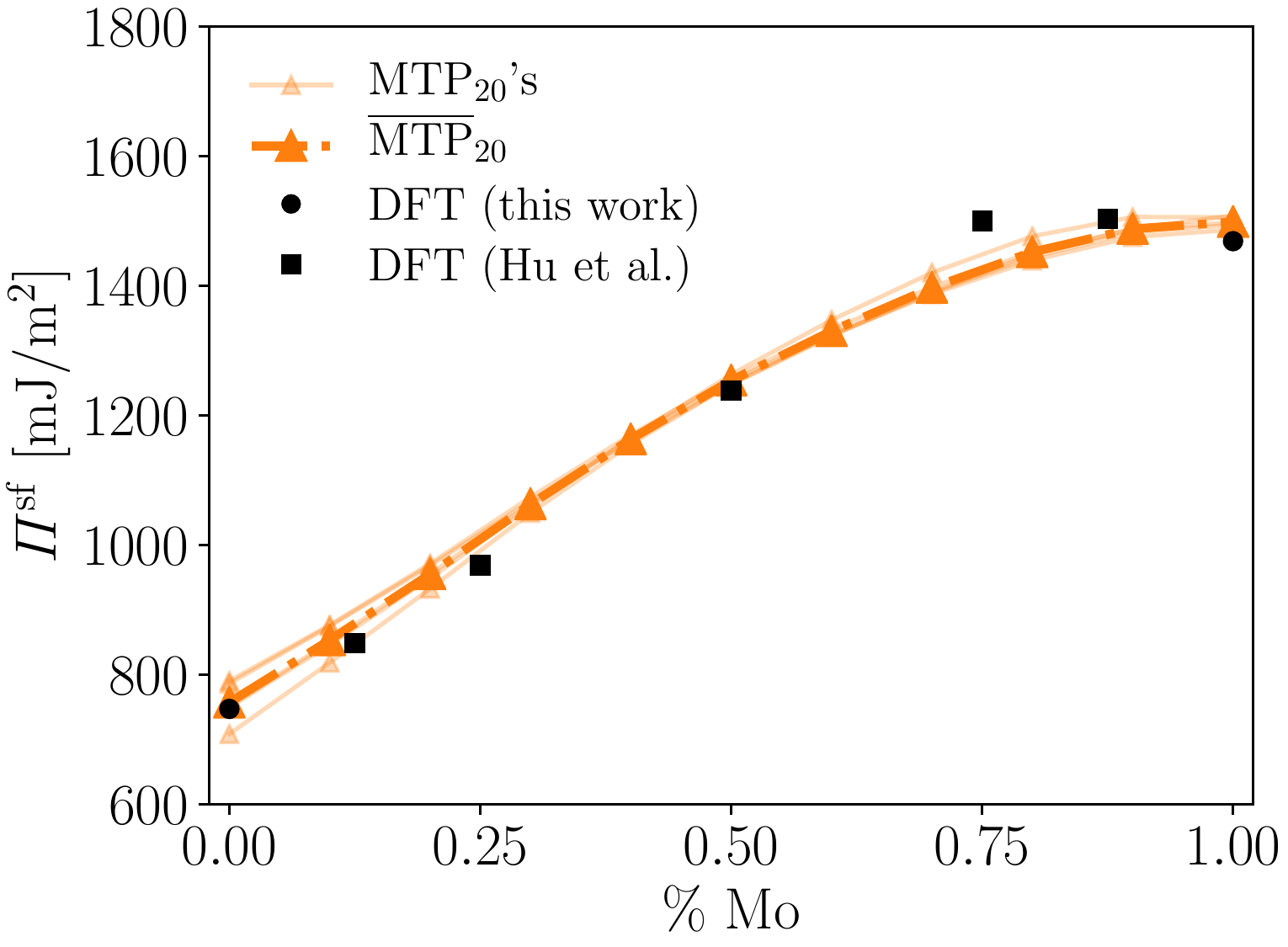}
 \end{minipage}
 \caption{SFE for MoTa as a function of the Mo content predicted by (a)--(c) level-16 MTPs trained on $\scT_\mrm{MoTa}^1$--$\scT_\mrm{MoTa}^3$, and by (d) level-20 MTPs trained on $\scT_\mrm{MoTa}^3$. The average predictions ($\overline{\text{MTP}}_X$) coincide with the DFT values up to a few percent (except for 100\,\%\,Mo in (b)). This accuracy is remarkable given the small size of the training set containing only 48 configurations}
 \label{fig:results_MoTa}
\end{figure}

The SFE for MoTa, predicted by the level-16 MTPs trained on $\scT_\mrm{MoTa}^1$--$\scT_\mrm{MoTa}^3$, is shown in Figure \ref{fig:results_MoTa} (a)--(c).
Overall, the agreement between the DFT and the average MTP values (the $\overline{\text{MTP}}_X$-curves) is very good with a maximum relative difference of just $\approx$\,6--7\% at Mo$_{0.75}$Nb$_{0.25}$ (except for the MTP$_{16}$($\scT_\mrm{MoTa}^2$)).
For MTPs trained on the grid-type-3 training sets $\scT_\mrm{MoTa}^1$, $\scT_\mrm{MoTa}^2$, the deviations between the different models are more pronounced in the corners.
Using a denser grid (grid type 5), the deviations over the compositional space become more balanced, likely due to the larger amount of candidate configurations with intermediate compositions.
Increasing the MTP level to 20 almost entirely removes any differences between the models (Figure \ref{fig:results_MoTa} (d)).

\begin{figure}[t]
 \centering
 \begin{minipage}{0.333333\textwidth}
  \centering
  (a)
 \end{minipage}
 \begin{minipage}{0.333333\textwidth}
  \centering
  (b)
 \end{minipage}\\[0.5em]
 \begin{minipage}{0.333333\textwidth}
  \centering
  \includegraphics[width=0.97\textwidth]{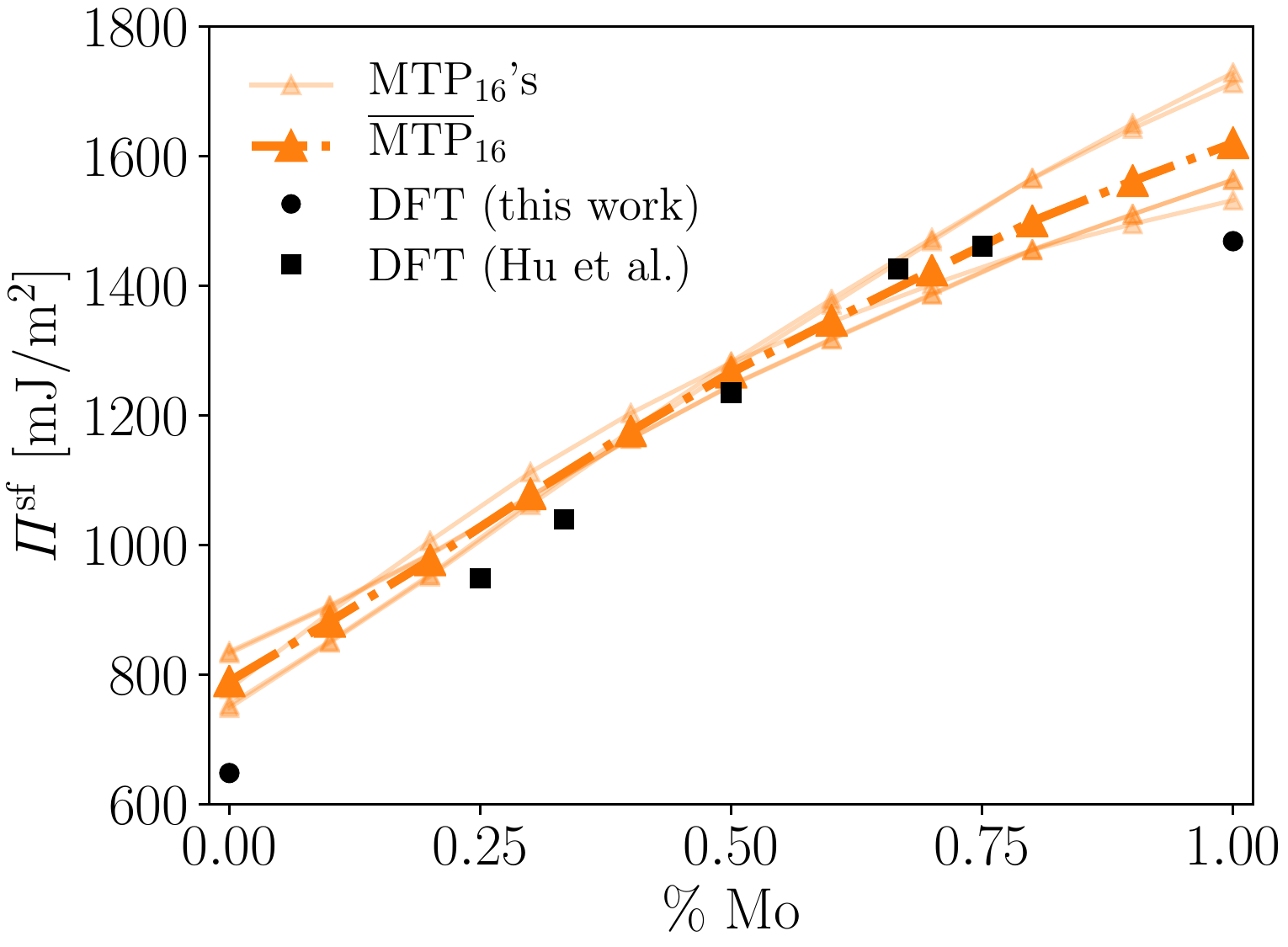}
 \end{minipage}
 \begin{minipage}{0.333333\textwidth}
  \centering
  \includegraphics[width=0.97\textwidth]{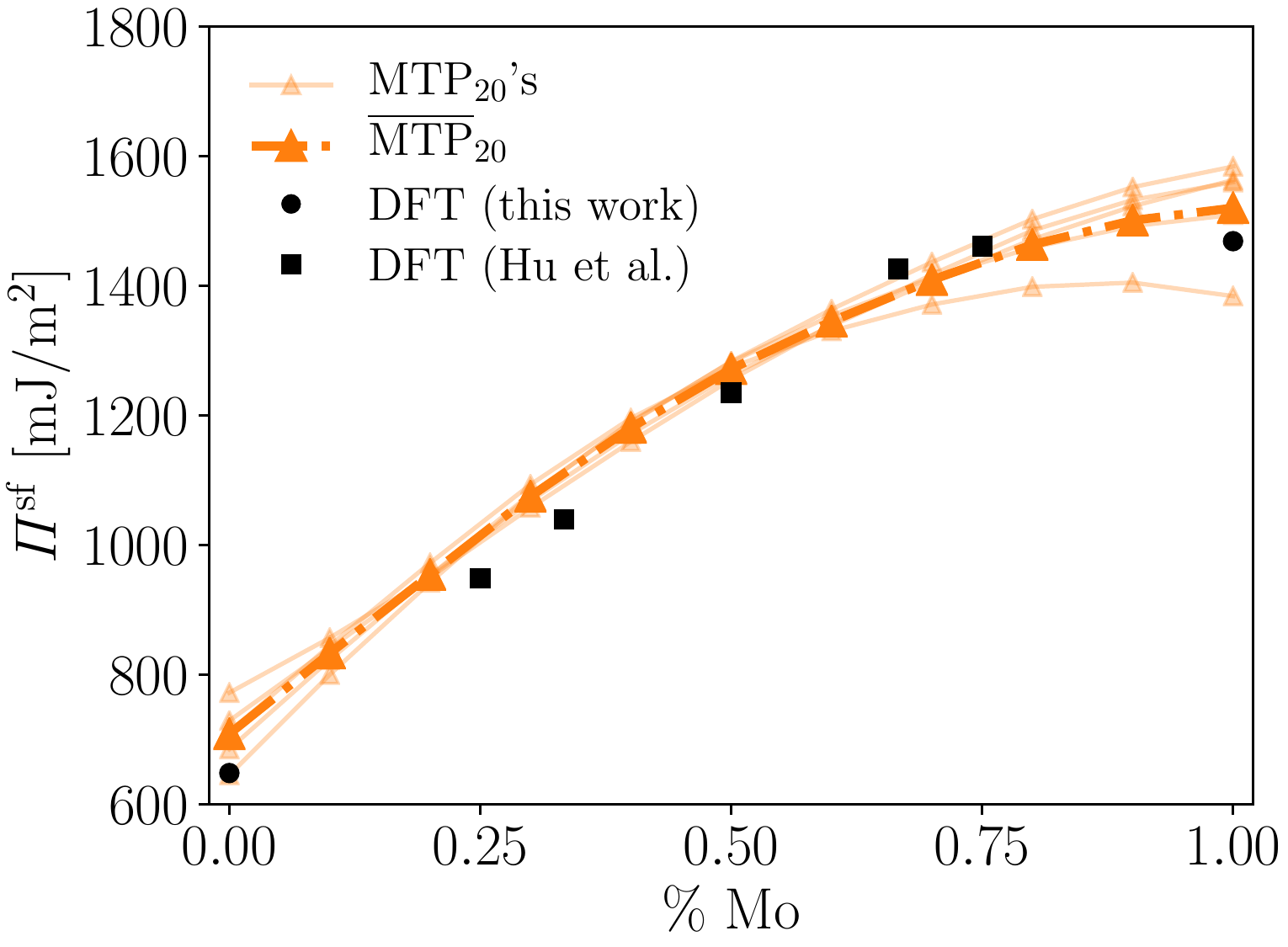}
 \end{minipage}
 \caption{SFE for MoNb as a function of the Mo content predicted by (a) level-16 MTPs, and by (d) level-20 MTPs (all trained on $\scT_\mrm{MoNb}$). The average predictions ($\overline{\text{MTP}}_X$) coincide with the DFT values up to a few percent (except for \%\;Mo\,$\lessapprox$\,0.25 and \%\;Mo\,$\gtrapprox$\,0.75 for the level-16 MTPs) which is very good, given that $\scT_\mrm{MoNb}$ contains only 44 configurations}
 \label{fig:results_MoNb}
\end{figure}

We then applied the active learning algorithm to the MoNb binary using grid type 5 and $n_\mrm{rand}$\,=\,100.
The training set $\scT_\mrm{MoNb}$, generated with this setup, is slightly smaller than for MoTa (see Table \ref{tab:ts_binaries}).
This might be the reason that the training errors, reported in Table \ref{tab:errors_MoNb_NbTa}, are slightly (but not significantly) smaller than for the MoTa potentials.
The smaller training set is likely also the reason that the accuracy of the level-16 MTP is rather poor for close-to unary compositions (Figure \ref{fig:results_MoNb} (a)).
Increasing the MTP level to 20, however, helped to overcome this deficiency, the deviations from the DFT SFEs now lying at most between $\approx$\,6--7\,\%, as for the MoTa potentials, which is very good.

We remark here that, instead of increasing the MTP level, adding more training data (e.g., by using a smaller $\gamma_\mrm{min}$) would be another possibility which improves the accuracy of the level-16 MTPs.
We will show this in the following section for the ternary alloy, where improving the accuracy by increasing the MTP level did not help.

\begin{table}[H]
 \centering
 \begin{tabular}{|c|c|c|c|}
  \hline
  Error & MTP$_{16}$($\scT_\mrm{MoNb}$) & MTP$_{20}$($\scT_\mrm{MoNb}$) & MTP$_{16}$($\scT_\mrm{NbTa}$) \\ \hline\hline
  $\mepsilon_\mrm{ave}^\mrm{atom}$ [meV]     & 1.21\,$\pm$\,0.15   & 0.74\,$\pm$\,0.05   & 0.19\,$\pm$\,0.03 \\ \hline
  $\mepsilon_\mrm{rms}^\mrm{atom}$ [meV]     & 1.69\,$\pm$\,0.09   & 0.97\,$\pm$\,0.02   & 0.28\,$\pm$\,0.04 \\ \hline
  $\mepsilon_\mrm{ave}^\mrm{force}$ [eV/\AA] & 0.059\,$\pm$\,0.002 & 0.049\,$\pm$\,0.002 & 0.029\,$\pm$\,0.002 \\ \hline
  $\mepsilon_\mrm{rms}^\mrm{force}$ [eV/\AA] & 0.067\,$\pm$\,0.002 & 0.055\,$\pm$\,0.002 & 0.031\,$\pm$\,0.003 \\ \hline
 \end{tabular}
 \caption{Mean training errors and deviations between the MTPs with different parameter initializations for the MoNb and NbTa binary systems. The errors, including the deviations, are very small (of the order of 1\,meV for per-atom energies and of the order of 0.1\,eV/{\AA} for per-atom forces---\emph{or smaller}), indicating overall reliability of the training}
 \label{tab:errors_MoNb_NbTa}
\end{table}

\begin{figure}[t]
 \centering
 \begin{minipage}{0.333333\textwidth}
  \centering
  \includegraphics[width=0.97\textwidth]{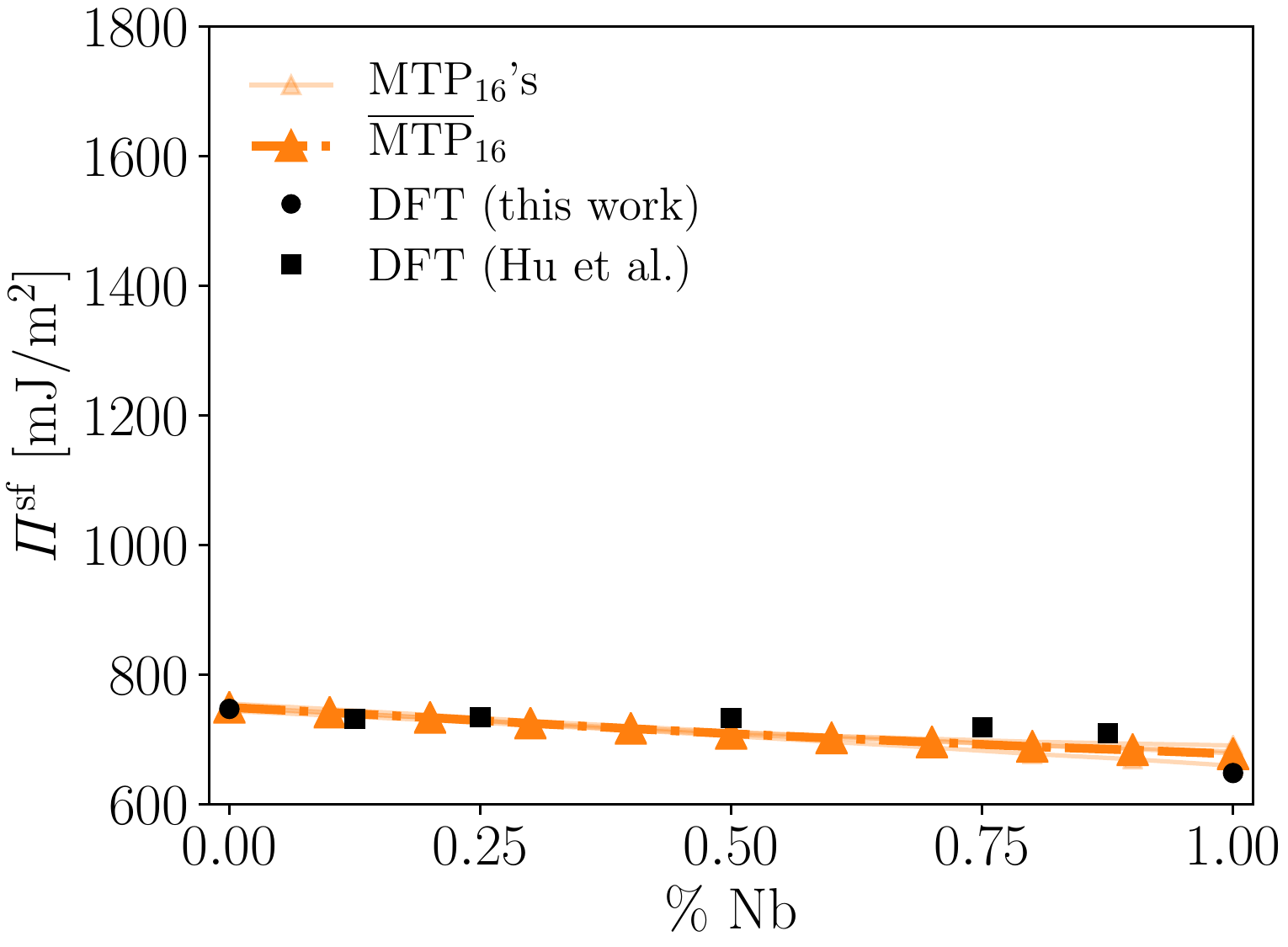}
 \end{minipage}
 \caption{SFE for NbTa as a function of the Nb content predicted by an ensemble of level-16 MTPs which reproduce the DFT values without any practical error}
 \label{fig:results_NbTa}
\end{figure}

Finally, we report the results for NbTa.
Since the configurational space, spanned by the set of training candidates, is smaller than for the other two binaries (since the lattice constants for Nb and Ta are roughly the same; cf. Table \ref{tab:a0_&_SFE_pure}) the training set is also smaller (see Table \ref{tab:ts_binaries}) and the training errors are, by far, the lowest among all binary MTPs (see Table \ref{tab:errors_MoNb_NbTa}).
Hence, it is not surprising that a level-16 MTP is sufficient to accurately predict the SFE of NbTa (Figure \ref{fig:results_NbTa}).

\subsection{MoNbTa ternary system}
\label{sec:results.ternary}

\subsubsection{Results using a random initialization of the training set}

\begin{figure}[t]
 \begin{minipage}{0.333333\textwidth}
  \centering
  (a)
 \end{minipage}\hfill
 \begin{minipage}{0.333333\textwidth}
  \centering
  (b)
 \end{minipage}\hfill
 \begin{minipage}{0.333333\textwidth}
  \centering
  (c)
 \end{minipage}\\
 \begin{minipage}{0.333333\textwidth}
  \centering
  \includegraphics[width=0.97\textwidth]{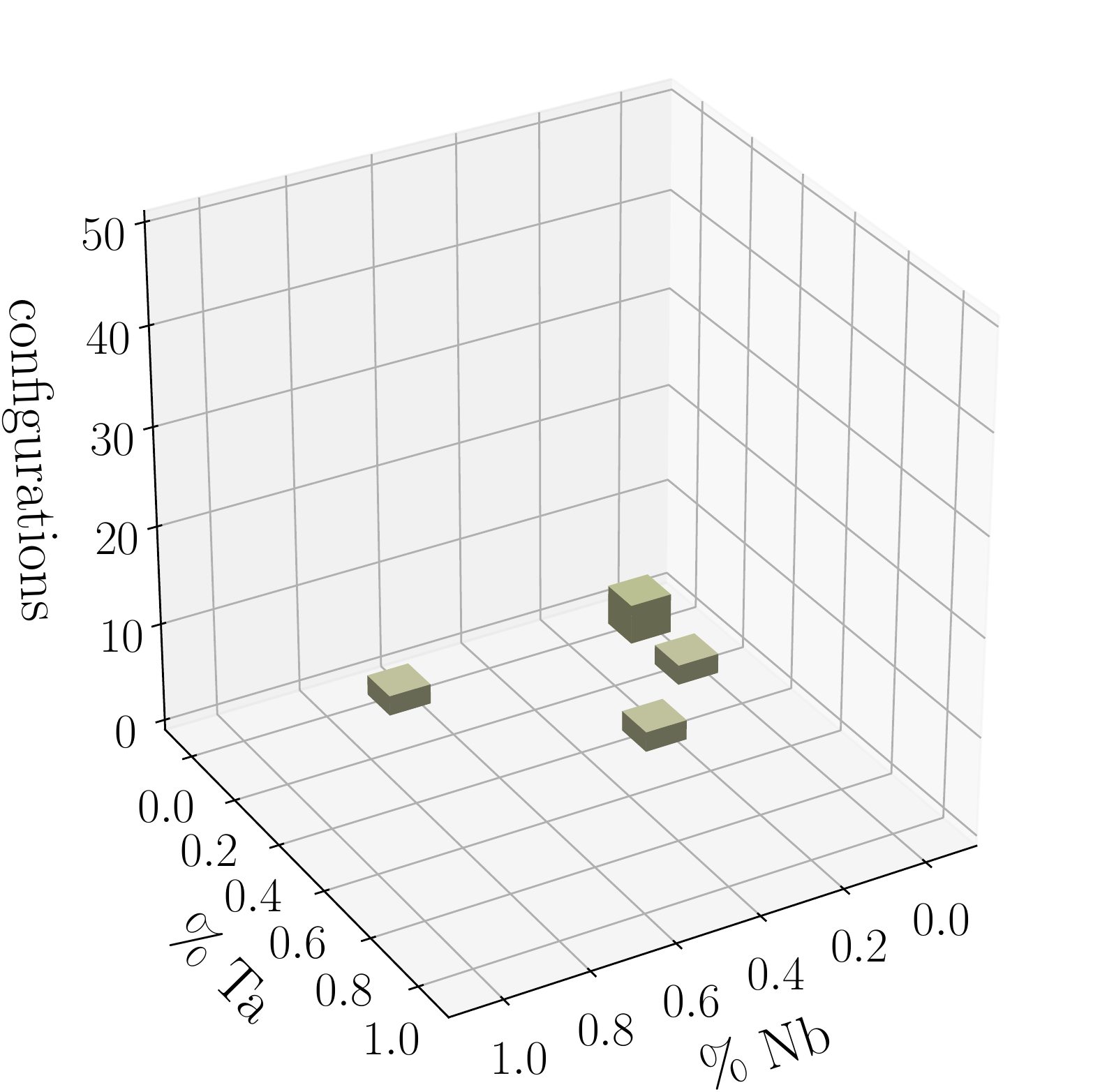}
 \end{minipage}\hfill
 \begin{minipage}{0.333333\textwidth}
  \centering
  \includegraphics[width=0.97\textwidth]{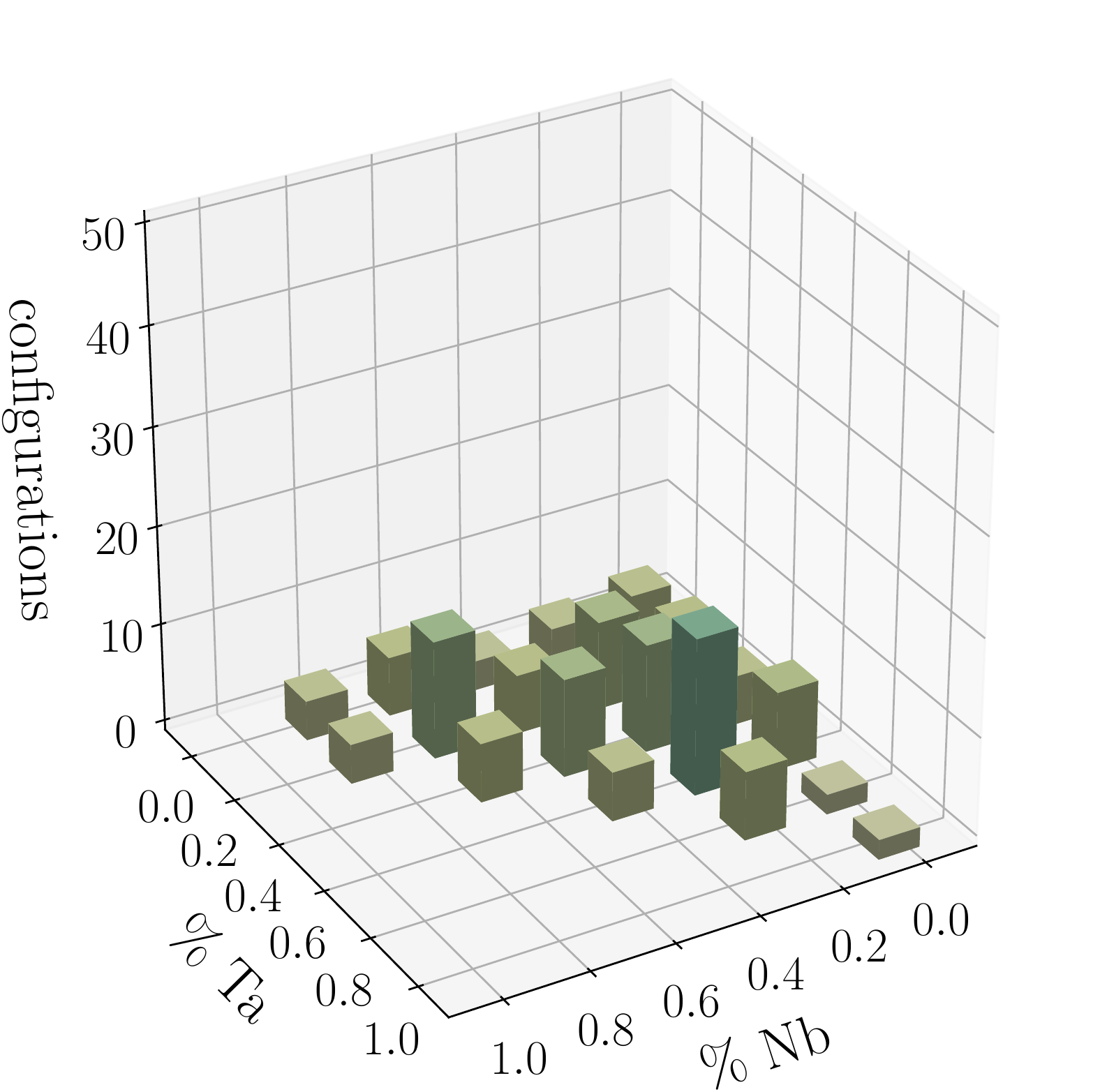}
 \end{minipage}\hfill
 \begin{minipage}{0.333333\textwidth}
  \centering
  \includegraphics[width=0.97\textwidth]{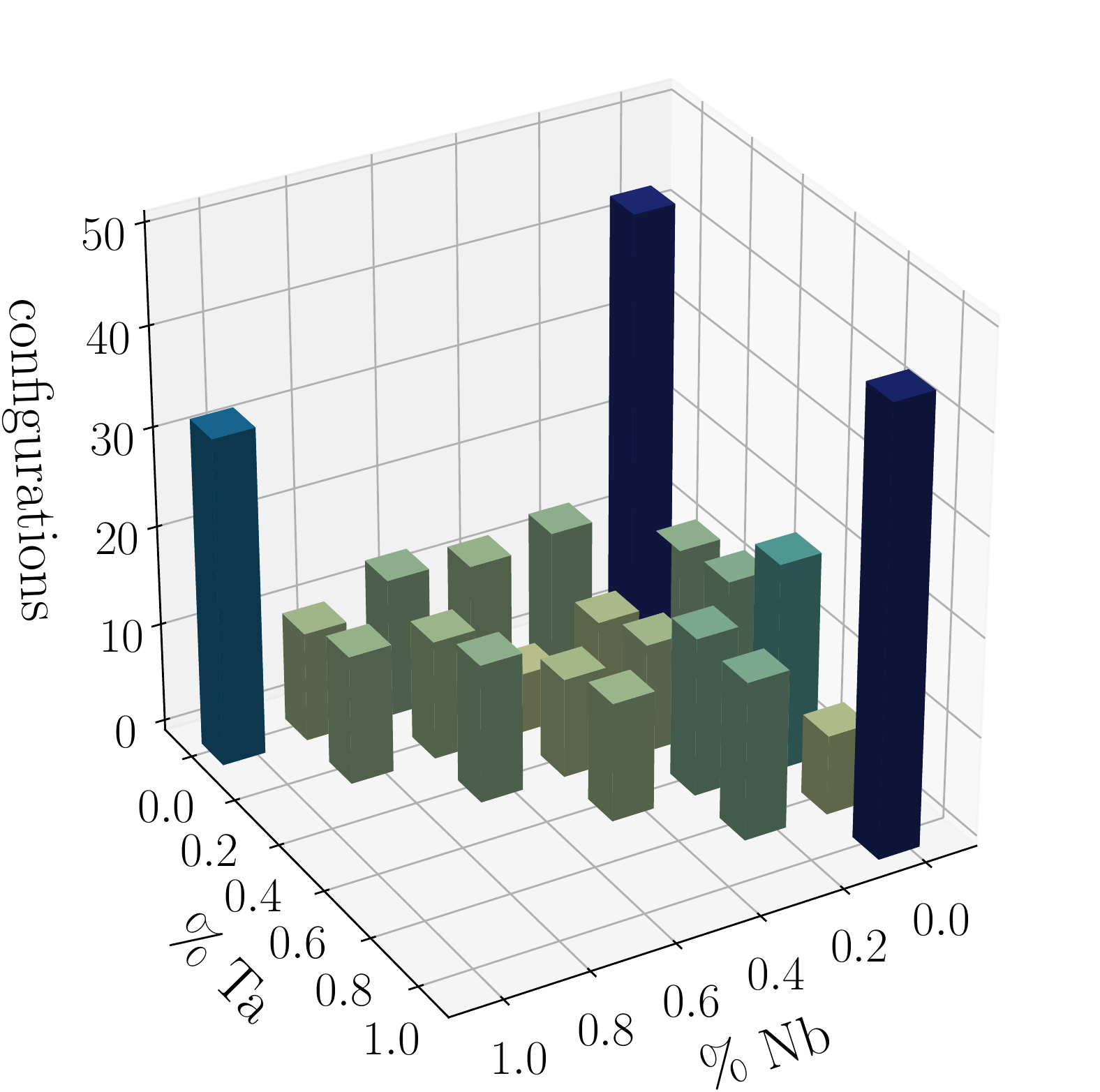}
 \end{minipage}
 \caption{Number of configurations per composition at various stages of the active learning algorithm. (a) Initial training set randomly selected from the full training candidate set after \textbf{Step\;2}. (b) Training set $\scT_\mrm{MoNbTa}^2$ generated by the active learning algorithm using $\gamma_\mrm{min}$\,$=$\,$1.4$. (c) Training set $\scT_\mrm{MoNbTa}^3$ after adding unary and binary configurations to $\scT_\mrm{MoNbTa}^2$.
 The added unary and binary configurations are necessary so that the MTPs predict the SFE over the entire compositional space (cf. Figure \ref{fig:results_MoNbTa_enriched_ts})}
 \label{fig:results_densities}
\end{figure}

To generate the training set for the MoNbTa ternary alloy, in the following denoted by $\scT_\mrm{MoNbTa}^1$, we essentially proceed as for the binaries, with the difference that we now use a \emph{three-dimensional} grid of type 5 (see Figure \ref{tab:grids}).
From this set of training candidates, we then randomly select ten configurations which serve as the initial training set (see Figure \ref{fig:results_densities} (a)).

Interestingly, the number of configurations in the training set is with 83 configurations only slightly larger than for the binaries (Table \ref{tab:ts_binaries}).
Presumably, active learning selects a more diverse set of configurations that interpolate on many of the configurations that would have been marked as extrapolative when using a smaller set of candidate configurations, as for the binaries.
The training errors (Table \ref{tab:errors_MoNbTa}) are in the same range as for the MoTa and MoNb binaries, indicating a good fit.

\begin{table}[b]
 \centering
 \begin{tabular}{|c|c|c|c|c|c|}
  \hline
  Error & MTP$_{16}$($\scT_\mrm{MoNbTa}^1$) & MTP$_{20}$($\scT_\mrm{MoNbTa}^1$) & MTP$_{16}$($\scT_\mrm{MoNbTa}^2$) & MTP$_{16}$($\scT_\mrm{MoNbTa}^3$) & MTP$_{20}$($\scT_\mrm{MoNbTa}^3$) \\ \hline\hline
  $\mepsilon_\mrm{ave}^\mrm{atom}$ [meV]     & 1.21\,$\pm$\,0.16   & 0.85\,$\pm$\,0.24   & 1.41\,$\pm$\,0.05 & 1.36\,$\pm$\,0.15 & 1.10\,$\pm$\,0.05 \\ \hline
  $\mepsilon_\mrm{rms}^\mrm{atom}$ [meV]     & 1.66\,$\pm$\,0.19   & 1.15\,$\pm$\,0.33   & 1.94\,$\pm$\,0.08 & 2.09\,$\pm$\,0.27 & 1.60\,$\pm$\,0.08\\ \hline
  $\mepsilon_\mrm{ave}^\mrm{force}$ [eV/\AA] & 0.068\,$\pm$\,0.001 & 0.059\,$\pm$\,0.003 & 0.072\,$\pm$\,0.003 & 0.071\,$\pm$\,0.006 & 0.061\,$\pm$\,0.002\\ \hline
  $\mepsilon_\mrm{rms}^\mrm{force}$ [eV/\AA] & 0.078\,$\pm$\,0.001 & 0.066\,$\pm$\,0.004 & 0.082\,$\pm$\,0.003 & 0.084\,$\pm$\,0.007 & 0.071\,$\pm$\,0.003\\ \hline
 \end{tabular}
 \caption{Training errors for the level-16 and level-20 MTPs that have been trained on the sets $\scT_\mrm{MoNbTa}^1$, $\scT_\mrm{MoNbTa}^2$, and $\scT_\mrm{MoNbTa}^3$, for the MoNbTa ternary as described in in Section \ref{sec:results.ternary}. The errors are similar to, or not significantly higher than, the errors for the binary MTPs (cf. Table \ref{tab:errors_MoTa} and \ref{tab:errors_MoNb_NbTa}) indicating a reliable fit}
 \label{tab:errors_MoNbTa}
\end{table}

\begin{figure}[t]
 \centering
 \includegraphics[width=0.5\textwidth]{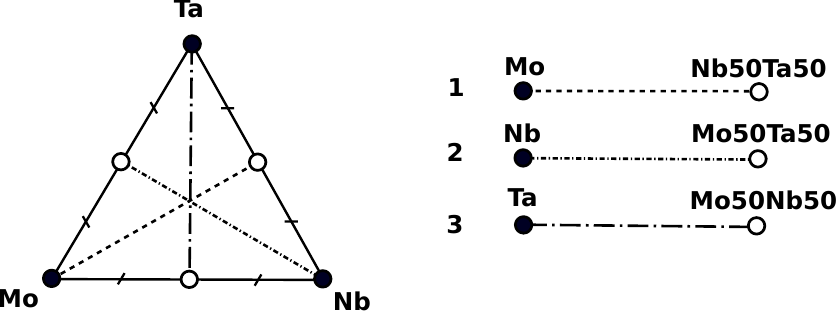}
 \caption{Three curves going through the MoNbTa ternary diagram for which the SFE is analyzed in Section \ref{sec:results.ternary}}
 \label{fig:spaghetti}
\end{figure}

We analyze the SFE predicted by MTPs for the three curves, \curveA, \curveB, \curveC, from Figure \ref{fig:spaghetti} that capture different parts of the ternary diagram.
Due to the lack of existing pure-DFT values for ternary compositions of MoNbTa, we compare our predictions to the values obtained with the surrogate model of \citet{hu_screening_2021}.
Their surrogate model is based on a set of descriptors which are functions of, i.a., atomic bond features and, of course, the alloy composition.
It was shown by \citet{hu_screening_2021} that their surrogate model is able to reproduce the SFE of a test set, consisting of up to quarternary random alloys, with a root-mean-square error of only 47\,mJ/m$^2$ (corresponding to relative errors of a few percent) and, so, provides a legitimate reference.

\begin{figure}[t]
 \begin{minipage}{0.333333\textwidth}
  \centering
  (a) \curveA
 \end{minipage}\hfill
 \begin{minipage}{0.333333\textwidth}
  \centering
  (b) \curveB
 \end{minipage}\hfill
 \begin{minipage}{0.333333\textwidth}
  \centering
  (c) \curveC
 \end{minipage}\\[0.5em]
 \begin{minipage}{0.333333\textwidth}
  \centering
  \includegraphics[width=0.97\textwidth]{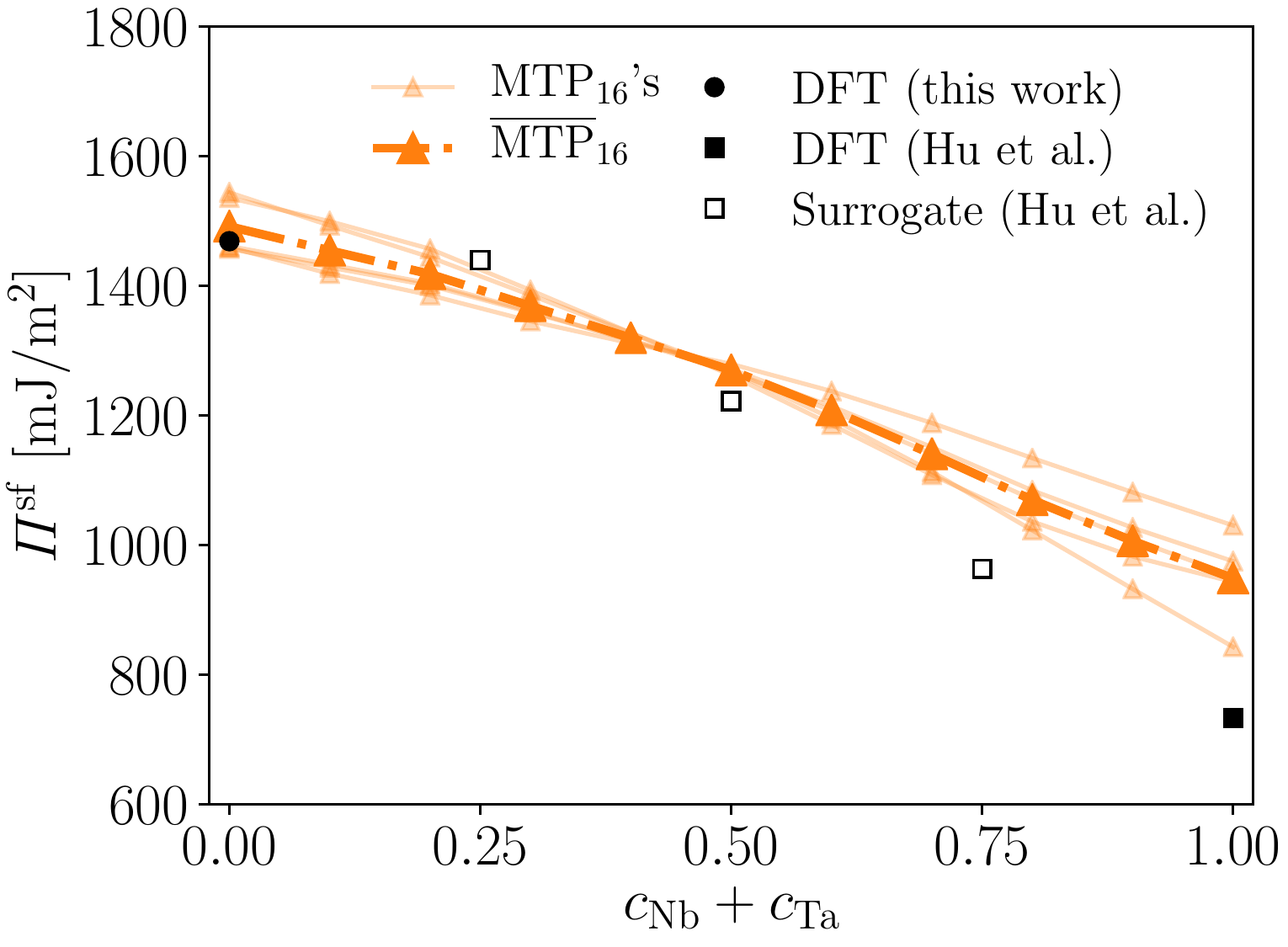}
 \end{minipage}\hfill
 \begin{minipage}{0.333333\textwidth}
  \centering
  \includegraphics[width=0.97\textwidth]{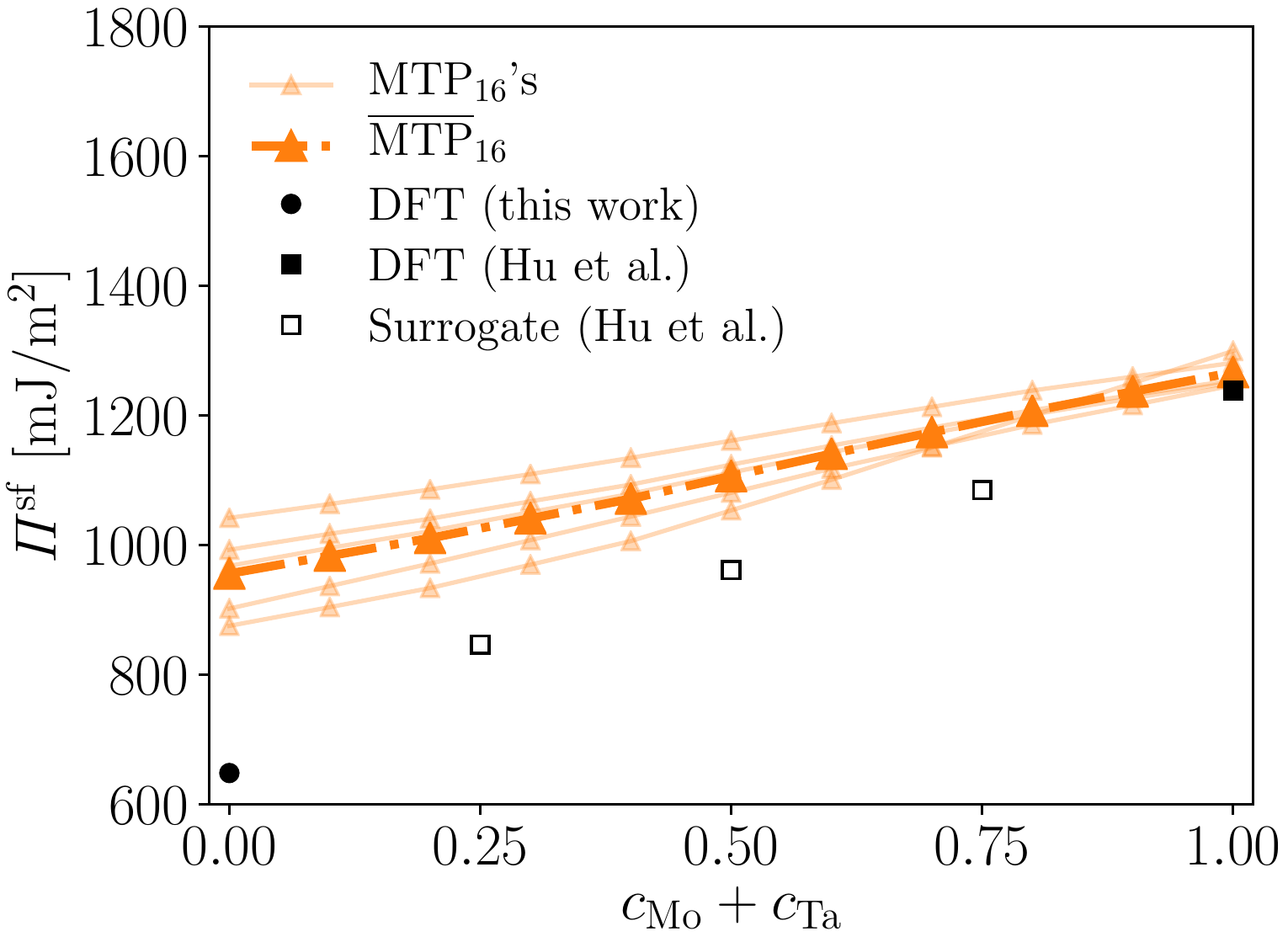}
 \end{minipage}\hfill
 \begin{minipage}{0.333333\textwidth}
  \centering
  \includegraphics[width=0.97\textwidth]{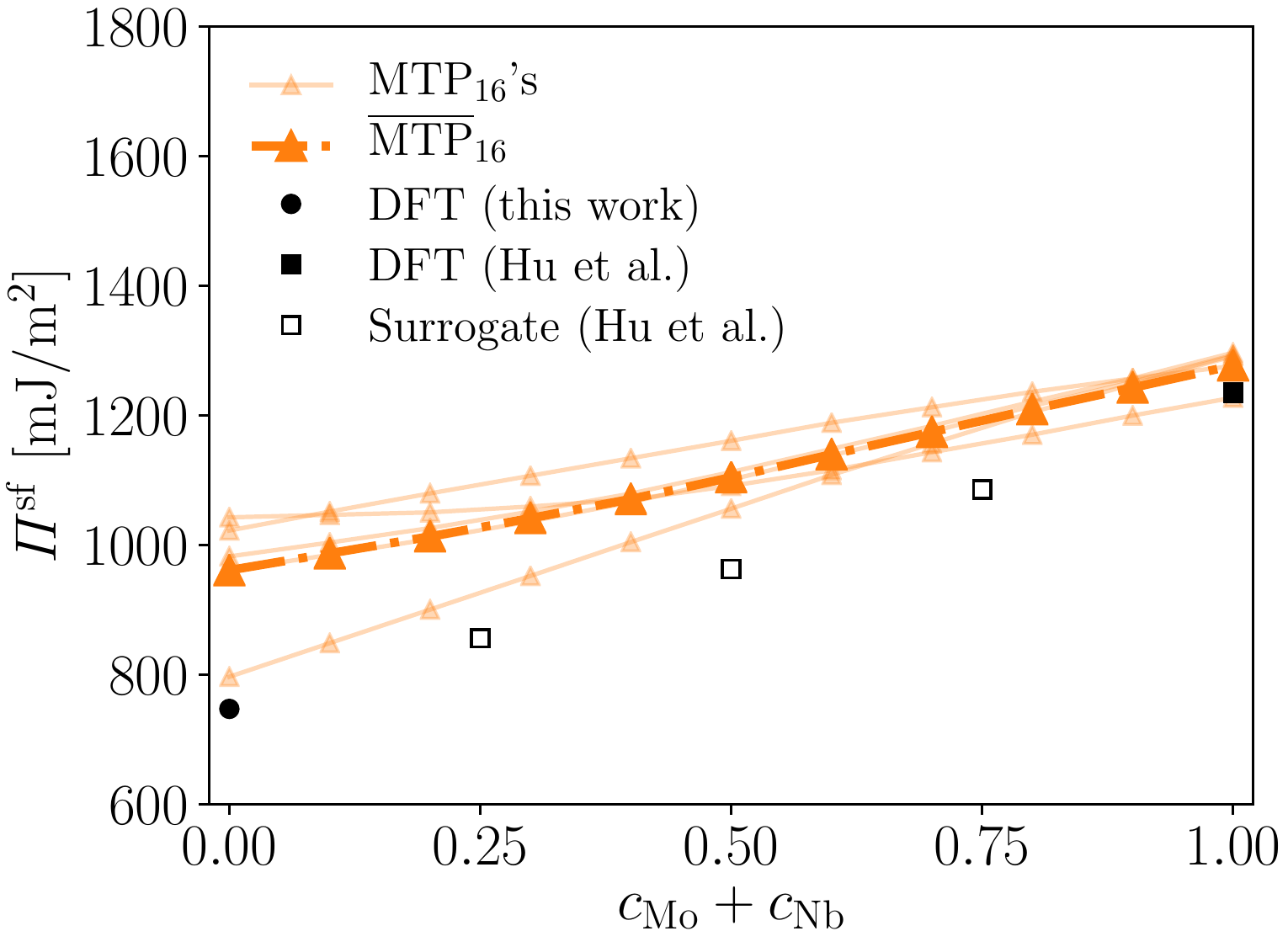}
 \end{minipage}\\[0.5em]
 \begin{minipage}{0.333333\textwidth}
  \centering
  (d) \curveA
 \end{minipage}\hfill
 \begin{minipage}{0.333333\textwidth}
  \centering
  (e) \curveB
 \end{minipage}\hfill
 \begin{minipage}{0.333333\textwidth}
  \centering
  (f) \curveC
 \end{minipage}\\[0.5em]
 \begin{minipage}{0.333333\textwidth}
  \centering
  \includegraphics[width=0.97\textwidth]{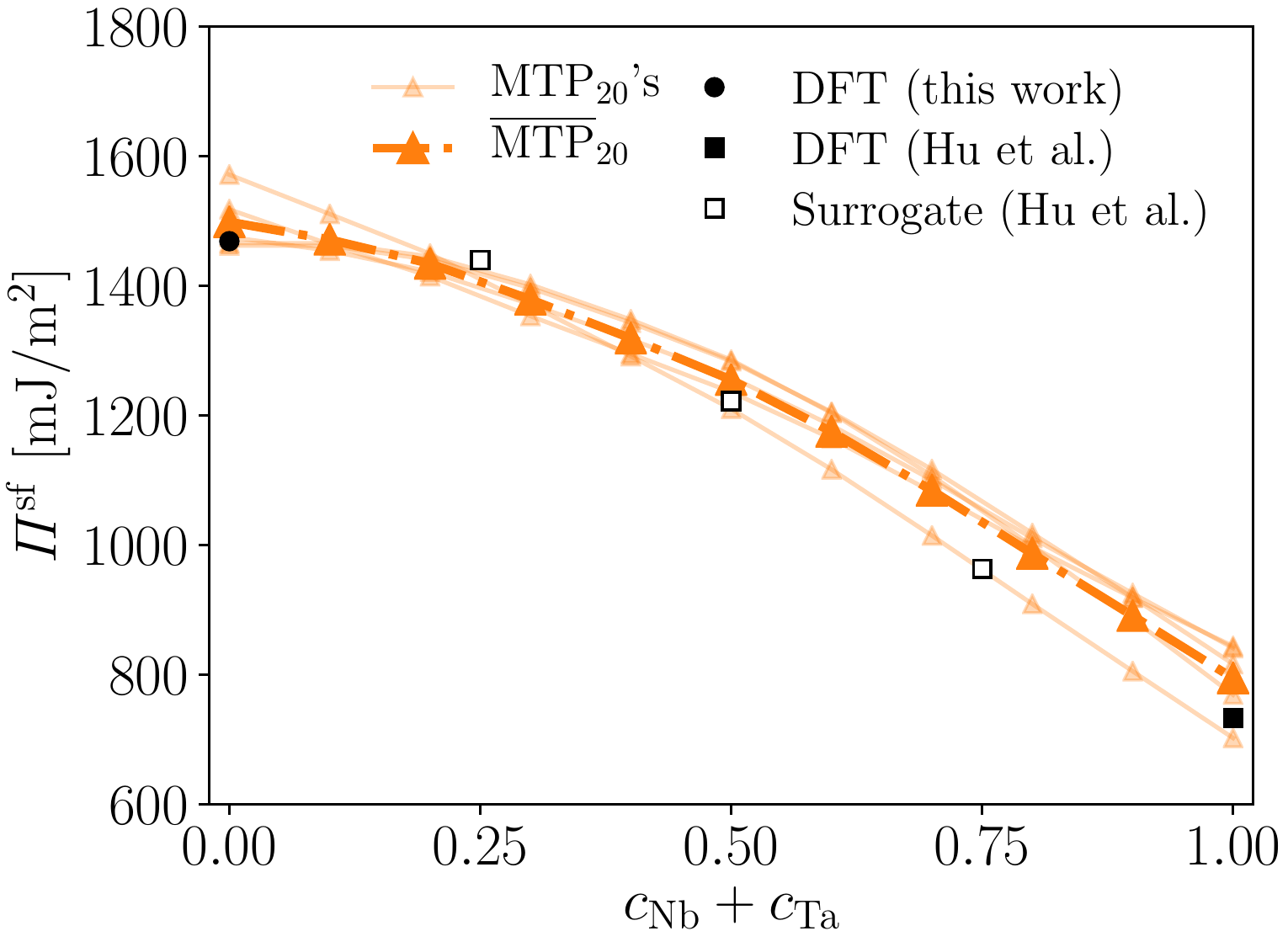}
 \end{minipage}\hfill
 \begin{minipage}{0.333333\textwidth}
  \centering
  \includegraphics[width=0.97\textwidth]{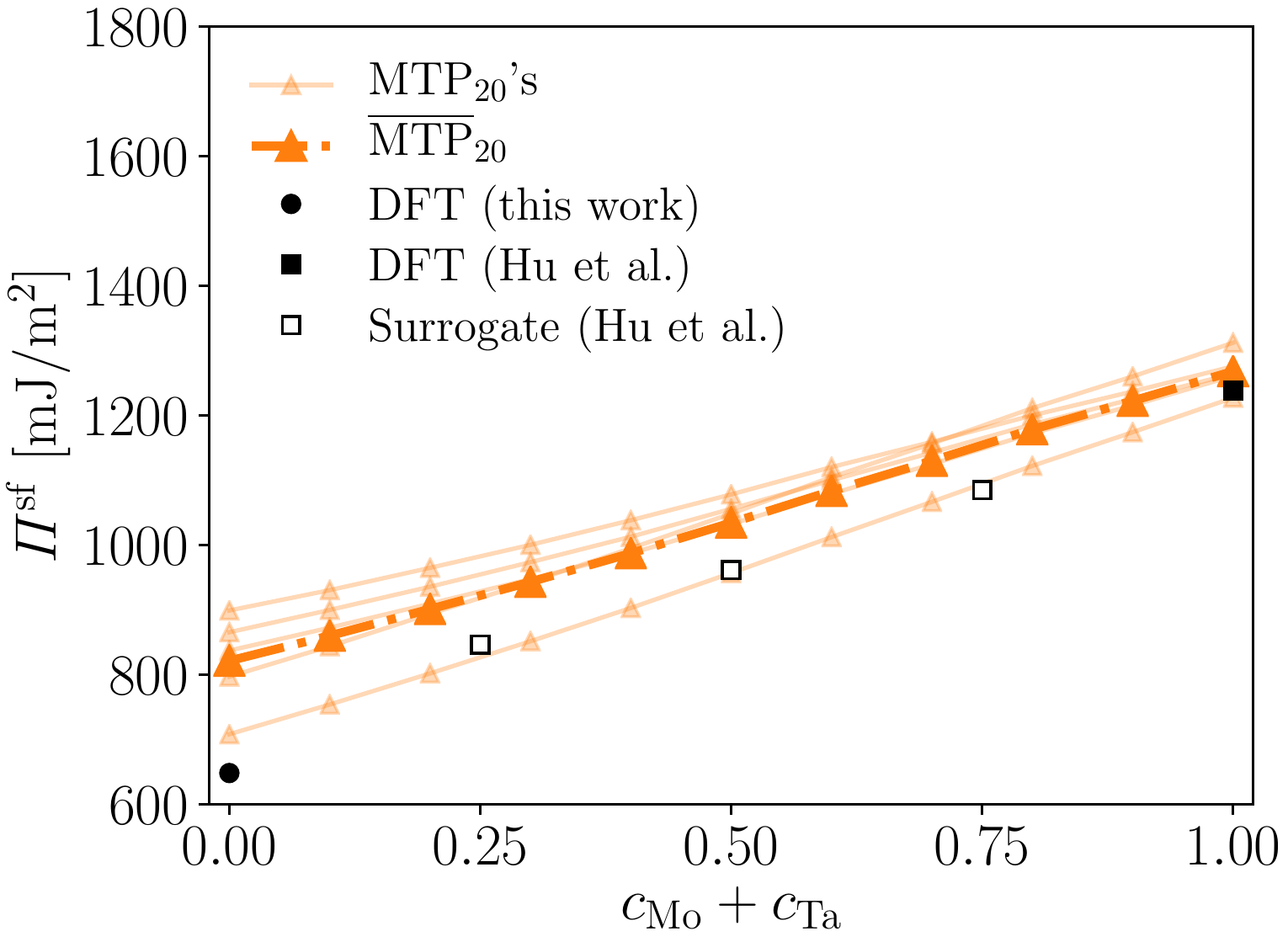}
 \end{minipage}\hfill
 \begin{minipage}{0.333333\textwidth}
  \centering
  \includegraphics[width=0.97\textwidth]{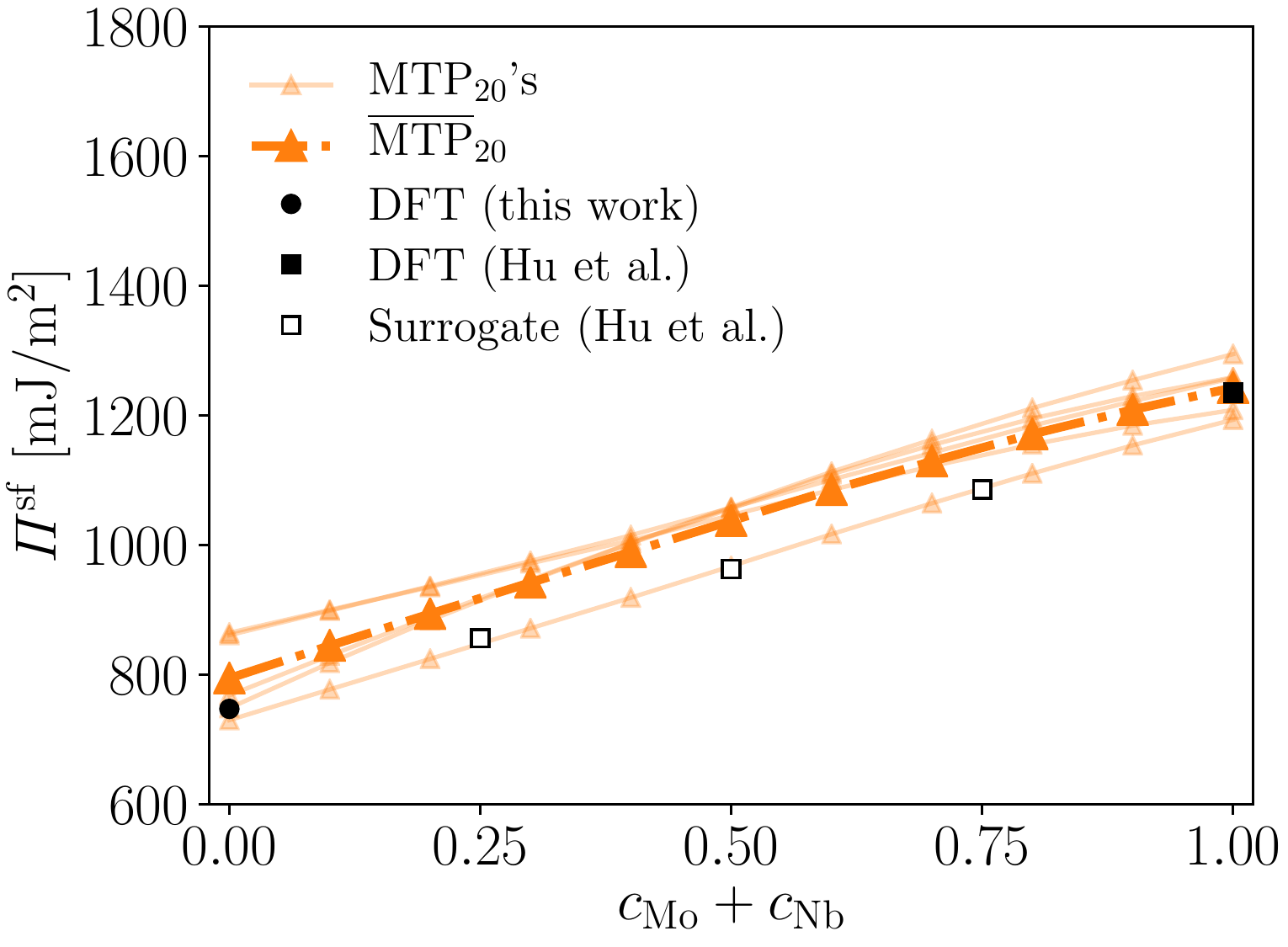}
 \end{minipage}
 \caption{SFEs along the three curves from Figure \ref{fig:spaghetti} as predicted by level-16 MTPs in (a)--(c) and level-20 MTPs in (d)--(f) trained on $\scT_\mrm{MoNbTa}^1$ that has been generated by the active learning algorithm described in Section \ref{sec:method.algo}. From (a)--(c) it can be observed that an MTP level of 16 is not sufficient to predict the SFE uniformly over the entire compositional space. The level-20 MTP is able to do so, except in regions with high Nb content in (e)}
 \label{fig:results_MoNbTa}
\end{figure}

However, despite the low training errors, the predictions with level-16 MTPs are rather poor, showing deviations from the DFT and surrogate values of up to $\sim$\,50\,\%, as shown in Figure \ref{fig:results_MoNbTa}, in particular near the corners when approaching binary and unary compositions.
Increasing the MTP level to 20 significantly improves the results for \curveA and \curveC, but for \curveB errors between 20 and 30\,\% still persist in regions with high Nb content.
Increasing the MTP level of the potential further did not resolve this problem.

\begin{figure}[t]
 \begin{minipage}{0.333333\textwidth}
  \centering
  (a) \curveA
 \end{minipage}\hfill
 \begin{minipage}{0.333333\textwidth}
  \centering
  (b) \curveB
 \end{minipage}\hfill
 \begin{minipage}{0.333333\textwidth}
  \centering
  (c) \curveC
 \end{minipage}\\[0.5em]
 \begin{minipage}{0.333333\textwidth}
  \centering
  \includegraphics[width=0.97\textwidth]{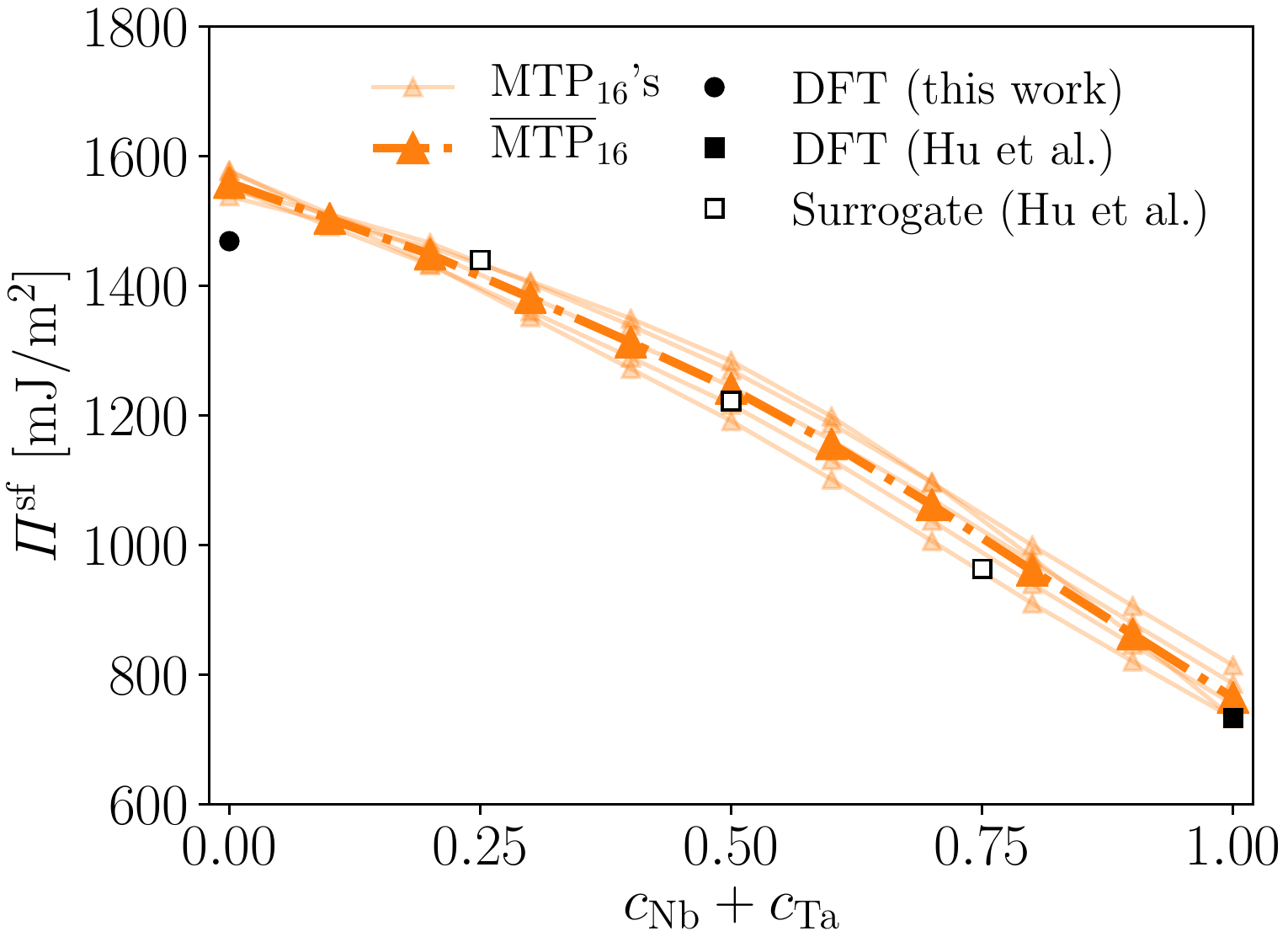}
 \end{minipage}\hfill
 \begin{minipage}{0.333333\textwidth}
  \centering
  \includegraphics[width=0.97\textwidth]{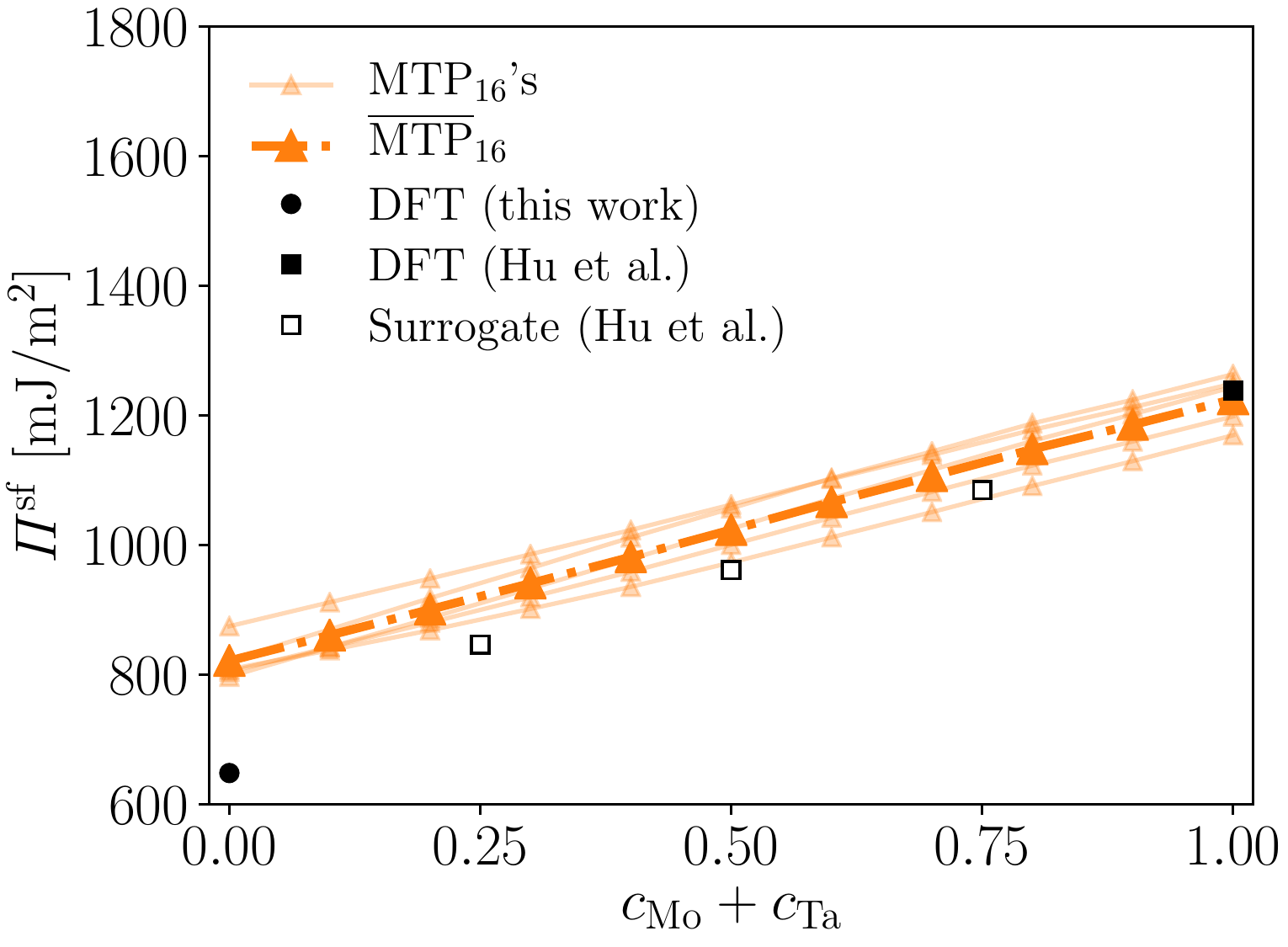}
 \end{minipage}\hfill
 \begin{minipage}{0.333333\textwidth}
  \centering
  \includegraphics[width=0.97\textwidth]{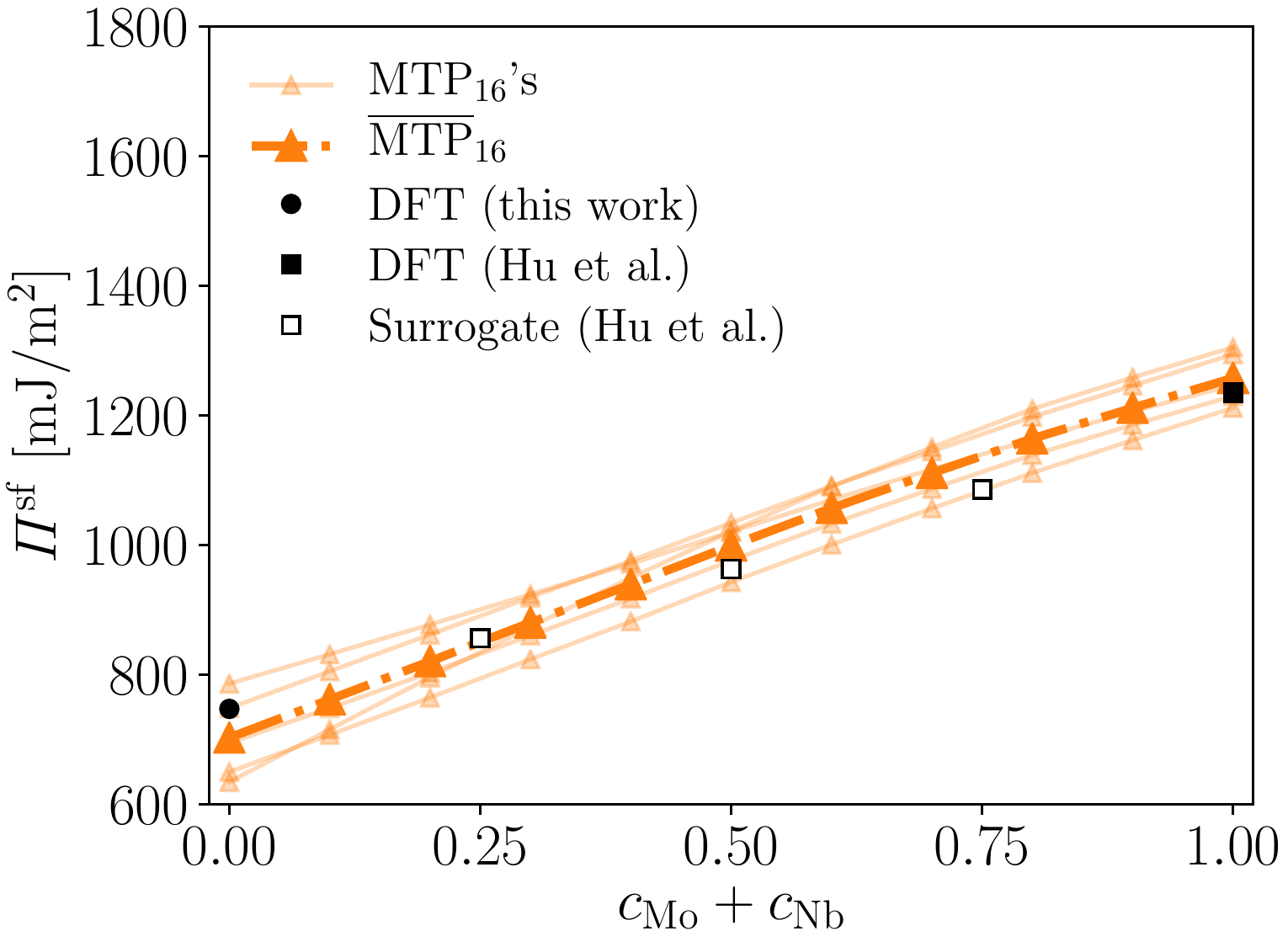}
 \end{minipage}
 \caption{SFEs along the three curves from Figure \ref{fig:spaghetti} as predicted by level-16 MTPs trained on $\scT_\mrm{MoNbTa}^2$ that has been generated by rerunning the active learning algorithm with a smaller minimum extrapolation grade than the one used to generate $\scT_\mrm{MoNbTa}^1$. The additional training data helps to improve the predictions heavily (compare with Figure \ref{fig:results_MoNbTa} (a)--(c)) but some discrepancy in the high-Nb regime is still present}
 \label{fig:results_MoNbTa_mingrade=1.4}
\end{figure}

We have thus tried to improve the accuracy of the MTPs by adding more training data and, therefore, generated a second training set,
\begin{itemize}[leftmargin=2.7cm,labelsep=1cm]
 \item[$\scT_\mathrm{MoNbTa}^2$,]
 by rerunning the active learning algorithm with a smaller minimum extrapolation grade of 1.4, and using $\scT_\mathrm{MoNbTa}^1$ as the initial training set.
\end{itemize}
This training set now contains 131 configurations, so more than 50\,\% more than $\scT_\mathrm{MoNbTa}^1$.
The training errors increased slightly when compared to MTP$_{16}$($\scT_\mathrm{MoNbTa}^1$) (see Figure \ref{tab:errors_MoNbTa}), which is expected.
The SFEs predicted by the MTP$_{16}$($\scT_\mathrm{MoNbTa}^2$)'s, shown in Figure \ref{fig:results_MoNbTa_mingrade=1.4}, improved significantly over the MTP$_{16}$($\scT_\mathrm{MoNbTa}^1$)'s and are now comparable to the ones obtained with the MTP$_{20}$($\scT_\mathrm{MoNbTa}^1$)'s.
However, an error of up to $\approx$\,20\,\% still remains in the left part of \curveB, i.e., for compositions with high Nb content.
Unfortunately, increasing the MTP level to 20 did not improve the results decisively.

By inspection of the training set, we observed that it contains slightly more Mo and Ta than Nb atoms.
Moreover, $\scT_\mrm{MoNbTa}^2$ does \emph{not} contain a \emph{pure-Nb} configuration, as shown in Figure \ref{fig:results_densities} (b).
This implies that active learning considers the pure-Nb from the training candidate set as interpolative between binary/ternary configurations.

At the moment, we cannot offer a definite explanation of this behavior.
However, we believe that it is related to the fact that potentials satisfying \eqref{eq:Epartition} are explicitly learning the dependence of the energy on the local atomic environment, but only implicitly learning the average impact of the far-field on the energy.
Because the binary/ternary configurational space is much larger than the unary one, the average far-field in the training database corresponds to binaries or ternaries.
But as far as the local environments are concerned, the D-optimality criterion asserts that the pure-Nb environments are interpolative.
We remark that this behavior is not unique to D-optimality and likely happens with other active learning algorithms as well.
This can be deduced from the small difference between the MTP predictions for pure Nb (Figure \ref{fig:results_MoNbTa_mingrade=1.4} (b)), so the popular query-by-committee method may also erroneously consider pure Nb as interpolative.

Even more interesting, we found that the maximum extrapolation grade computed for the pure-Nb configuration is less than 1.
Hence, adding more training data by rerunning the active learning algorithm with a minimum extrapolation grade of less than 1.4 is likely not going to improve the results.

In practice, this behavior becomes problematic when the initial training set---as in our case---does not contain a sufficient amount of Nb (Figure \ref{fig:results_densities} (a)).
The active learning algorithm may then iteratively consider Nb as less interpolative between Nb-rich binary and ternary configurations with higher stacking fault energies.

Therefore, to overcome this shortcoming of our algorithm, we created another training set,
\begin{itemize}[leftmargin=2.7cm,labelsep=1cm]
 \item[$\scT_\mathrm{MoNbTa}^3$,]
 by adding additional unary and binary training data, more precisely, the training sets $\scT_\mathrm{MoTa}^3$, $\scT_\mathrm{MoNb}$, and $\scT_\mathrm{NbTa}$ (cf. previous section), to $\scT_\mathrm{MoNbTa}^2$.
\end{itemize}
This training set contains 351 configurations, including several pure-Nb bulk configurations with 54 atoms, and pure-Nb configurations with 72 atoms containing a stacking fault (see Figure \ref{fig:results_densities} (c)).

\begin{figure}[t]
 \begin{minipage}{0.333333\textwidth}
  \centering
  (a) \curveA
 \end{minipage}\hfill
 \begin{minipage}{0.333333\textwidth}
  \centering
  (b) \curveB
 \end{minipage}\hfill
 \begin{minipage}{0.333333\textwidth}
  \centering
  (c) \curveC
 \end{minipage}\\[0.5em]
 \begin{minipage}{0.333333\textwidth}
  \centering
  \includegraphics[width=0.97\textwidth]{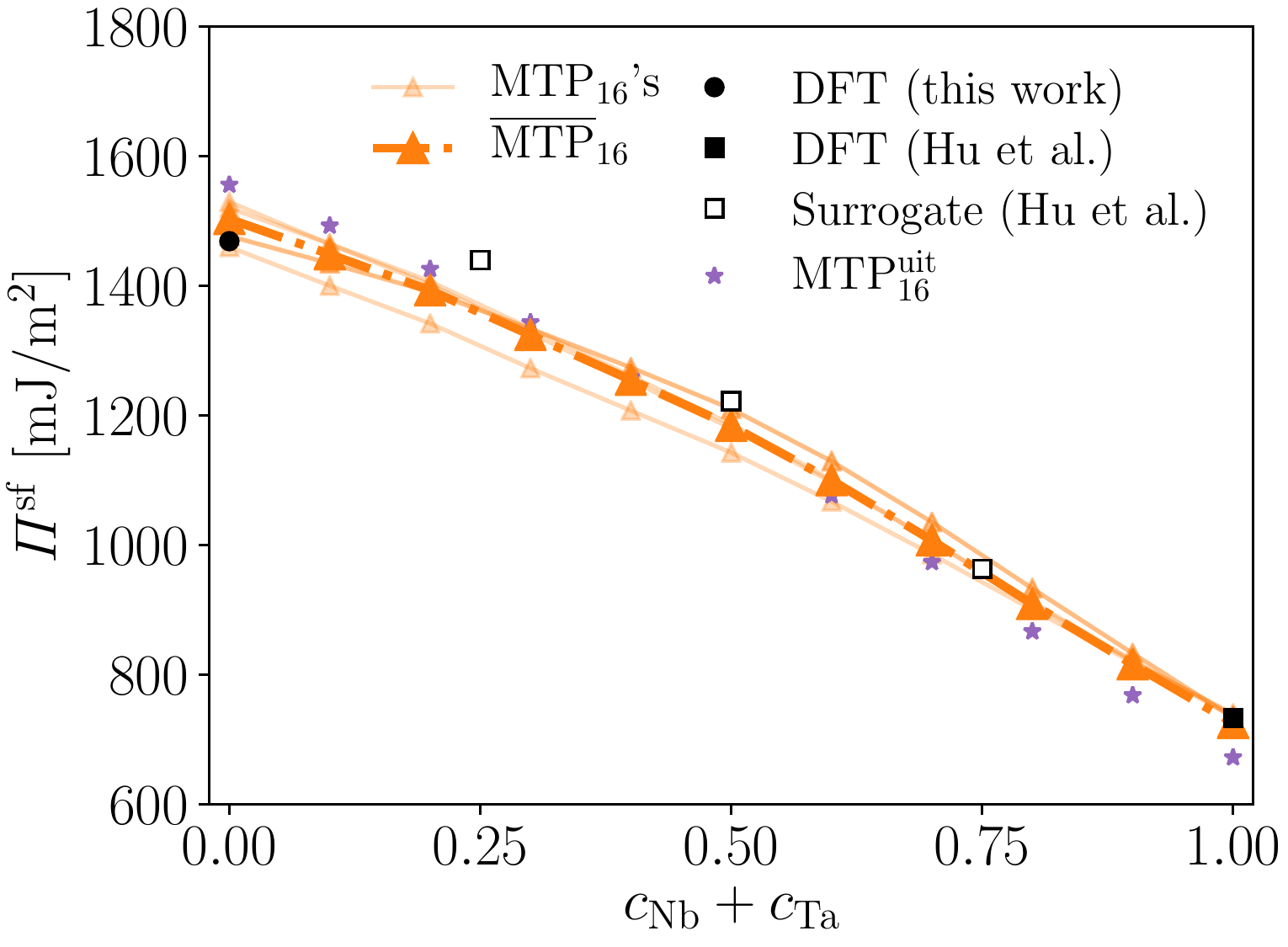}
 \end{minipage}\hfill
 \begin{minipage}{0.333333\textwidth}
  \centering
  \includegraphics[width=0.97\textwidth]{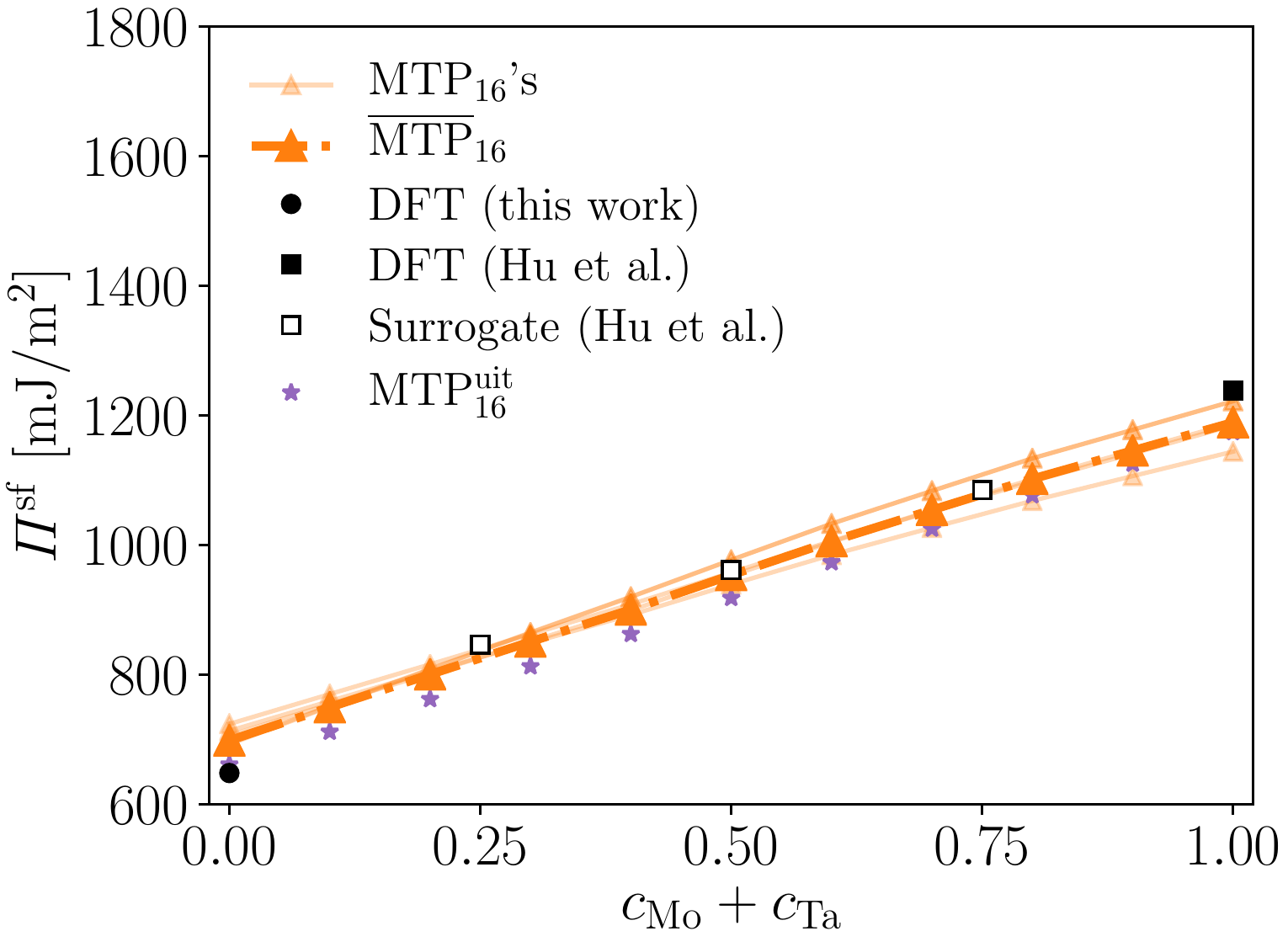}
 \end{minipage}\hfill
 \begin{minipage}{0.333333\textwidth}
  \centering
  \includegraphics[width=0.97\textwidth]{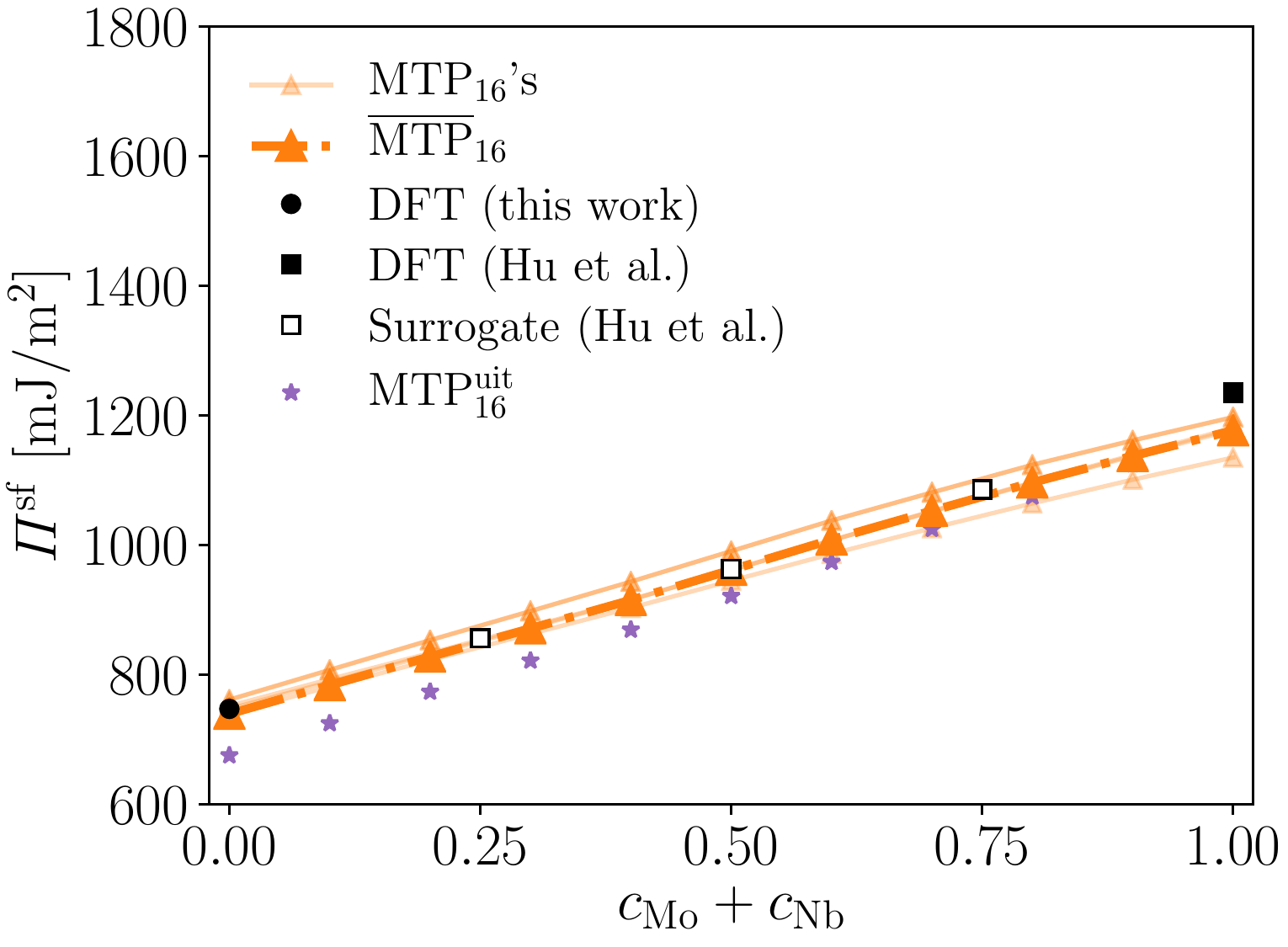}
 \end{minipage}\\[0.5em]
 \begin{minipage}{0.333333\textwidth}
  \centering
  (d) \curveA
 \end{minipage}\hfill
 \begin{minipage}{0.333333\textwidth}
  \centering
  (e) \curveB
 \end{minipage}\hfill
 \begin{minipage}{0.333333\textwidth}
  \centering
  (f) \curveC
 \end{minipage}\\[0.5em]
 \begin{minipage}{0.333333\textwidth}
  \centering
  \includegraphics[width=0.97\textwidth]{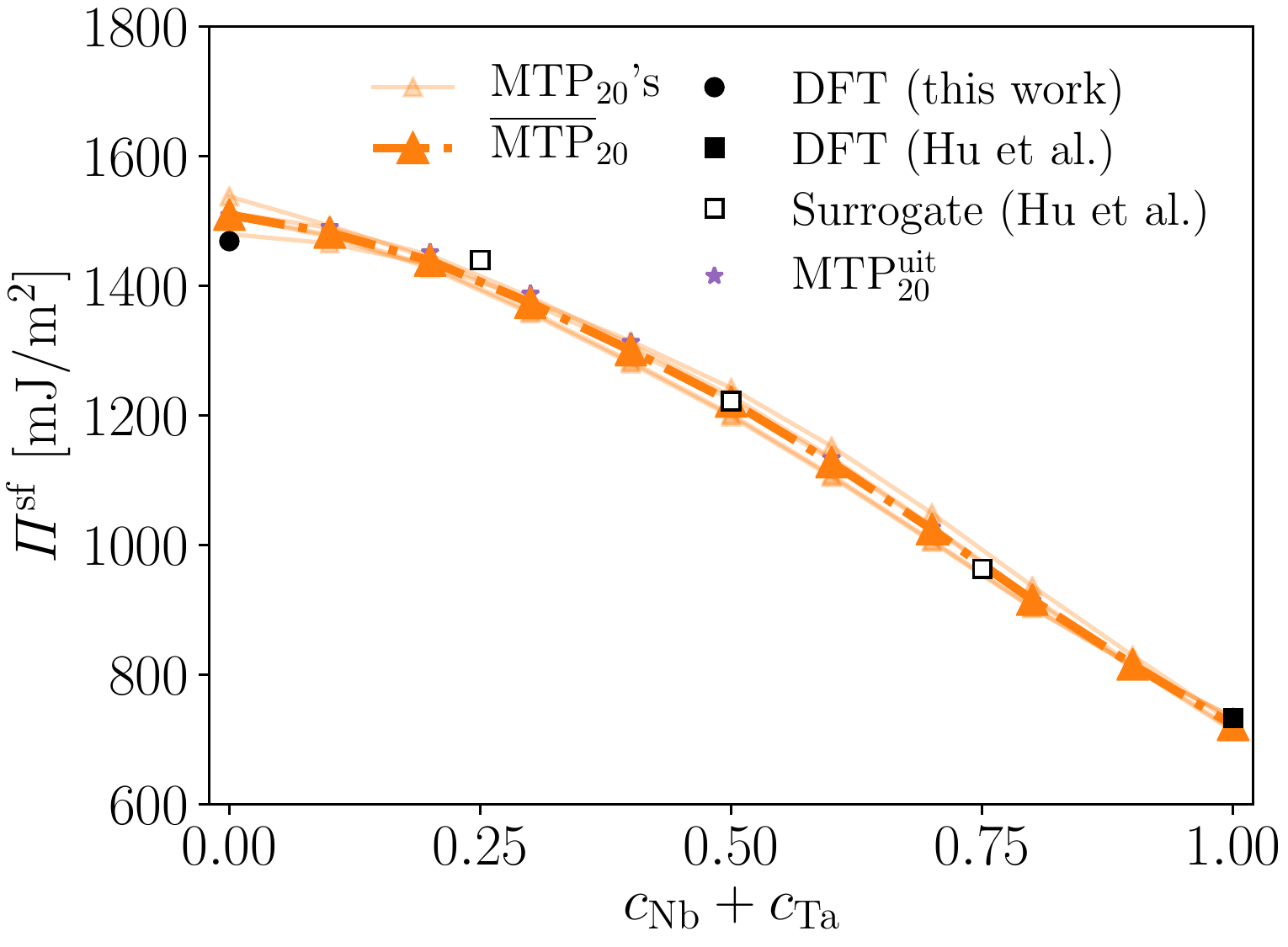}
 \end{minipage}\hfill
 \begin{minipage}{0.333333\textwidth}
  \centering
  \includegraphics[width=0.97\textwidth]{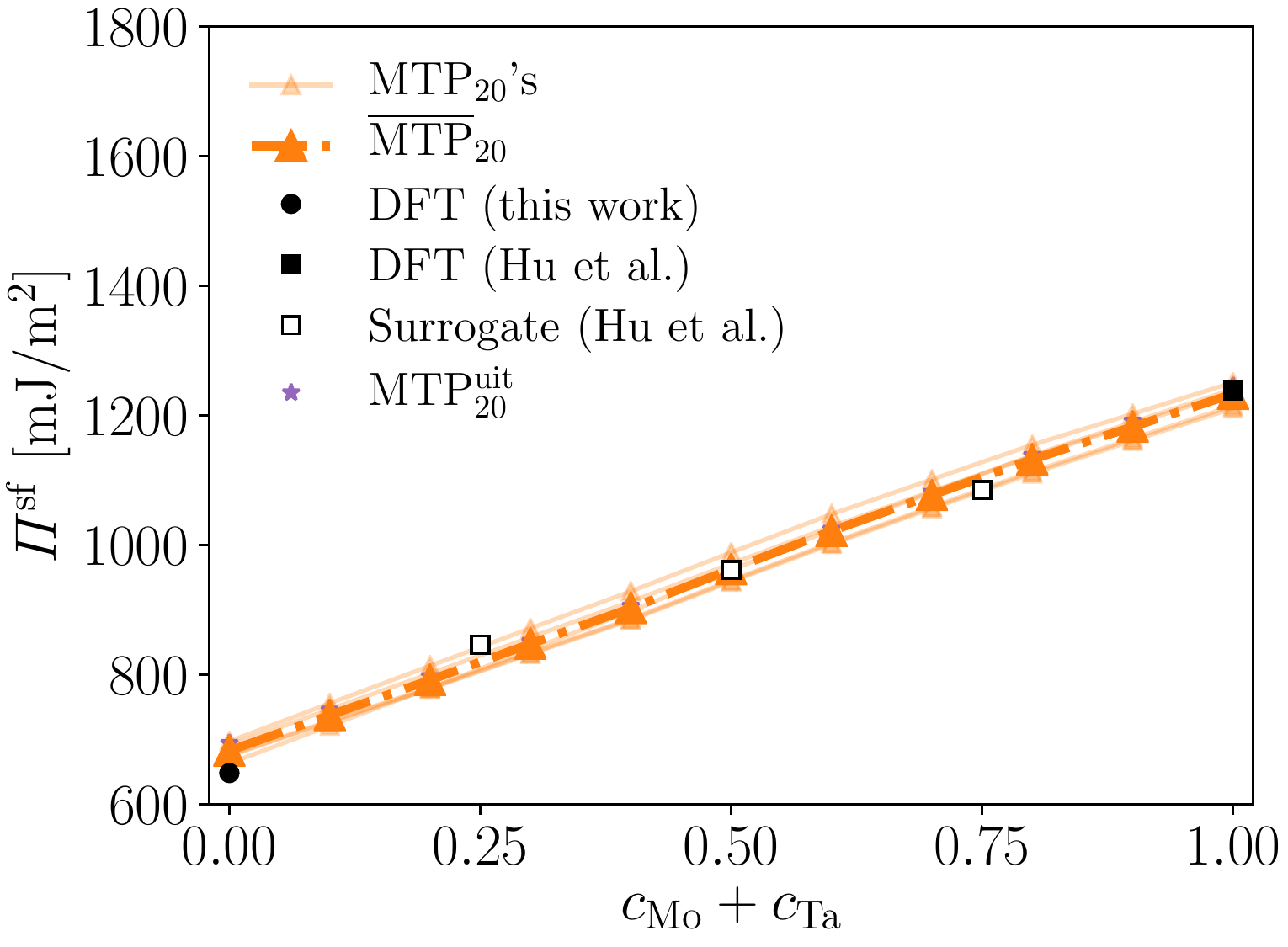}
 \end{minipage}\hfill
 \begin{minipage}{0.333333\textwidth}
  \centering
  \includegraphics[width=0.97\textwidth]{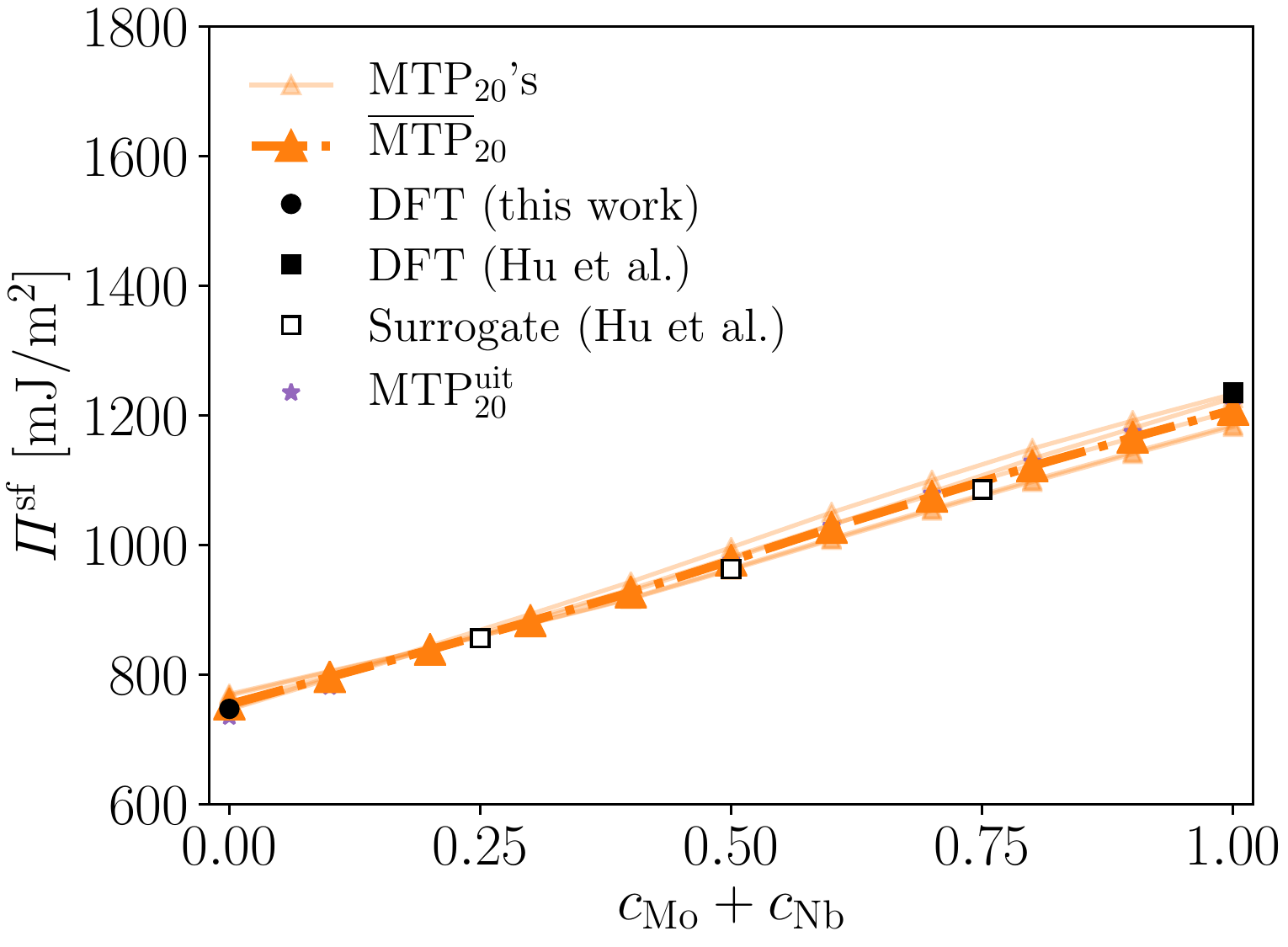}
 \end{minipage}
 \caption{SFEs along the three curves from Figure \ref{fig:spaghetti} as predicted by level-16 MTPs in (a)--(c) and level-20 MTPs in (d)--(f) trained on $\scT_\mrm{MoNbTa}^3$. Both types of MTPs are able to reproduce the SFE, the $\overline{\text{MTP}}_{20}$ practically without error;
 The MTP$_{16}^\mrm{uit}$ and the MTP$_{20}^\mrm{uit}$ refer to the potentials from Section \ref{sec:results.ternary.influence_init_ts} fitted with respect to the training set that has been generated with a uniform initialization (Figure \ref{fig:results_densities_uniform&random} (f))}
 \label{fig:results_MoNbTa_enriched_ts}
\end{figure}

\begin{figure}[t!]
 \centering
 \includegraphics[width=0.5\textwidth]{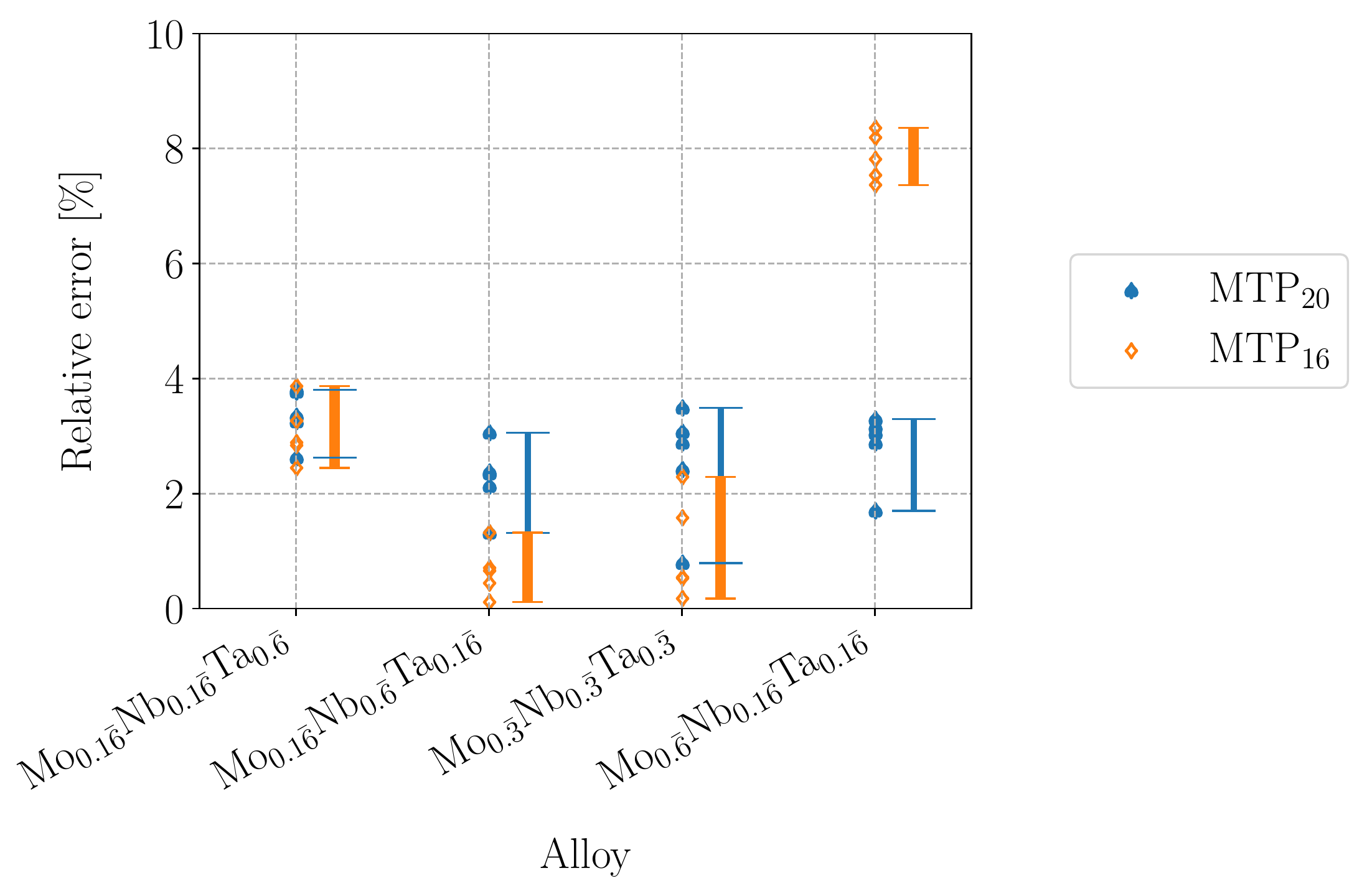}
 \caption{Relative errors in the stacking fault energy of the MTPs trained on $\scT_\mathrm{MoNbTa}^3$ with respect to DFT for several samples of random ternary alloys; The DFT energies were computed by performing single-point calculations on the bulk and stacking fault configurations that have been relaxed with a level-20 MTP.
 The errors are not worse than for the unaries and binaries indicating that also the SFE of \emph{random} ternary configurations (not included in the training set) is well-reproduced by the MTPs}
 \label{fig:results_mtp-dft_comparison_for_ternaries}
\end{figure}

Manually adding unary and binary data fully resolved the problem as shown in Figure \ref{fig:results_MoNbTa_enriched_ts}.
Even the level-16 MTP is now very accurate, showing a maximum difference of not more than 5--10\,\% with respect to the DFT and surrogate values.
An MTP of level 20, trained on $\scT_\mathrm{MoNbTa}^3$, essentially reproduces the DFT and surrogate values without noteworthy error.

\begin{figure}[t!]
 \begin{minipage}{0.333333\textwidth}
  \centering
  (a)
 \end{minipage}\hfill
 \begin{minipage}{0.333333\textwidth}
  \centering
  (b)
 \end{minipage}\hfill
 \begin{minipage}{0.333333\textwidth}
  \centering
  (c)
 \end{minipage}\\
 \begin{minipage}{0.333333\textwidth}
  \centering
  \includegraphics[width=0.97\textwidth]{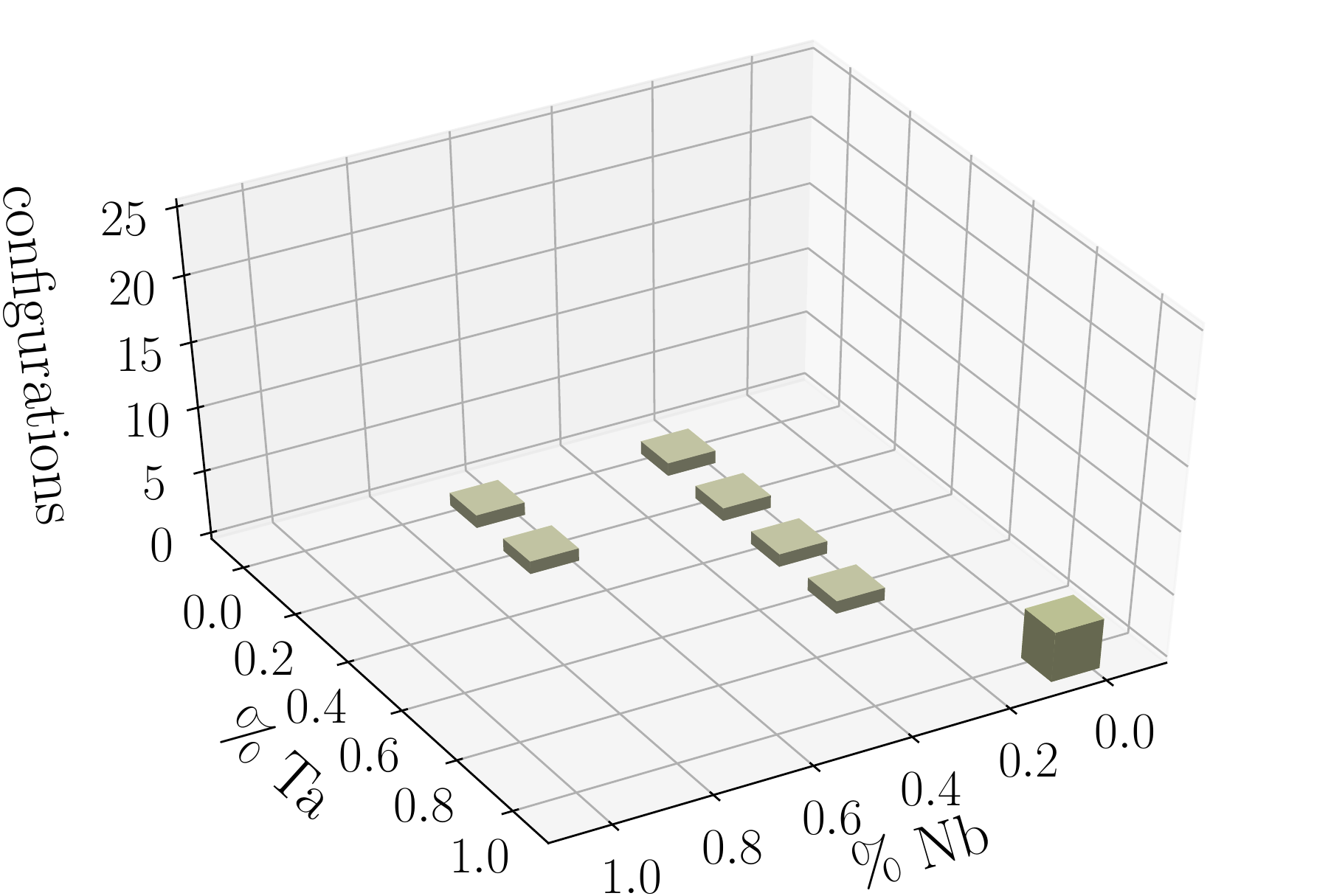}
 \end{minipage}\hfill
 \begin{minipage}{0.333333\textwidth}
  \centering
  \includegraphics[width=0.97\textwidth]{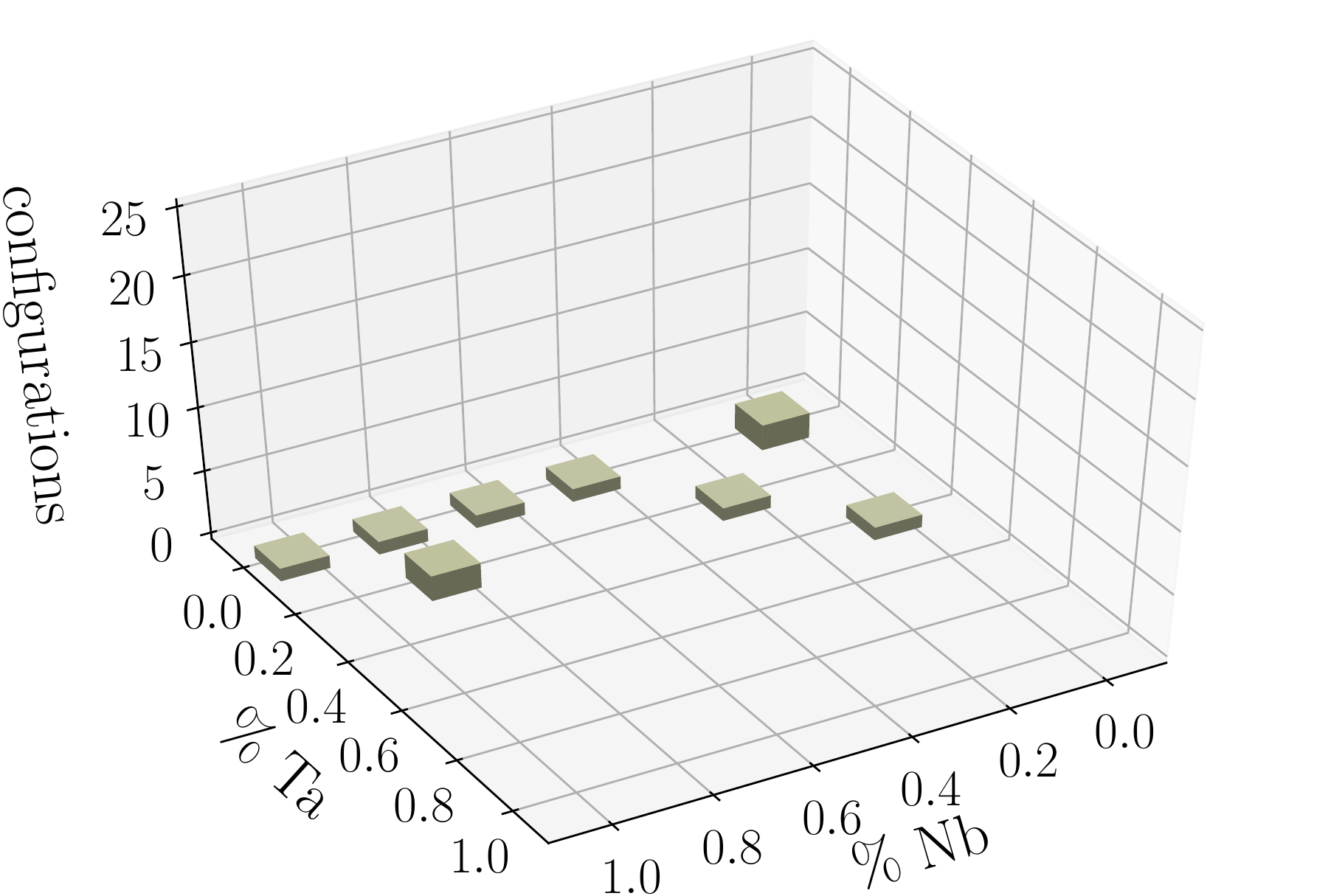}
 \end{minipage}\hfill
 \begin{minipage}{0.333333\textwidth}
  \centering
  \includegraphics[width=0.97\textwidth]{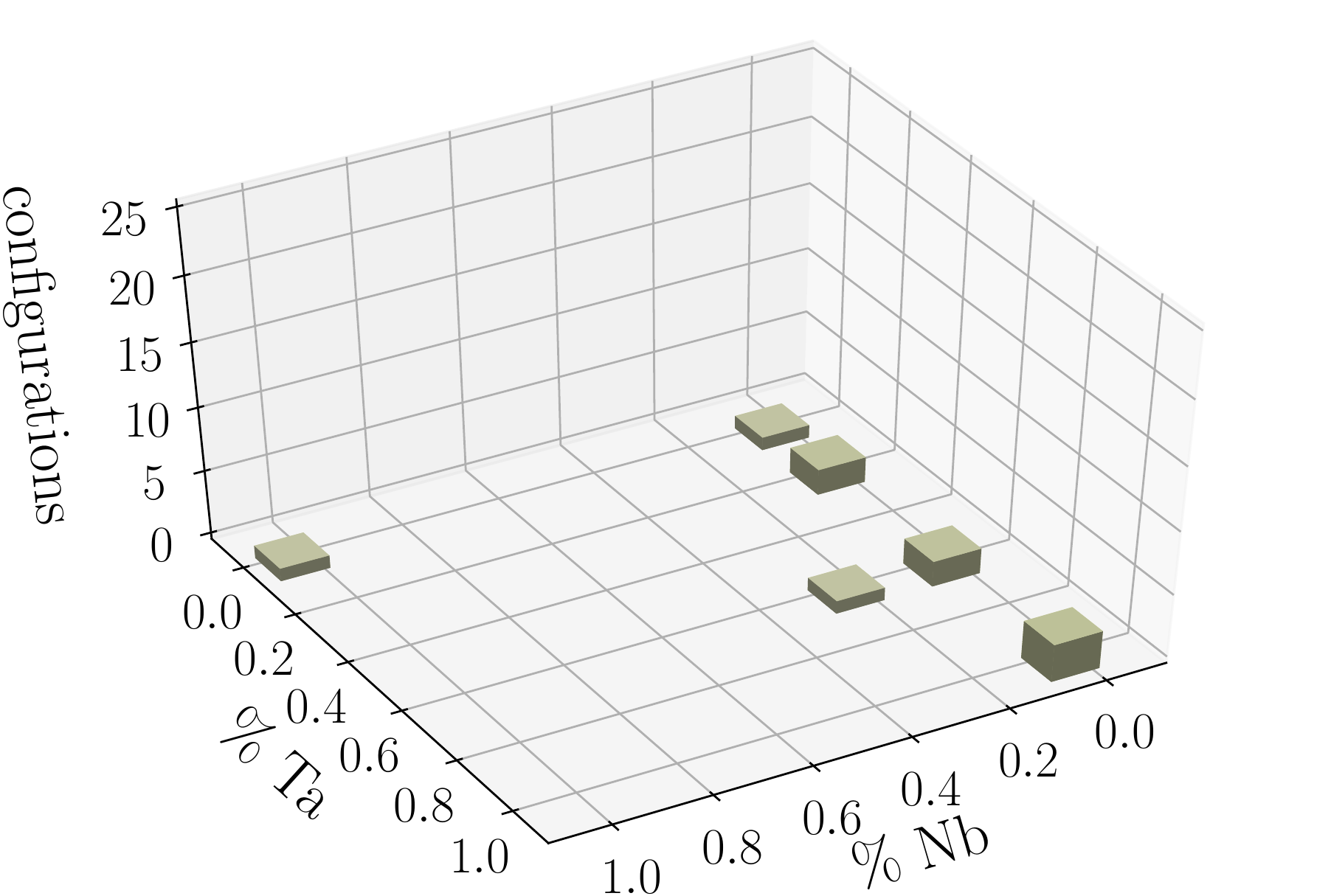}
 \end{minipage}\\[0.5em]
 \begin{minipage}{0.333333\textwidth}
  \centering\large\bf
  \rotatebox[origin=c]{-90}{\ding{224}}
  \,{Active Selection}\,
  \rotatebox[origin=c]{-90}{\ding{224}}
 \end{minipage}\hfill
 \begin{minipage}{0.333333\textwidth}
  \centering\large\bf
  \rotatebox[origin=c]{-90}{\ding{224}}
  \,{Active Selection}\,
  \rotatebox[origin=c]{-90}{\ding{224}}
 \end{minipage}\hfill
 \begin{minipage}{0.333333\textwidth}
  \centering\large\bf
  \rotatebox[origin=c]{-90}{\ding{224}}
  \,{Active Selection}\,
  \rotatebox[origin=c]{-90}{\ding{224}}
 \end{minipage}\\
 \begin{minipage}{0.333333\textwidth}
  \centering
  \includegraphics[width=0.97\textwidth]{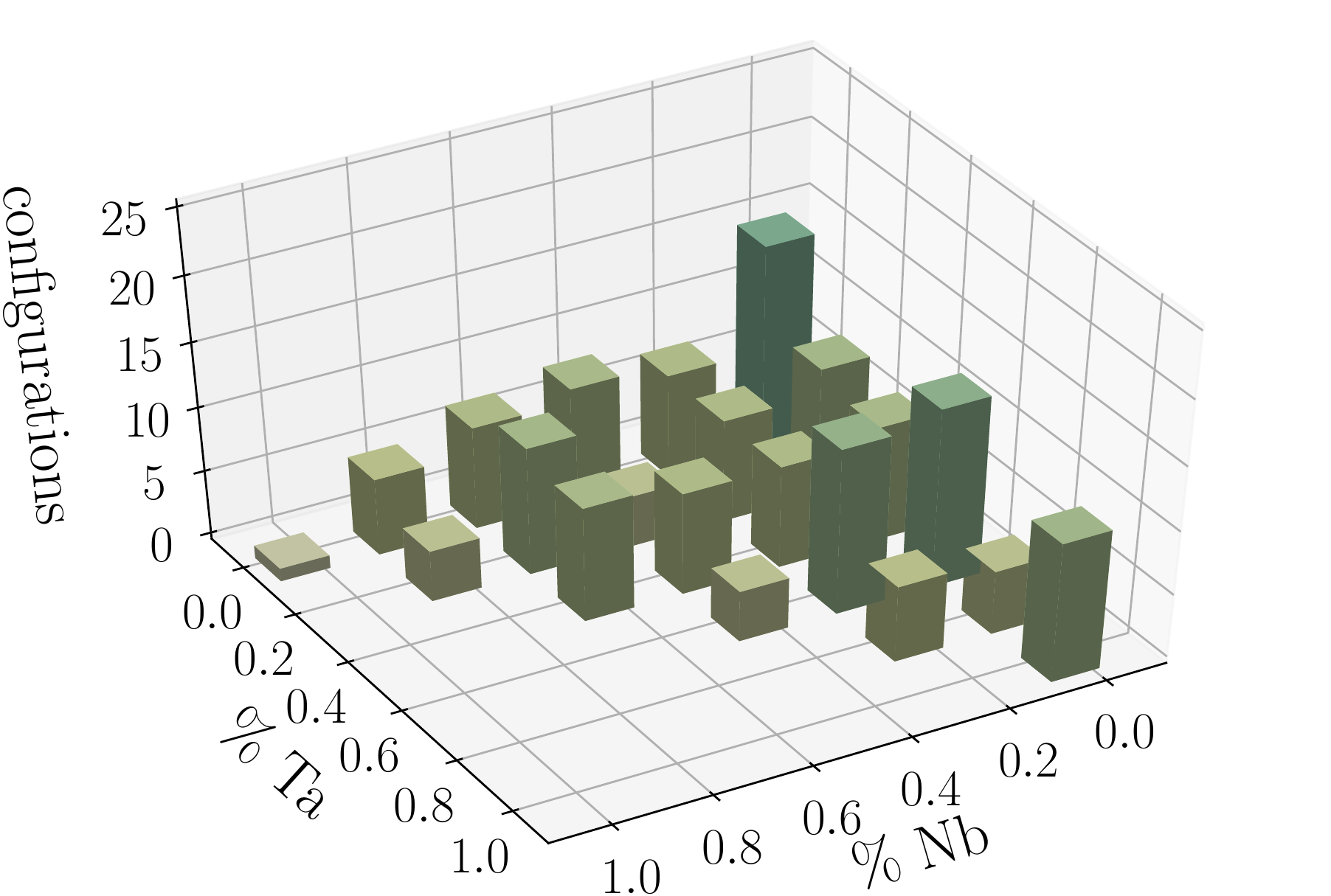}
 \end{minipage}\hfill
 \begin{minipage}{0.333333\textwidth}
  \centering
  \includegraphics[width=0.97\textwidth]{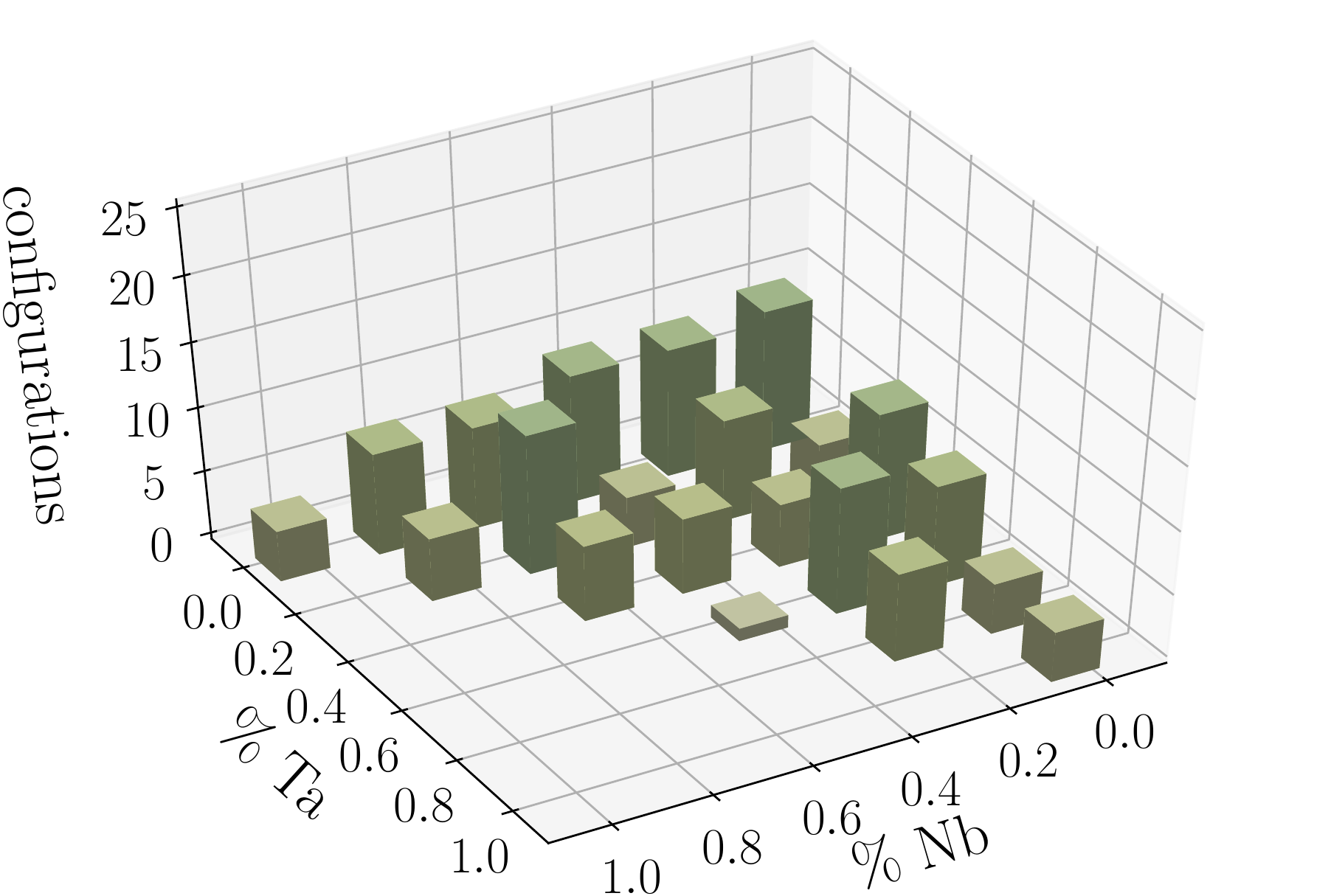}
 \end{minipage}\hfill
 \begin{minipage}{0.333333\textwidth}
  \centering
  \includegraphics[width=0.97\textwidth]{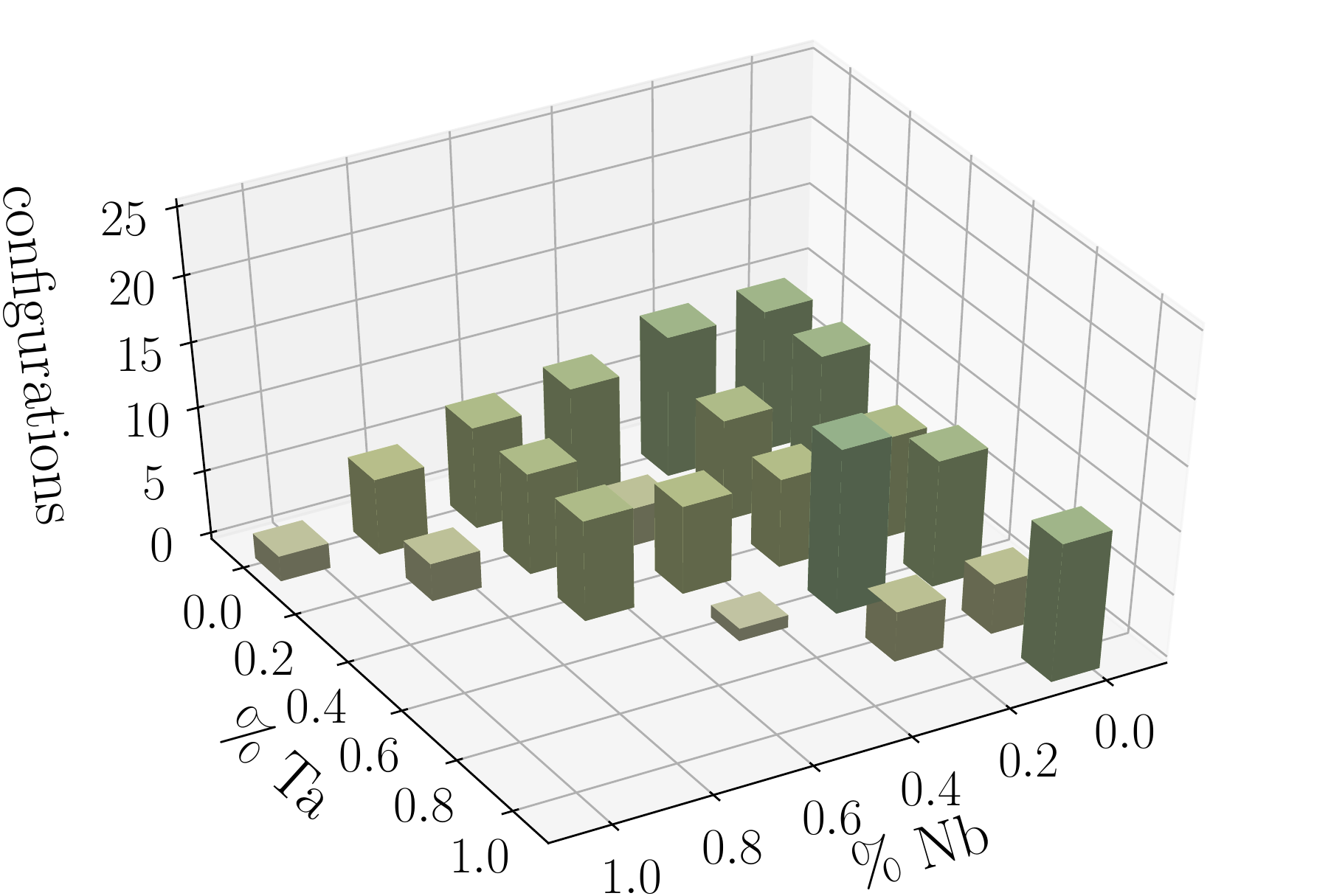}
 \end{minipage}\\[1em]
 \begin{minipage}{0.333333\textwidth}
  \centering
  (d)
 \end{minipage}\hfill
 \begin{minipage}{0.333333\textwidth}
  \centering
  (e)
 \end{minipage}\hfill
 \begin{minipage}{0.333333\textwidth}
  \centering
  (f)
 \end{minipage}\\
 \begin{minipage}{0.333333\textwidth}
  \centering
  \includegraphics[width=0.97\textwidth]{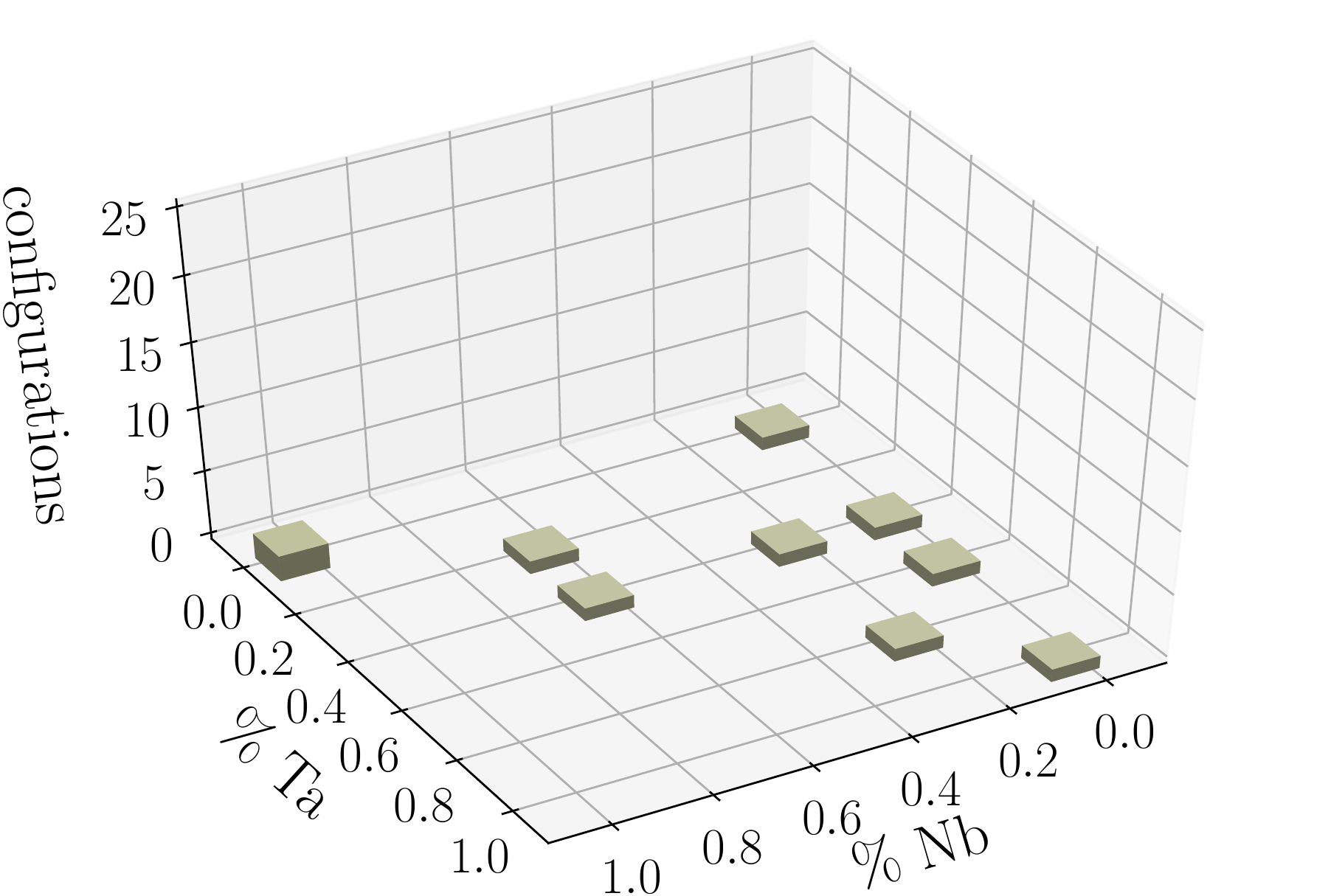}
 \end{minipage}\hfill
 \begin{minipage}{0.333333\textwidth}
  \centering
  \includegraphics[width=0.97\textwidth]{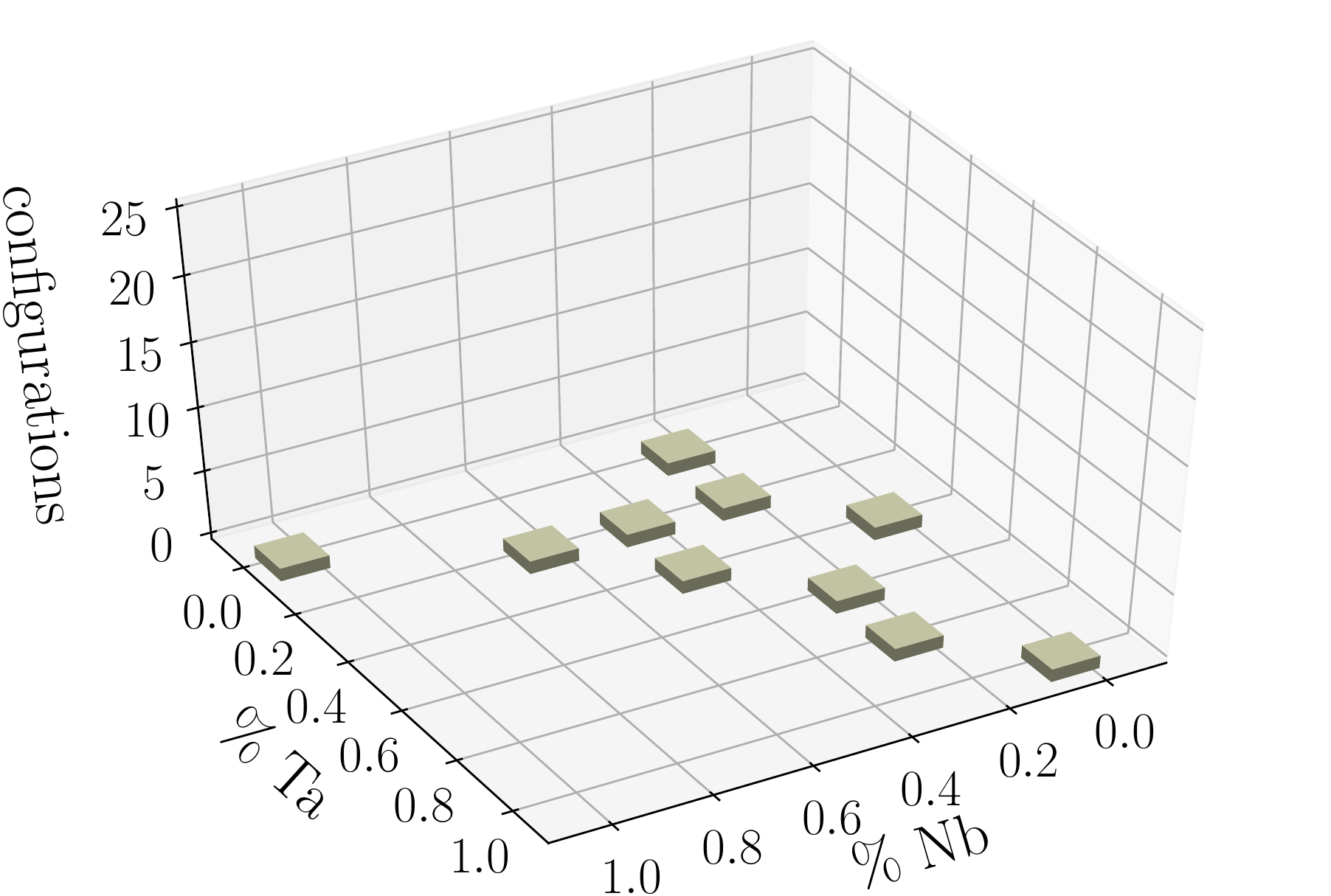}
 \end{minipage}\hfill
 \begin{minipage}{0.333333\textwidth}
  \centering
  \includegraphics[width=0.97\textwidth]{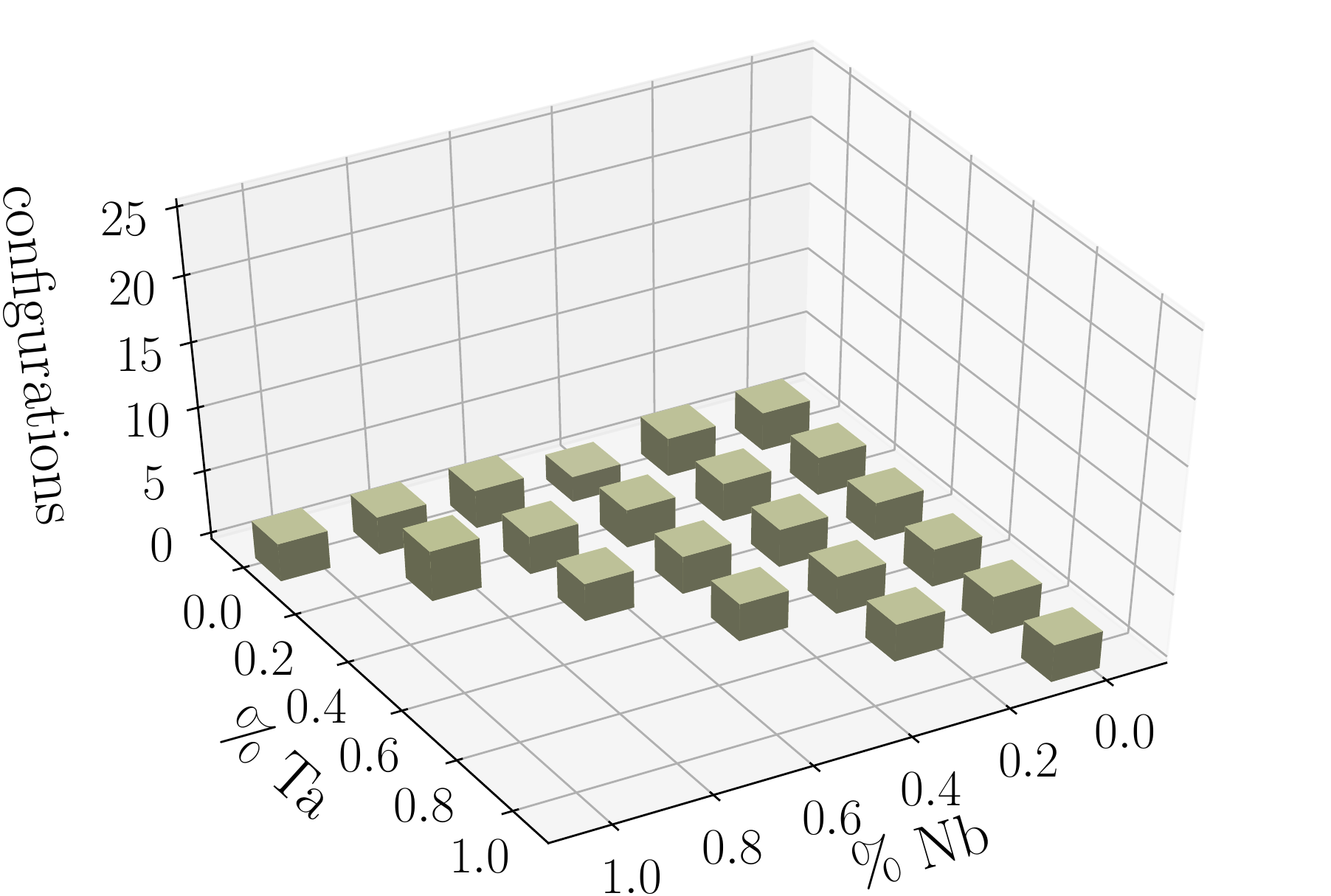}
 \end{minipage}\\[0.5em]
 \begin{minipage}{0.333333\textwidth}
  \centering\large\bf
  \rotatebox[origin=c]{-90}{\ding{224}}
  \,{Active Selection}\,
  \rotatebox[origin=c]{-90}{\ding{224}}
 \end{minipage}\hfill
 \begin{minipage}{0.333333\textwidth}
  \centering\large\bf
  \rotatebox[origin=c]{-90}{\ding{224}}
  \,{Active Selection}\,
  \rotatebox[origin=c]{-90}{\ding{224}}
 \end{minipage}\hfill
 \begin{minipage}{0.333333\textwidth}
  \centering\large\bf
  \rotatebox[origin=c]{-90}{\ding{224}}
  \,{Active Selection}\,
  \rotatebox[origin=c]{-90}{\ding{224}}
 \end{minipage}\\
 \begin{minipage}{0.333333\textwidth}
  \centering
  \includegraphics[width=0.97\textwidth]{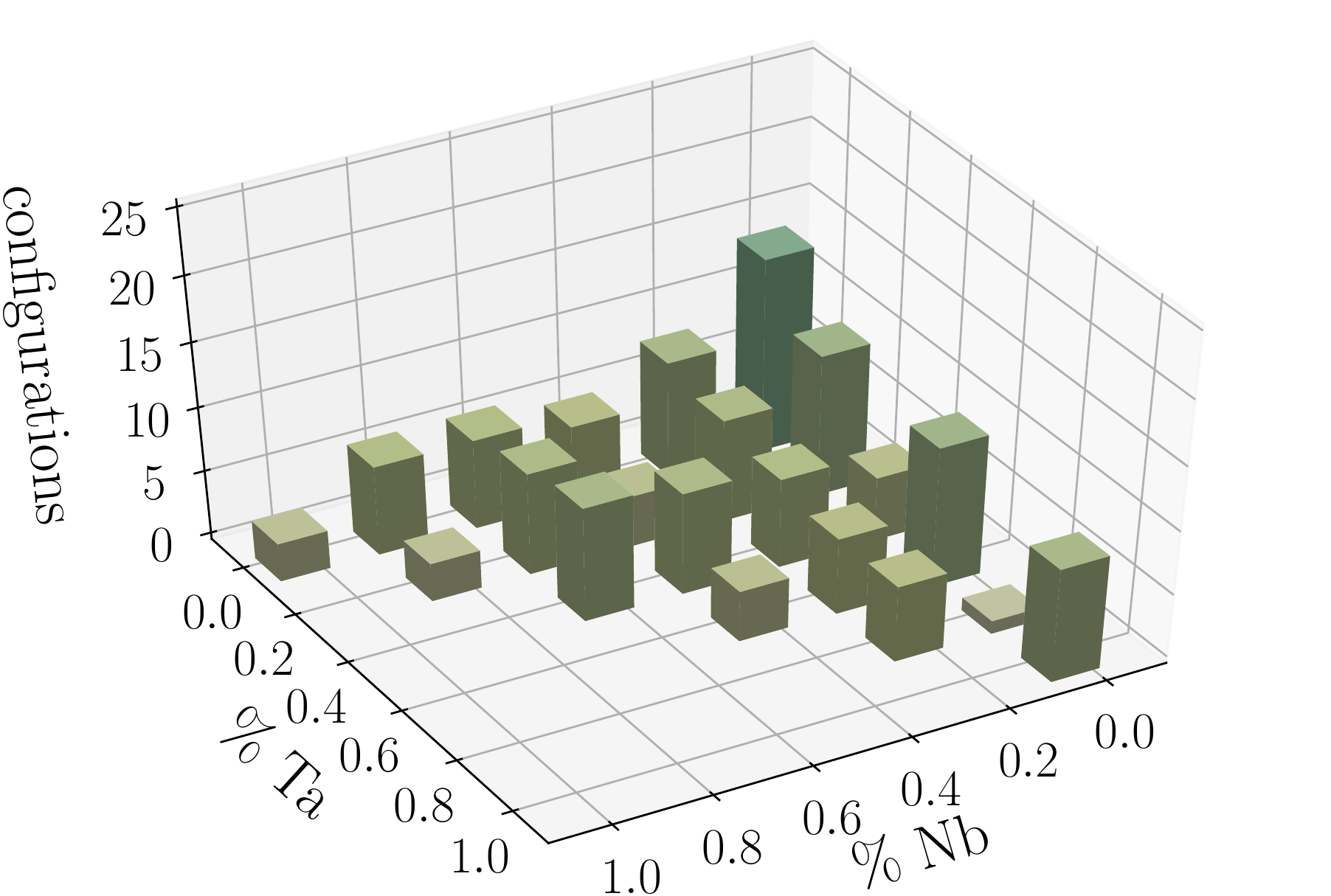}
 \end{minipage}\hfill
 \begin{minipage}{0.333333\textwidth}
  \centering
  \includegraphics[width=0.97\textwidth]{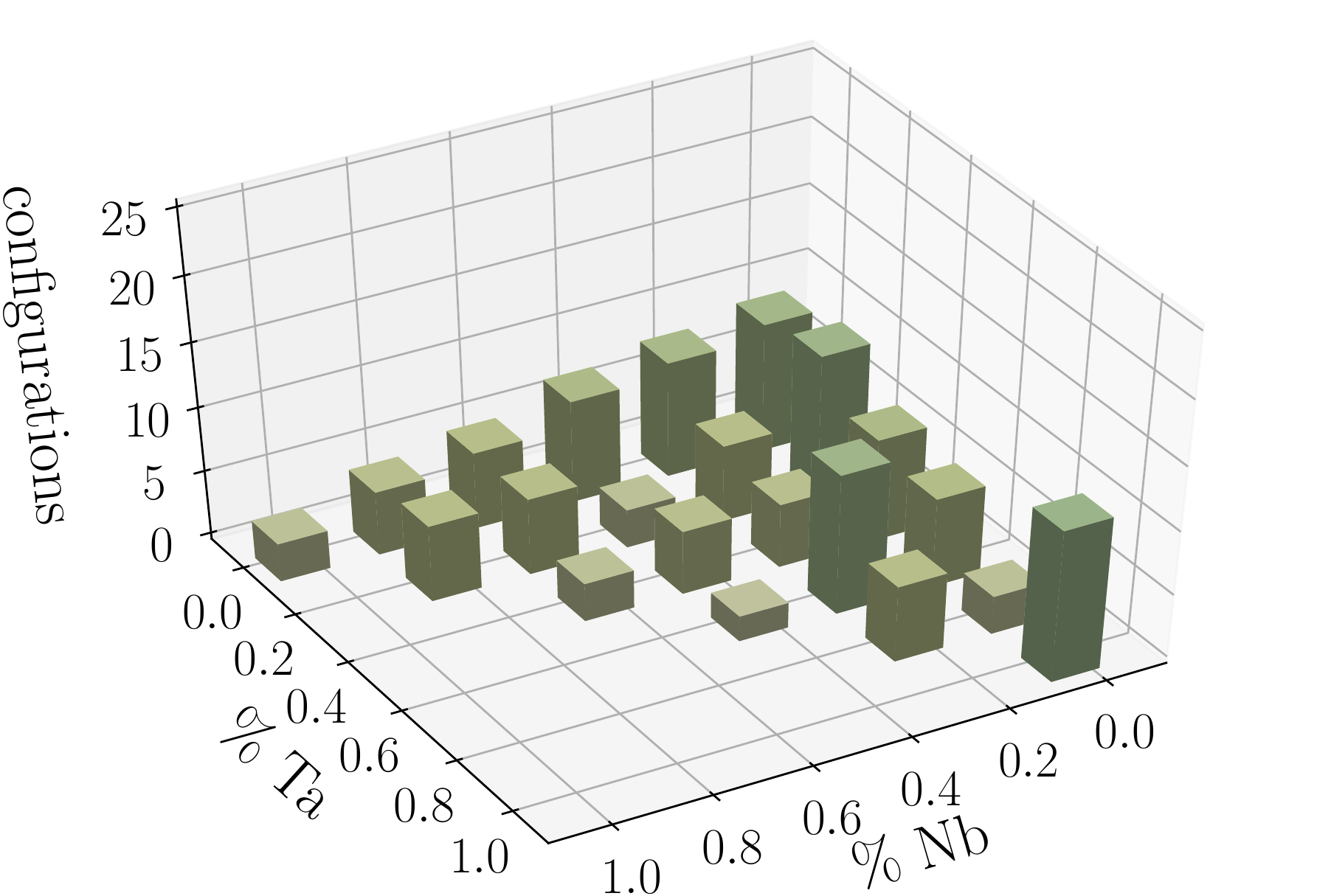}
 \end{minipage}\hfill
 \begin{minipage}{0.333333\textwidth}
  \centering
  \includegraphics[width=0.97\textwidth]{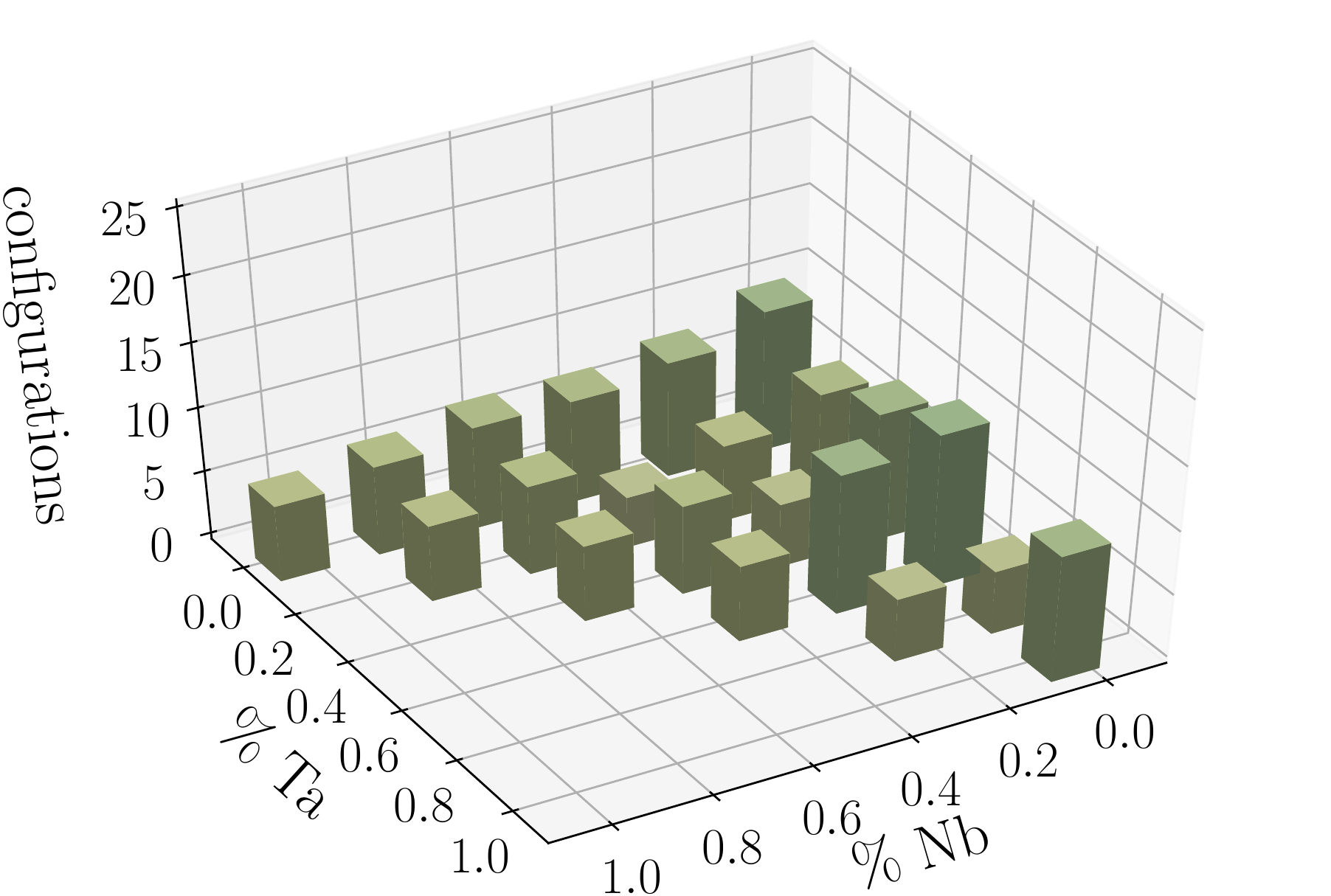}
 \end{minipage}
 \caption{Initial training sets and corresponding final training sets obtained after an active selection of configurations from $\scT_\mathrm{MoNbTa}^3$.
 Random initial sampling (a)--(e) raises the danger of undersampling one or more unaries, e.g., in (a).
 Uniform initial sampling (f) resolves this issue leading to a reliable final training set: an MTP trained on the final training set in (f) predicts the SFE over the entire compositional space (see Figure \ref{fig:results_MoNbTa_enriched_ts}).}
 \label{fig:results_densities_uniform&random}
\end{figure}

In addition, we have validated the MTPs trained on $\scT_\mathrm{MoNbTa}^3$ with respect to single-point DFT calculations for several ternary compositions.
For this purpose, we have generated several random samples for various ternary bulk and stacking fault configurations that were not included in the active learning algorithm and relaxed them with a level-20 MTP.
We then performed single-point DFT calculations on these relaxed configurations and computed the relative error of the MTP with respect to the DFT stacking fault energies.
As shown in Figure \ref{fig:results_mtp-dft_comparison_for_ternaries}, the errors of the MTPs are of a few percent, so, in the same range as for the unaries and binaries.
This further demonstrates the effectiveness of the MTPs in predicting the stacking fault energies of \emph{random} configurations---in addition to predicting average energies.
This should be important for performing follow-up studies, such as investigating preferences for chemical ordering.

\subsubsection{Influence of the initialization on the training set}
\label{sec:results.ternary.influence_init_ts}

We now investigate the effect of initializing the active learning algorithm with different training sets on the final training set.
To that end, we considered $\scT_\mathrm{MoNbTa}^3$ as our training candidate set, randomly generated five different initial training sets, and then run an active selection of extrapolative configurations from $\scT_\mathrm{MoNbTa}^3$.
The initial and final training sets are shown in Figure \ref{fig:results_densities_uniform&random} (a)--(e).
While there appears to be a trend of leaving out more Nb than Mo or Ta, the precise amount indeed depends on the initial training set.
For example, an Nb-rich initial training set, as the one in Figure \ref{fig:results_densities_uniform&random} (b), yields a final training set with more added Nb configurations than using an Nb-poor initial training set, e.g., the one in (a).
This behavior is in agreement with our hypothesis from the previous section relating the absence of Nb to potentially missing far-field contributions to the atomic energies in our active learning algorithm.
In fact, it can also be observed for Ta (compare the Ta-poor training sets in \ref{fig:results_densities_uniform&random} (b) with the other ones), although less pronounced.

To ensure that this behavior does not yield erroneous predictions, another possibility---as opposed to adding additional unary and binary data---is to start from a uniformly sampled initial training set.
In this case, the active learning algorithm selects a much more uniform training set, as shown in Figure \ref{fig:results_densities_uniform&random} (f).
While this training set still contains more Mo and Ta atoms, the MTPs trained on this training set perform comparable to those trained on $\scT_\mathrm{MoNbTa}^3$, as shown in Figure \ref{fig:results_MoNbTa_enriched_ts}.
We thus conclude that running our algorithm with uniform initial training sets, or manually supplying the training set with unary and binary configurations, as done in the previous section, are equally valid possibilities to construct MTPs that reproduce the DFT stacking fault energy over the entire compositional space.

\section{Concluding remarks}
\label{sec:discussion}

\subsection{Discussion and future usage}

We have presented an efficient, automatized method for predicting stacking fault energies of random alloys using moment tensor potentials (MTPs).
In particular, we have shown that a relatively small number of configurations of the order of 100 suffices to be computed with \emph{single-point} DFT calculations in order to construct a training set for MTPs which predict the stacking fault energy of the ternary MoNbTa random alloy at arbitrary composition.
The method is thus not only much more efficient than (pure-DFT) SQS-based methods, but also appears tractable to be applied to a whole series of new problems as we discuss below.

Our algorithm of generating the training set is based on the D-optimality-based active sampling, plus a correction so that enough unary and binary configurations are present in the training set.
The necessity of the latter is demonstrated by the fact that D-optimality did not automatically add the pure-Nb configurations to the training set, asserting that pure-Nb atomic environments are interpolative with respect to the binary and ternary environments---however, the error for pure Nb was larger than for other compositions.
We have attributed this to the far-field contribution to atomic energies which is currently missing in our algorithm.
Fortunately, we were able to overcome this by going beyond the active learning algorithm and merging the ternary data with the binary and unary ones.
We point out that this procedure \emph{can be automatized} by performing several iterations of the algorithm, i.e., by first running it for all the binaries, and then the ternary.
Another option would be to start with an initial training set consisting of configurations that are uniformly sampled from the entire compositional space, instead of randomly selecting them from the training candidate set.

Motivated by these promising results, we think that our method potentially allows for what cannot be achieved thus far with state-of-the-art DFT-based methods:
a rapid screening of random alloy properties over the \emph{entire compositional space}.
That is, we anticipate that our method is general and can be applied to predict many other interesting properties of random alloys, for example:
\begin{itemize}
 \item
 Generalized stacking fault curves\\
 \textit{In addition to the configurations with $\frac{1}{4}[111]$ stacking faults, one needs to add configurations with intermediate $X[111]$ stacking faults, where $X$ is some value between $0$ and $\frac{1}{4}$.}
 \item
 Surface energies\\
 \textit{One needs to consider configurations with free surfaces.
 This can simply be done by adding vacuum regions to the bulk configurations.}
 \item
 Grain boundary energies\\
 \textit{One needs to consider configurations containing grain boundaries.
 If the grain boundary angle is too large and, therefore, requires configurations with 300--500 atoms or more to be considered, one possibility would be to first create a training set using smaller angles, and then try to enrich it with a small number of larger configurations with higher angles.}
 \item
 Chemical ordering\\
 \textit{With MTPs it is computationally feasible to incorporate Monte Carlo methods in our algorithm to investigate the effects of short-range order after computing the property of interest in {\rm\textbf{Step\;5}}.}
\end{itemize}

We are planning to explore this in future work and, in tandem, investigate the performance of our algorithm when using random alloys with more than three chemical elements.

\subsection{Application to large-scale problems}

While the MTPs fitted to the training sets generated in the present work mainly serve for modeling stacking faults,
our algorithm can be integrated into automatized training protocols for constructing MTPs for general large-scale simulations, possibly containing arbitrarily many arrangements of defects.
This can be achieved by enriching the set of training candidates with other types of configurations in the training candidate set, e.g., configurations containing vacancies, grain boundaries, etc.

For example, for the particular case of modeling screw dislocations in bcc random alloys, one possibility we envision is the following two-step algorithm.
In the first step, we add configurations with intermediate stacking faults along the slip plane to the training candidate set---in addition to the bulk configurations and the configurations containing unstable stacking faults.
We then run our algorithm using this training candidate set in order to generate the (initial) training set for the MTP.
In the second step, we start the simulation of a screw dislocation using this MTP and, while running the simulation, continue to measure its extrapolation grade.
Should the extrapolation grade exceed some threshold, we extract the few extrapolative neighborhoods, complete them to periodic configurations using our ``in operando active learning'' technique \citep{hodapp_operando_2020}, retrain the MTP on the periodic configurations, and restart the simulation.
We think that such an algorithm has the potential to become a very accurate and efficient means to compute properties of dislocations over the entire compositional space, inaccessible with existing methods, which can then be fed into strengthening \citep{varvenne_theory_2016,maresca_theory_2020} or ductility \citep{mak_ductility_2021} models.


\section*{Acknowledgements}

M.\;Hodapp acknowledges the financial support from the Swiss National Science Foundation (project 191680).
A.\;Shapeev acknowledges the financial support from the Russian Science Foundation (grant number 18-13-00479).

\section*{Data availability}

The implementation of the MTPs is included in the MLIP package which is publicly available for academic use at \url{https://mlip.skoltech.ru/download/} upon registration.
Additional scripts, necessary to run our algorithm, as well as the training data, are available from the authors upon reasonable request.


\begin{appendix}

\section{Functional form of MTPs and training procedure}
\label{sec:appdx.mtp}

The MTP basis functions from \eqref{eq:Eatom} are given by \citep{shapeev_moment_2016,gubaev_accelerating_2019}
\begin{equation}
 B_\alpha(\Neigh;\{\mtheta_\beta\}) = \prod_{l=1}^k M_{\mu_l(\alpha),\nu_l(\alpha)}(\Neigh;\{\mtheta_\beta\}),
\end{equation}
with the moment tensor descriptors
\begin{equation}
 M_{\mu_l,\nu_l}(\Neigh;\{\mtheta_\beta\}) = \sum_{\Atom_{ij} \in \Neigh} f_{\mu_l}(\abs{\bmr_{ij}},X_i,X_j;\{\mtheta_\beta\})
 \, \underbrace{\bmr_{ij} \otimes \cdots \otimes \bmr_{ij}}_{\nu_l \; \text{times}}.
\end{equation}
The moment tensor descriptors depend on Chebyshev radial basis functions $f_{\mu_l}(\abs{\bmr_{ij}},X_i,X_j;\{\mtheta_\beta\})$, where $X_i,X_j$ denotes the type of atom $i,j$, respectively.
The radial basis functions vanish beyond a cut-off radius that we have universally set to 5\,{\AA} in the present work.

The body order of an MTP is characterized by its level.
More precisely, an MTP of level $d$ implies that all possible basis functions which satisfy $(2\mu_1 + \nu_1) + (2\mu_2 + \nu_2) + ... \le d$ are taken into account.
For an MTP of level $d$ we may also use the shorthand notation MTP$_d$.

Training is performed with respect to DFT energies and forces.
That is, given a training set $\scT$ consisting of configurations $\{\Atom_i\}^{\rm tr}$, we compute the parameters $\umtheta$ by minimizing the loss functional
\begin{equation}
 L(\umtheta) =
 C_\rme \sum_{\{ \Atom_i \}^{\rm tr} \in \scT} \left( \Etot(\{ \Atom_i \}^{\rm tr};\umtheta) - \Etot^{\rm qm}(\{ \Atom_i \}^{\rm tr}) \right)^2 +
 C_\rmf \sum_{\{ \Atom_i \}^{\rm tr} \in \scT}\sum_{\Atom_i \in \{ \Atom_i \}^{\rm tr}} \left\| \bforce_{\atom_i}(\{ \Atom_i \}^{\rm tr};\umtheta) - \bforce^{\rm qm}_{\atom_i}(\{ \Atom_i \}^{\rm tr}) \right\|^2,
\end{equation}
where $\bforce_{\atom_i}$, $\bforce^{\rm qm}_{\atom_i}$ are the MTP and quantum mechanical forces on an atom $\Atom_i$ corresponding to a specific configuration $\{ \Atom_i \}^{\rm tr}$ in $\scT$, respectively.
The $C$-constants are regularization parameters which we have set to $C_\rme$\,=\,1 and $C_\rmf$\,=\,0.01\,\AA$^{-2}$.
We may then denote an MTP of level $d$ that has been trained on a training set $\scT$ by MTP$_d(\scT)$.

\section{VASP calculations}
\label{sec:appdx.vasp}

The parameters we have used in all our VASP calculations of bcc Mo, Nb, and Ta, are given in Table \ref{tab:VASP}. The energy cut-off was chosen to be 1.5 times the default value. Self-consistent electronic relaxation was performed using the preconditioned minimal residual method, as implemented in VASP, and terminated when the energy difference between two subsequent iterations was less than 10$^{-4}$\,eV.

\begin{table}[hbt]
 \centering
 \begin{tabular}{|c||c|c|c|}
  \hline
  \multirow{2}{*}{Option} & \multicolumn{3}{c|}{Element} \\
  & Mo & Nb & Ta \\ \hline\hline
  Exchange-correlation & \multicolumn{3}{c|}{PE generalized gradient approximation \citep{perdew_generalized_1996}} \\ \hline
  PAW potential & PAW\_PBE Mo 08Apr2002 & PAW\_PBE Nb\_pv 08Apr2002 & PAW\_PBE Ta 17Jan2003 \\ \hline
  Energy cut-off & 336.876\,eV & 312.912\,eV & 335.501\,eV \\ \hline
  $\bmk$-point spacing & \multicolumn{3}{c|}{0.15\,\AA$^{-1}$} \\ \hline
  Smearing width & \multicolumn{3}{c|}{0.06\,eV} \\ \hline
 \end{tabular}
 \caption{VASP parameters}
 \label{tab:VASP}
\end{table}

\end{appendix}


\section*{References}
\bibliographystyle{elsarticle-harv}
\bibliography{references}

\end{document}